\documentclass[11pt]{article}
\usepackage{amsmath,amsthm,latexsym,amssymb,amsfonts,epsfig}

\newcommand{\Q}{\sqrt{\det{Q}}}
\newcommand{\Qb}{\sqrt{\det{\overline{Q}}}}

\newcommand{\ov}[1]{\overline{#1}}

\makeatletter
\@addtoreset{equation}{section}
\makeatother

\def\be{\begin{equation}}
\def\ee{\end{equation}}
\def\ba{\begin{eqnarray}}
\def\ea{\end{eqnarray}}
\def\Nl{{\mathchoice
{\setbox0=\hbox{$\displaystyle\rm N$}\hbox{\hbox to0pt
{\kern0.4\wd0\vrule height0.9\ht0\hss}\box0}}
{\setbox0=\hbox{$\textstyle\rm N$}\hbox{\hbox to0pt
{\kern0.4\wd0\vrule height0.9\ht0\hss}\box0}}
{\setbox0=\hbox{$\scriptstyle\rm N$}\hbox{\hbox to0pt
{\kern0.4\wd0\vrule height0.9\ht0\hss}\box0}}
{\setbox0=\hbox{$\scriptscriptstyle\rm N$}\hbox{\hbox to0pt
{\kern0.4\wd0\vrule height0.9\ht0\hss}\box0}}}}
\def\Zl{{\mathchoice
{\setbox0=\hbox{$\displaystyle\rm Z$}\hbox{\hbox to0pt
{\kern0.4\wd0\vrule height0.9\ht0\hss}\box0}}
{\setbox0=\hbox{$\textstyle\rm Z$}\hbox{\hbox to0pt
{\kern0.4\wd0\vrule height0.9\ht0\hss}\box0}}
{\setbox0=\hbox{$\scriptstyle\rm Z$}\hbox{\hbox to0pt
{\kern0.4\wd0\vrule height0.9\ht0\hss}\box0}}
{\setbox0=\hbox{$\scriptscriptstyle\rm Z$}\hbox{\hbox to0pt
{\kern0.4\wd0\vrule height0.9\ht0\hss}\box0}}}}
\def\Ql{{\mathchoice
{\setbox0=\hbox{$\displaystyle\rm Q$}\hbox{\hbox to0pt
{\kern0.4\wd0\vrule height0.9\ht0\hss}\box0}}
{\setbox0=\hbox{$\textstyle\rm Q$}\hbox{\hbox to0pt
{\kern0.4\wd0\vrule height0.9\ht0\hss}\box0}}
{\setbox0=\hbox{$\scriptstyle\rm Q$}\hbox{\hbox to0pt
{\kern0.4\wd0\vrule height0.9\ht0\hss}\box0}}
{\setbox0=\hbox{$\scriptscriptstyle\rm Q$}\hbox{\hbox to0pt
{\kern0.4\wd0\vrule height0.9\ht0\hss}\box0}}}}
\def\Rl{{\mathchoice
{\setbox0=\hbox{$\displaystyle\rm R$}\hbox{\hbox to0pt
{\kern0.4\wd0\vrule height0.9\ht0\hss}\box0}}
{\setbox0=\hbox{$\textstyle\rm R$}\hbox{\hbox to0pt
{\kern0.4\wd0\vrule height0.9\ht0\hss}\box0}}
{\setbox0=\hbox{$\scriptstyle\rm R$}\hbox{\hbox to0pt
{\kern0.4\wd0\vrule height0.9\ht0\hss}\box0}}
{\setbox0=\hbox{$\scriptscriptstyle\rm R$}\hbox{\hbox to0pt
{\kern0.4\wd0\vrule height0.9\ht0\hss}\box0}}}}
\def\Cl{{\mathchoice
{\setbox0=\hbox{$\displaystyle\rm C$}\hbox{\hbox to0pt
{\kern0.4\wd0\vrule height0.9\ht0\hss}\box0}}
{\setbox0=\hbox{$\textstyle\rm C$}\hbox{\hbox to0pt
{\kern0.4\wd0\vrule height0.9\ht0\hss}\box0}}
{\setbox0=\hbox{$\scriptstyle\rm C$}\hbox{\hbox to0pt
{\kern0.4\wd0\vrule height0.9\ht0\hss}\box0}}
{\setbox0=\hbox{$\scriptscriptstyle\rm C$}\hbox{\hbox to0pt
{\kern0.4\wd0\vrule height0.9\ht0\hss}\box0}}}}
\def\Hl{{\mathchoice
{\setbox0=\hbox{$\displaystyle\rm H$}\hbox{\hbox to0pt
{\kern0.4\wd0\vrule height0.9\ht0\hss}\box0}}
{\setbox0=\hbox{$\textstyle\rm H$}\hbox{\hbox to0pt
{\kern0.4\wd0\vrule height0.9\ht0\hss}\box0}}
{\setbox0=\hbox{$\scriptstyle\rm H$}\hbox{\hbox to0pt
{\kern0.4\wd0\vrule height0.9\ht0\hss}\box0}}
{\setbox0=\hbox{$\scriptscriptstyle\rm H$}\hbox{\hbox to0pt
{\kern0.4\wd0\vrule height0.9\ht0\hss}\box0}}}}
\def\Ol{{\mathchoice
{\setbox0=\hbox{$\displaystyle\rm O$}\hbox{\hbox to0pt
{\kern0.4\wd0\vrule height0.9\ht0\hss}\box0}}
{\setbox0=\hbox{$\textstyle\rm O$}\hbox{\hbox to0pt
{\kern0.4\wd0\vrule height0.9\ht0\hss}\box0}}
{\setbox0=\hbox{$\scriptstyle\rm O$}\hbox{\hbox to0pt
{\kern0.4\wd0\vrule height0.9\ht0\hss}\box0}}
{\setbox0=\hbox{$\scriptscriptstyle\rm O$}\hbox{\hbox to0pt
{\kern0.4\wd0\vrule height0.9\ht0\hss}\box0}}}}
\DeclareMathOperator{\HF}{\boldsymbol{\mathsf {H}}}
\DeclareMathOperator{\LF}{\boldsymbol{\mathsf {L}}}

\title{{\sf Manifestly Gauge-Invariant General Relativistic}\\{\sf  Perturbation Theory:
 II. FRW Background and First Order}}

\author{{\sf K. Giesel$^1$}\thanks{{\sf gieskri@aei.mpg.de}}, {\sf S.
Hofmann$^{2,3}$}\thanks{{\sf stefan@nordita.org}},
{\sf T.
Thiemann$^{1,2}$}\thanks{{\sf
thiemann@aei.mpg.de,tthiemann@perimeterinstitute.ca}},
{\sf O. Winkler$^2$}\thanks{{\sf owinkler@perimeterinstitute.ca}} \\
\\
{\sf $^1$ MPI f. Gravitationsphysik, Albert-Einstein-Institut,} \\
{\sf Am M\"uhlenberg 1, 14476 Potsdam, Germany}\\
\\
{\sf $^2$ Perimeter Institute for Theoretical Physics,}\\
{\sf 31 Caroline Street N, Waterloo, ON N2L 2Y5, Canada}\\
\\
{\sf $^3$ NORDITA,}\\
{\sf Roslagstullsbacken 23, SE--10691 Stockholm, Sweden}}
\date{{\small\sf Preprint AEI-2007-151}}

\begin{document}

\maketitle

\begin{abstract}
{\sf In our companion paper we identified a complete set of
manifestly gauge-invariant observables for general relativity. This
was possible by coupling the system of gravity and matter to
pressureless dust which plays the role of a dynamically coupled observer. The
evolution of those observables is governed by a physical Hamiltonian
and we derived the corresponding equations of motion. Linear
perturbation theory of those equations of motion around a general
exact solution in terms of manifestly gauge invariant perturbations
was then developed.

In this paper we specialise our previous results to an FRW
background which is also a solution of our modified equations of
motion. We then compare the resulting equations with those derived
in standard cosmological perturbation theory (SCPT). We exhibit the
precise relation between our manifestly gauge-invariant
perturbations and the linearly gauge-invariant variables in SCPT. We
find that our equations of motion can be cast into SCPT form plus
corrections. These corrections are the trace that the dust leaves on
the system in terms of a conserved energy momentum current density.
It turns out that these corrections decay, in fact, in the late
universe they are negligible whatever the value of the conserved
current.

We conclude that the addition of dust which serves as a test
observer medium, while implying modifications of Einstein's
equations without dust, leads to acceptable agreement with known
results, while having the advantage that one now talks about
manifestly gauge-invariant, that is measurable, quantities, which
can be used even in perturbation theory at higher orders. }
\end{abstract}

\newpage

\tableofcontents

\newpage

\section{Introduction}
\label{s1}

In our companion paper \cite{1} we combined the framework of
relational observables \cite{2,3} with the Brown -- Kucha{\v r}
mechanism \cite{4} in order to cast general relativity (including
all known physical matter) into the form of an ordinary Hamiltonian
system, with a true gauge invariant Hamiltonian which only depends
on the gauge invariant geometry and matter degrees of freedom. From
that foundation we then developed a manifestly gauge-invariant
perturbation theory.

The motivation \cite{5} was the trivial observation that general
relativity is a gauge theory (the gauge group being the
diffeomorphism group). Therefore, neither metric nor matter fields
are directly observable since they are not gauge invariant and
likewise Einstein's equations do not describe physical evolution of
observables but rather the behaviour of non -- observables under
gauge transformations. On the other hand, we would like to think of
Einstein's equations as describing the physical time evolution of
observables. The latter point of view is often justified by the
argument that the universe could be filled, in principle, by
geodesic test observers who serve as a material reference system.
The achievement of \cite{4} is to implement that idea field
theoretically by adding pressureless dust to the system which serves
as the medium of test observers. What we did in \cite{1} is to
extend that idea further and thus to arrive at a fully gauge
invariant description with a true dynamics of gauge invariant
observables which directly correspond to the usual metric and matter
fields.

Instead of Einstein's equations, we now have the Hamiltonian
equations for those fundamental fields. The interesting question is
then how close these equations come to the Einstein equations. In
\cite{1} we showed that by identifying suitable fields, one can
almost exactly match our equations of motion with Einstein's.
However, there are corrections which are due to the presence of the
dust. More precisely, the dust itself no longer plays a role at the
level of the physical observables, however it implies the existence
of a conserved energy momentum current density with appropriate
conservation laws which substitute the usual initial value
constraints of General Relativity. This is potentially dangerous
since, due to the absence of the constraints, our theory has four
more physical configuration degrees of freedom than it would have
without dust. Hence our theory should predict that these modes are
either not excited or have decayed in order to agree with standard
General Relativity (two physical degrees of freedom) plus the
observationally confirmed matter of the standard model. Of course,
it would be intriguing if one would find those additional modes in
an experiment which then would provide indirect experimental
evidence for the existence of the dust.

How much these corrections and additional degrees of freedom affect the
results on the solutions of
Einstein's equations is difficult to decide in full generality and
analytically, therefore we will resort as usual to perturbation theory.
In \cite{1} we applied linear Hamiltonian perturbation theory to
our system of Hamiltonian equations of motion. That is, one takes a
general exact solution and then perturbs the equations of motion to
linear order in the perturbations. Notice that in contrast to usual
perturbation theory our perturbations are manifestly gauge invariant,
they are full (diffeomorphism invariant) observables.

In this paper we specialise the formulae found in \cite{1} to an FRW
background which is an exact solution to our equations of motion.
This allows us to study linear cosmological perturbation theory.
Needless to say, this is a topic of fundamental importance in modern
cosmology which goes back all the way to the pioneering work by
Lifshitz \cite{oliverlif}. Early on, it was realized that the
problem of gauge freedom poses serious challenges for the use and
interpretation of perturbation theory in general relativity, in
general, and cosmology, in particular. A significant breakthrough
was the construction of quantities that are gauge-invariant to
linear order \cite{oliverbar}. Based on these, a satisfactory
framework for linear perturbation theory for cosmology was developed
and then applied by many authors. Authoritative review articles can
be found in \cite{oliverkod,6}. For the purpose of this paper, we
will call the standard approach as presented in these references
"standard cosmological perturbation theory" and abbreviate it as
SCPT. In that context, we will always use the notation of \cite{6}.
While there seems to be little room left for significant
improvements at the linear level, developing perturbation theory to
higher orders is far from fully established, due to the challenge of
finding gauge-invariant quantities. Some recent approaches can be
found in \cite{oliver15,oliver7,oliver8,oliver9,oliver13}. That the
question of higher order perturbations is far from academic can be
seen, for instance, from the growing interest in the issue of
non-Gaussianity of cosmological perturbations \cite{oliver20}. This
is where we expect our approach to lead to major progress, due to
its method of implementing gauge-invariance. Before following this
line of investigation, however, we should see the results that our
framework delivers at linear order and compare them with SCPT.

We find that\\
1. We can match our gauge invariant geometry and matter observables
with the linearly gauge invariant ones of SCPT.\\
2. We can cast our equations of motion into SCPT form plus
corrections
which are proportional to the conserved energy momentum tensor current\\
3. The corrections decay in the late universe at least as $1/a$
where $a$ is the physical scale factor and are thus negligible
whatever their value is. This is crucial because otherwise our
theory would predict, even without matter, the coexistence of vector
and scalar modes apart from
the gravitational waves.\\
We conclude that, at least in cosmological settings, our theory is
in agreement with the usual results. As a special case this also
extends to the Minkowski background (standard model plus gravity, in
vacuum gravitational waves). Hence our theory has passed a first
important consistency check. It should be mentioned here that there
is an alternative approach to cosmological perturbation theory,
based on the use of covariant quantities \cite{oliver10,oliver11}.
We already compared it with our framework in \cite{1}, so here we
will limit ourselves to a detailed comparison between our results
and SCPT.
\\
\\
The plan of the current paper is as follows:\\
\\
In section two we specialise the gauge-invariant linear perturbation
theory of \cite{1} to the FRW background. In appendix B, as a
calculational check, we also follow a shorter route which is to
first linearise the first time derivative equations and then to
derive the second time derivative equations for the perturbations,
which, as we know from appendix D of \cite{1}, must give the same
result.

In section three we review the SCPT framework from a conceptual
point of view. We then define a map between the linearly gauge
invariant SCPT variables for gravitational perturbations and our
fully gauge invariant variables. We show that under this map our
equations of motion can be cast into SCPT form plus corrections
which, as already stated, decay.

In section four we repeat this analysis for the case where dust is
included in SCPT. We then show that the corresponding SCPT
perturbation variables precisely match our gauge invariant
perturbation variables when we expand them up to first order in the
{\it non gauge invariant} perturbations. This demonstrates that the
identification performed in section three is
correct.\\

In section five we summarise and conclude.

In appendix A we give a concise review of SCPT from a technical
point of view.

In appendix B we compute the equations derived in section two by an
alternative and faster route, thus providing a consistency check. As
a side result, we show that the ordinary FRW background is a
solution also to our theory (with a modified energy density) and
thus provides a viable background.

For the benefit of the reader, we list our notation once more on the
next page.

\newpage

{\bf \large Notation}\\
\\
\\
As a rule of thumb, gauge non invariant quantities are denoted by lower
case letters, gauge invariant quantities by capital letters. The only
exceptions from this rule are the dust fields $T,S^j,\rho,W_j$, their
conjugate momenta $P, P_j, I, I^j$ and their associated primary
constraints $Z_j,Z,Z^j$ which however
disappear in the final picture. Partially gauge invariant quantities
(with respect to spatial diffeomorphisms) carry a tilde. Background
quantities carry a bar. Our
signature
convention is that of relativists, that is, mostly plus.\\
\\
\be \nonumber
\begin{array}{cl}
\mbox{symbol} & \mbox{meaning}  \\
 & \\
G_N & \mbox{Newton constant}\\
\kappa=16\pi G_N & \mbox{gravitational coupling constant}\\
\lambda & \mbox{scalar coupling constant}\\
\Lambda & \mbox{cosmological constant}\\
M & \mbox{spacetime manifold}\\
{\cal X} & \mbox{spatial manifold}\\
{\cal T} & \mbox{dust time manifold}\\
{\cal S} & \mbox{dust space manifold}\\
\mu,\nu,\rho,..=0,..,3 & \mbox{tensor indices on }M\\
a,b,c,..=1,2,3 & \mbox{tensor indices on }{\cal X}\\
i,j,k,..=1,2,3 & \mbox{tensor indices on }{\cal S}\\
X^\mu & \mbox{coordinates on }M\\
x^a & \mbox{coordinates on }{\cal X}\\
\sigma^j & \mbox{coordinates on }{\cal S}\\
t & \mbox{foliation parameter}\\
\tau & \mbox{dust time coordinate}\\
Y_t^\mu & \mbox{one parameter family of embeddings }{\cal S} \to M\\
{\cal S}_t=Y_t({\cal S}) & \mbox{leaves of the foliation}\\
g_{\mu\nu} & \mbox{metric on }M\\
q_{ab} & \mbox{(pullback) metric on }{\cal X}\\
\tilde{q}_{ij} & \mbox{(pullback) metric on }{\cal S}\\
Q_{ij} & \mbox{Dirac observable associated to } q_{ab}\\
p^{ab} & \mbox{momentum conjugate to }q_{ab}\\
\tilde{p}^{ij} & \mbox{momentum conjugate to }\tilde{q}_{ij}\\
P^{ij} & \mbox{momentum conjugate to }Q_{ij}\\
\zeta & \mbox{scalar field on }M\\
\xi & \mbox{scalar field on }{\cal X}\\
\tilde{\xi} & \mbox{pullback scalar field on }{\cal S}\\
\Xi & \mbox{Dirac observable associated to }\xi\\
\pi & \mbox{momentum conjugate to }\xi\\
\tilde{\pi} & \mbox{momentum conjugate to }\tilde{\xi}\\
\Pi & \mbox{momentum conjugate to }\Xi\\
v & \mbox{potential of }\zeta,\;\xi,\;\tilde{\xi},\;\Xi\\
T & \mbox{dust time field on }{\cal X}\\
\tilde{T} & \mbox{dust time field on }{\cal S}\\
S^j & \mbox{dust space fields on }{\cal X}\\
\rho & \mbox{dust energy density on }M,\;{\cal X}\\
W_j & \mbox{dust Lagrange multiplier field on }M,{\cal X}\\
U=-dT+W_j dS^j & \mbox{dust deformation covector field on }M\\
J=\det(\partial S/\partial x) & \mbox{dust field spatial density
on}{\cal X}\\
P & \mbox{momentum conjugate to }T\\
\tilde{P} & \mbox{momentum conjugate to }\tilde{T}\\
P_j & \mbox{momentum conjugate to }S^j\\
I & \mbox{momentum conjugate to }\rho\\
I^j & \mbox{momentum conjugate to }W_j\\
Z_j,\;Z,\;Z^j & \mbox{dust primary constraints on }{\cal X}\\
\mu^j,\;\mu,\;\mu_j & \mbox{dust primary constraint Lagrange
multipliers on }{\cal X}
\end{array}
\ee

\newpage

\be \nonumber\\
\begin{array}{cl}
\varphi & \mbox{diffeomorphism of } \cal X\\
n^\mu & \mbox{unit normal of spacelike hypersurface on } M\\
n & \mbox{coordinate lapse function on }{\cal X}\\
n^a & \mbox{coordinate shift function on }{\cal X}\\
p & \mbox{momentum conjugate to n}\\
p_a & \mbox{momentum conjugate to }n^a\\
z,\;z_a & \mbox{primary constraint for lapse, shift}\\
\nu,\;\nu^a & \mbox{lapse and shift primary constraint Lagrange
multipliers}\\
\phi,\psi, B, E & \mbox{SCPT scalars on }{\cal X},\;{\cal S}\\
S_a, F_a & \mbox{SCPT transversal vectors on }{\cal X}\\
S_j, F_j & \mbox{SCPT transversal vectors on }{\cal S}\\
h_{ab} & \mbox{SCPT transverse tracefree tensor on }{\cal X}\\
h_{jk} & \mbox{SCPT transverse tracefree tensor on }{\cal S}\\
\Phi,\Psi & \mbox{linear gauge invariant completions of }\phi,\psi\\
V_a & \mbox{linear gauge invariant completions of }F_a\\
V_j & \mbox{linear gauge invariant completions of }F_j\\
c^{\rm tot}_a & \mbox{total spatial diffeomorphism constraint on }{\cal X}\\
c^{\rm tot}_j=S^a_j c^{\rm tot}_a & \mbox{total spatial diffeomorphism
constraint on }{\cal X}\\
c^{\rm tot} & \mbox{total Hamiltonian constraint on }{\cal X}\\
c_a & \mbox{non -- dust contribution to spatial diffeomorphism
constraint on }{\cal X}\\
c_j=S^a_j c_a & \mbox{non -- dust contribution to spatial diffeomorphism
constraint on }{\cal X}\\
\tilde{c}_j & \mbox{non -- dust contribution to spatial diffeomorphism
constraint on }{\cal S}\\
C_j\not=\tilde{c}_j & \mbox{momentum density: Dirac observable
associated to
}\tilde{c}_j\\
c & \mbox{non -- dust contribution to Hamiltonian
constraint on }{\cal X}\\
\tilde{c} & \mbox{non -- dust contribution to Hamiltonian
constraint on }{\cal S}\\
C\not=\tilde{c} & \mbox{Dirac observable associated to }\tilde{c}\\
h & \mbox{energy density on }{\cal X}\\
\tilde{h} & \mbox{energy density on }{\cal S}\\
H=\tilde{h} & \mbox{energy density: Dirac observable associated to
}\tilde{h}\\
h_j=c^{\rm tot}_j-P_j & \mbox{auxiliary density on }{\cal X}\\
\epsilon & \mbox{numerical energy density on }{\cal S}\\
\epsilon_j & \mbox{numerical momentum density on }{\cal S}\\
\HF=\int_{{\cal S}}\;d^3\sigma\; H & \mbox{physical Hamiltonian,
energy}\\
L & \mbox{Lagrange density associated to }H\\
\LF=\int_{{\cal S}}\;d^3\sigma\; L & \mbox{physical Lagrangian}\\
V_{jk} & \mbox{velocity associated to }Q_{jk}\\
\Upsilon & \mbox{velocity associated to }\Xi\\
N=C/H & \mbox{dynamical lapse function on }{\cal S}\\
N_j=-C_j/H & \mbox{dynamical shift function on }{\cal S}\\
N^j=Q^{jk} N_k & \mbox{dynamical shift function on }{\cal S}\\
\nabla_\mu & g_{\mu\nu} \mbox{ compatible covariant differential on }
M\\
D_a & q_{ab} \mbox{ compatible covariant differential on } {\cal X}\\
\tilde{D}_j & \tilde{q}_{jk} \mbox{ compatible
covariant differential on } {\cal S}\\
D_j & Q_{jk} \mbox{ compatible covariant differential on } {\cal S}\\
\overline{Q}_{jk} &\mbox{ background spatial metric}\\
\overline{P}^{jk} &\mbox{ background momentum conjugate to }\overline{Q}_{jk}\\
\overline{\Xi}& \mbox{ background scalar field}\\
\overline{\Pi} & \mbox{ background momentum conjugate to }\overline{\Xi}\\
\overline{\rho}=\frac{1}{2\lambda}[\dot{\overline{\Xi}}^2+v(\overline{\Xi})]
& \mbox{
background scalar energy density}\\
\overline{p}=\frac{1}{2\lambda}[\dot{\overline{\Xi}}^2-v(\overline{\Xi})]
& \mbox{
background scalar pressure}
\end{array}
\ee

\newpage

\be \nonumber\\
\begin{array}{cl}
G_{jkmn}=Q_{j(m}Q_{n)k}-\frac{1}{2}Q_{jk} Q_{mn} & \mbox{ physical DeWitt
bimetric}\\
{[}G^{-1}]^{jkmn}=Q^{j(m}Q^{n)k}-Q^{jk} Q^{mn} & \mbox{
inverse physical DeWitt bimetric}\\
\overline{G}_{jkmn}=\delta_{j(m}\delta_{n)k}-\frac{1}{2}\delta_{jk} \delta_{mn}
&
\mbox{
flat background DeWitt
bimetric}\\
{[}\overline{G}^{-1}]^{jkmn}=\delta^{j(m}\delta^{n)k}-\delta^{jk}
\delta^{mn} &
\mbox{
inverse flat background DeWitt bimetric}
\end{array}
\ee

\newpage
\section{Specialisation of Linear Perturbation Theory to FRW Background}
\label{s2}
In this section we want to specialise the perturbed equations that
were derived in our companion paper \cite{1} for an arbitrary
background to the case of an FRW spacetime. The perturbed matter
scalar field $\delta\Xi$, as well as the perturbed three metric
$\delta Q_{jk}$ are manifestly gauge invariant. This gauge
invariance is not restricted to linear perturbations only, but
extends to arbitrarily high orders. The underlying reason is the
fact that we applied perturbation theory to an already manifestly
gauge invariant physical system, since the quantities $Q_{jk}$ and
$\Xi$ are gauge invariant by construction. Hence, any function of
$\delta\Xi,\delta Q_{jk}$ is still a gauge invariant quantity. For
this reason higher powers of the perturbation that occur in higher
order perturbation theory will in our framework not destroy gauge
invariance. Therefore, in principle, it is possible to analyse
cosmological perturbation theory also in arbitrarily high orders.
However, in this section we will focus on cosmological perturbation
theory up to linear order, the discussion on higher order equations
will be left for future work.
\\
\\
Due to the homogeneity and isotropy associated with FRW, the final
second order equations of motion, shown in equation (6.13) in
\cite{1} for the linear perturbation of the matter scalar field
$\delta\Xi$, and in equation (6.31) for the perturbed three metric
$\delta Q_{jk}$, simplify drastically. A similar simplification
occurs for the coefficient functions that were introduced in section
6.3 in \cite{1} in order to write those second order equations in a
less complicated form. Before actually discussing the specialisation
to FRW for the perturbed equations of motions, we will first
consider the unperturbed second order time derivative equations of motion for $\Xi$
and $Q_{jk}$, derived in \cite{1} in equation (4.24) and (4.25),
respectively, and check that they yield a gauge invariant version of
the ordinary FRW equation of standard cosmology. An alternative
derivation of the gauge invariant FRW equation can be found in
appendix \ref{sb} in section \ref{sb.1}, where already the first
order time derivative Hamiltonian equations are specialised to FRW before the second
order time derivative equations are calculated.

\subsection{Gauge Invariant FRW Equations}
 In the remaining part of this paper we will indicate quantities which are specialised to FRW spacetimes with a bar.
Due to the homogeneity of an FRW spacetime, all spatial derivatives
vanish. For this reason the matter as well as the geometry part of
the (gauge invariant) diffeomorphism constraint, denoted by
$\ov{C}^{\mathrm{matter}}_j$ and $\ov{C}^{\mathrm{geo}}_j$, vanish as can be
seen from equation (\ref{CundCj}): \ba \label{CundCj} C_j(\sigma)
&=& \big[-2 \ov{Q}_{jk} (\ov{D}_k \ov{P}^{kl})+\ov{\Pi}\; \ov{\Xi}_{,j}\big](\tau,\sigma)
\nonumber\\
&=:&\ov{C}^{\mathrm{geo}}_j(\tau,\sigma)+\ov{C}^{\mathrm{matter}}_j(\tau,
\sigma)\nonumber\\
\ov{C}(\tau,\sigma) &=&
\frac{1}{\kappa}\Big[\frac{1}{\Qb}\big[\ov{Q}_{jm} \ov{Q}_{kn}-\frac{1}{2}
\ov{Q}_{jk}
\ov{Q}_{mn}\big]\ov{P}^{jk} \ov{P}^{mn}- \Qb \;\ov{R}^{(3)}[\ov{Q}] +2\Lambda
\Qb\Big](\tau,\sigma)
\nonumber\\
&&+ \frac{1}{2\lambda}\Big[\frac{\ov{\Pi}^2}{\Qb}+\Qb
\big[\ov{Q}^{jk}\; (\ov{D}_j\Xi)\;(\ov{D}_k \ov{\Xi})+v(\ov{\Xi})\big]\Big](\tau,\sigma)
\nonumber\\
&=:& \ov{C}^{\mathrm{geo}}(\tau,\sigma)+\ov{C}^{\mathrm{matter}}(\tau,\sigma),
\ea where $\ov{D}_j$ is the covariant differential compatible with
$\ov{Q}_{jk}$ and we used that the Christoffel symbols with spatial
indices vanish in the case of FRW. The results in \cite{1} showed
that one of the effects when dust is used as a clock is that we
obtain a phase space dependent and thus dynamical lapse function and
shift vector, which we denoted by $\ov{N}$ and $\ov{N}_j$, respectively. Their
explicit form is given by \be
\ov{N}=\frac{\ov{C}(\tau,\sigma)}{\ov{H}(\sigma)}\quad\mathrm{and}\quad
\ov{N}_j=-\frac{\ov{C}_j(\tau,\sigma)}{\ov{H}(\sigma)} \ee where \be
\label{Hdensity}
\ov{H}(\sigma)=\sqrt{\ov{C}(\tau,\sigma)^2-\ov{Q}^{jk}(\tau,\sigma)\;\ov{C}_j(\sigma)
\;C_k(\sigma)} \ee is the physical Hamiltonian density. This is the
density of the physical Hamiltonian $\ov{\HF}$ that is generating the
(dust) time evolution  for all observables, in particular for $\ov{\Xi}$
and $\ov{Q}_{jk}$. Knowing now that $\ov{C}_j$ vanishes for FRW, we get
immediately that the shift vector vanishes, too. Hence in the case
of FRW, all terms proportional to $\ov{N}_j$ can be omitted in equations
(4.17) and (4.18). Furthermore from equation (\ref{Hdensity}) and
the definition of $\ov{N}$ and $\ov{N}_j$ we get for FRW \be
\ov{N}=\frac{\ov{C}}{\ov{H}}=\frac{\sqrt{\ov{H}^2+\ov{Q}^{jk}\ov{C}_j\ov{C}_k}}{\ov{H}}=\sqrt{1+\ov{Q}^{jk}\ov{N}_j\ov{N}_k}=1.
\ee We observe that $\ov{N}$ and $\ov{N}_j$ are not independent quantities and
that for the special case of FRW the lapse function takes the
constant value +1. Keeping this in mind and taking into account that
all terms containing spatial derivatives vanish, the equations
(4.17) and (4.18) for $\ddot{\Xi}$ and $\ddot{Q}_{jk}$ in our
companion paper \cite{1} reduce to \ba \label{XiQEOM}
\ddot{\ov{\Xi}}&=&-\frac{(\Qb)^{\bf\dot{}}}{\Qb}\,\dot{\ov{\Xi}}-\frac{1}{2}v^{\prime}
(\ov{\Xi})\nonumber\\
\ddot{\ov{Q}}_{jk}&=&-\frac{(\Qb)^{\bf\dot{}}}{\Qb}\,\dot{\ov{Q}}_{jk}+\ov{Q}^{mn}\dot{\ov{Q}}_{mj}\dot{\ov{Q}}_{nk}+\ov{Q}_{jk}
\Big(-\frac{\kappa}{2\Qb}\,\ov{C}+2\Lambda +\frac{\kappa}{2\lambda}v(\ov{\Xi})\Big).
\ea Using that $\ov{Q}_{jk}=A^2\delta_{jk}$ where $A$ is the physical
scale factor, that is the gauge invariant extension of the usual non
gauge invariant scale factor $a$ used in standard cosmology, we
obtain for $\ddot{\ov{\Xi}}$ \be \label{XiFRWfinal}
\ddot{\ov{\Xi}}=-3\Big(\frac{\dot{A}}{A}\Big)\dot{\ov{\Xi}}-\frac{1}{2}v^{\prime}(\ov{\Xi}).
\ee This agrees (formally) with the usual FRW equation for a scalar
field. The agreement is formal in the sense that here we have an
evolution equation for a gauge-invariant and thus physical scalar
field $\ov{\Xi}$. Also the equation contains the physical scale factor
$A$, not $a$ which usually occurs in this equation. Furthermore the
dot refers to a derivative with respect to dust time $\tau$ in the
equation above. For $\ddot{\ov{Q}}_{jk}$ we need the explicit form of the
matter, as well as the geometry part of the Hamiltonian constraint,
whose sum is denoted by $\ov{C}$. Specialised to FRW it is given by \be
\label{CFRW} \ov{C}=A^3\Big[\frac{1}{\kappa}\Big( 2\Lambda
-6\Big(\frac{\dot{A}}{A}\Big)^2\Big)+\frac{1}{2\lambda}\big(\dot{\ov{\Xi}}
^2+v(\ov{\Xi})\big)\Big]. \ee Inserting this back into equation
(\ref{XiQEOM}) and performing the derivatives yields 
\be
\label{DerQFRW1}
\Big(\frac{\ddot{A}}{A}\Big)=-\frac{1}{2}\Big(\frac{\dot{A}}{A}\Big)^2+\frac{1}{
2}\Lambda-\frac{\kappa}{4\lambda}\frac{1}{2}\big(\dot{\Xi}^2-v(\ov{\Xi})\big).
\ee As shown in our companion paper \cite{1}, the Hamiltonian
density $\ov{H}$ in equation (\ref{Hdensity}) is a constant of motion. In
the case of FRW, $\ov{H}$ reduces to \be
\ov{H}=\sqrt{\ov{C}^2-\ov{Q}^{jk}\ov{N}_j\ov{N}_k}=\sqrt{\ov{C}^2}=\ov{C} \ee due to the vanishing of
the shift vector. Denoting the energy density by $\epsilon$, we have
the conservation law \be \ov{H}=\ov{C}=A^3\Big[\frac{1}{\kappa}\Big( 2\Lambda
-6\Big(\frac{\dot{A}}{A}\Big)^2\Big)+\frac{1}{2\lambda}\big(\dot{\ov{\Xi}}
^2+v(\ov{\Xi})\big)\Big]=\epsilon. \ee Note that $\ov{C}$ is not a constraint
here, since only the total sum
$\ov{C}^{\mathrm{tot}}=\ov{C}^{\mathrm{geo}}+\ov{C}^{\mathrm{matter}}+\ov{C}^{\mathrm{dust}}$,
with the dust included, is constrained to vanish. The conservation
law above can be equivalently written as \be
3\Big(\frac{\dot{A}}{A}\Big)^2=-\frac{\epsilon\kappa}{2A^3}+\Lambda+\frac{\kappa
}{4\lambda}\big(\dot{\ov{\Xi}}^2+v(\ov{\Xi})\big). \ee Replacing
$(\dot{A}/A)^2$ in equation (\ref{DerQFRW1}) by the expression in
the equation above, we end up with \be \label{DerQFRW2}
3\Big(\frac{\ddot{A}}{A}\Big)=\Lambda-\frac{\kappa}{4}\Big[\frac{1}{2\lambda}
\Big(\dot{\ov{\Xi}}^2+v(\ov{\Xi})\Big)+3\frac{1}{2\lambda}\Big(\dot{\ov{\Xi}}
^2-v(\ov{\Xi})\Big)-\frac{\epsilon}{A^3}\Big]. \ee Introducing the matter density and pressure as well as the dust density \ba
\label{rhoMD}
\ov{\rho}^{\mathrm{matter}}&=&\frac{\ov{C}^{\mathrm{matter}}}{A^3}=\frac{1}{2\lambda}
\big(\dot{\ov{\Xi}}^2+v(\ov{\Xi})\Big)\nonumber\\
\ov{\rho}^{\mathrm{dust}}&=&\frac{\ov{C}^{\mathrm{dust}}}{A^3}=-\frac{\ov{C}^{\mathrm{matter}}
+\ov{C}^{\mathrm{matter}}}{A^3}
-\frac{\ov{C}}{A^3}=-\frac{\epsilon}{A^3}\nonumber\\
\ov{p}^{\mathrm{matter}}&=&\frac{1}{2\lambda}\big(\dot{\ov{\Xi}}^2-v(\ov{\Xi})\big),
\ea we can rewrite equation (\ref{DerQFRW2}) as \be
\label{QFRWfinal}
3\Big(\frac{\ddot{A}}{A}\Big)=\Lambda-\frac{\kappa}{4}\Big(\ov{\rho}^{\mathrm{matter}
}+\ov{\rho}^{\mathrm{dust}}+3\ov{p}^{\mathrm{matter}}\Big). \ee This again
coincides formally with the usual FRW equation for the gravitational
part if the system "gravity + scalar field + dust" is considered and
the dust is assumed to be pressureless as is the case in our
framework. The minus sign in $\ov{\rho}^{\mathrm{dust}}$ in equation
(\ref{rhoMD}) reflects the phantom nature of the dust which was
discussed in detail in section 2.3.4 of our companion paper
\cite{1}.
\\
\\
Summarising, when (phantom) dust is employed in order to
deparametrise the constraints of General Relativity, the resulting
physical Hamiltonian generates equations of motion that formally
agree with the standard FRW equations. For this reason, the dust
clock seems to be the clock closest to the framework of standard
cosmology where the Hamiltonian constraint is taken as a true
Hamiltonian. In the next section we will move on to the discussion
of the perturbed equations of motion for $\delta\Xi$ and $\delta
Q_{jk}$.
\subsection{Specialisation of the Perturbed Equations of Motion to an FRW Background}
Let us now specialise the perturbed equations for $\delta\ddot{\Xi}$
and $\delta\ddot{Q}_{jk}$, derived in our companion paper \cite{1},
to the case of an FRW background. As discussed in the last section,
in this case the lapse function takes the constant value $\ov{N}=1$ and
the shift vector $\ov{N}_j$ vanishes. Taking into account that all
spatial derivatives of background quantities (these are indicated
with a bar) vanish, the coefficient functions for $\delta\Xi$ that
can be found in equation (6.33), (6.34) and (6.35) of \cite{1}
simplify to \ba \label{XiCoeffFRW}
[C_{\Xi}]
&=&\frac{\partial^2}{\partial\tau^2}+\frac{(\Qb)^{\bf\dot{}}}{\Qb}\frac{\partial}{
\partial\tau}
-\Delta+\frac{1}{2}v^{\prime\prime}(\ov{\Xi})
=\frac{\partial^2}{\partial\tau^2}+3\Big(\frac{\dot{A}}{A}\Big)\frac{\partial}{
\partial\tau}-\frac{1}{A^2}\delta^{jk}\frac{\partial}{\partial
x^j}\frac{\partial}{\partial
x^k}+\frac{1}{2}v^{\prime\prime}(\ov{\Xi})\nonumber\\
{[C_{\Xi}]}^{jk}&=&-\frac{1}{2}\dot{\ov{\Xi}}\frac{\partial}{\partial\tau}
\big(\ov{Q}^{jk}\big)
=-\frac{1}{2}\dot{\ov{\Xi}}\,\dot{\ov{Q}}^{jk}-\frac{1}{2}\dot{\ov{\Xi}}\,\ov{Q}
^{jk}\frac{\partial}{\partial\tau}
=+\frac{1}{A^2}\Big(\frac{\dot{A}}{A}\Big)\dot{\ov{\Xi}}\delta^{jk}-\frac{1}{2}
\frac{1}{A^2}\dot{\ov{\Xi}}\delta^{jk}\frac{\partial}{\partial\tau}\nonumber\\
{[C_{\Xi}]}^{j}&=&\dot{\ov{\Xi}}\,\ov{Q}^{jk}\frac{\partial}{\partial
x^k}=\frac{1}{A^2}\dot{\ov{\Xi}}\delta^{jk}\frac{\partial}{\partial
x^k}. \ea
Using these coefficients, we obtain the following
perturbed equation \ba
\Big[\frac{\partial^2}{\partial\tau^2}+3\Big(\frac{\dot{A}}{A}\Big)\frac{\partial}{
\partial\tau}-\frac{1}{A^2}\delta^{jk}\frac{\partial}{\partial
x^j}\frac{\partial}{\partial
x^k}+\frac{1}{2}v^{\prime\prime}(\ov{\Xi})\Big]\delta\Xi
&=&
\Big[+\frac{1}{A^2}\Big(\frac{\dot{A}}{A}\Big)\dot{\ov{\Xi}}\delta^{jk}-\frac{1}{2}
\frac{1}{A^2}\dot{\ov{\Xi}}\delta^{jk}\frac{\partial}{\partial\tau}\Big]\delta Q_{jk}
+\Big[\frac{1}{2}\frac{1}{A^2}\dot{\ov{\Xi}}\delta^{jk}\Big]\delta
N_j.\nonumber\\
\ea When performing the derivatives and rearranging the terms, the
equation above can be rewritten as \ba \label{PerXifinal}
\delta\ddot{\Xi}&=&-3\Big(\frac{\dot{A}}{A}\Big)\delta\dot{\Xi}+\frac{1}{A^2}
\delta^{jk}\delta\Xi_{,jk}-\frac{1}{2}v^{\prime\prime}(\ov{\Xi})\delta\Xi
+\frac{1}{A^2}\Big(\frac{\dot{A}}{A}\Big)\dot{\ov{\Xi}}\delta
Q_{jj}-\frac{1}{2}\frac{1}{A^2}\dot{\ov{\Xi}}\delta\dot{Q}_{jj}+\frac{1}{A^2}
\dot{\ov{\Xi}}\delta N_{j,j}. \ea This equation agrees with equation
(\ref{b.20}) in appendix \ref{sb} where an alternative route is
taken to obtain the second order equation of motion. There we first
perturb the Hamiltonian first order equation and derived then the
second order perturbed equations. For linear perturbation theory it
is proven in appendix D of \cite{1} that these two ways of deriving the
second order equation of motion are equivalent. Compared to the
perturbed equations in SCPT we obtain an additional term, namely the
last term on the right-hand side of equation (\ref{PerXifinal}).
This term arises from the interaction of the clock, namely the dust,
with the physical system. However, in the next section where we
compare our results in detail with the results from SCPT, we will
show that the physical predictions of both frameworks are in
agreement.
\\
Let us now consider the coefficients for $\delta Q_{jk}$  in
equation (6.37)-(6.41) of \cite {1} and their specialisation to FRW.
Taking again into account that $N=1$, that all terms proportional to
$N_j$ can be omitted and that all spatial derivatives of background
quantities are zero, we obtain \ba \label{QCoeffFRW}
{[C_{Q}]}&=&\frac{\partial^2}{\partial\tau^2}+\frac{\kappa}{2\Qb}\,\ov{C}
-\big(2\Lambda+\frac{\kappa}{2\lambda}v(\Xi)\big)-\ov{D}_m\ov{D}_n\ov{Q}^{mn}
+\frac{(\Qb)^{\bf\dot{}}}{\Qb}\frac{\partial}{\partial\tau}\nonumber\\
&=&
\frac{\partial^2}{\partial\tau^2}
-3\Big(\frac{\dot{A}}{A}\Big)^2-\Lambda+\frac{\kappa}{4\lambda}\big(\dot{\ov{\Xi}}^2-v(\ov{\Xi})\big)
-\ov{D}_m\ov{D}_n\ov{Q}^{mn}
+3\Big(\frac{\dot{A}}{A}
\Big)\frac{\partial}{\partial\tau}\nonumber\\
{[A_Q]}_{jk}&=&-\ov{Q}_{jk}\frac{\kappa}{2\lambda}\dot{\ov{\Xi}}\frac{\partial}{
\partial\tau}+\ov{Q}_{jk}\frac{\kappa}{4\lambda}v^{\prime}(\ov{\Xi})
=\Big[-\frac{1}{A^2}\frac{\kappa}{2\lambda}\dot{\ov{\Xi}}\frac{\partial}{\partial\tau
}+\frac{1}{A^2}\frac{\kappa}{4\lambda}v^{\prime}(\ov{\Xi})\Big]\delta_{jk}\nonumber\\
{[B_Q]}_{jk}&=&0\nonumber\\
{[C_Q]}^m_{(k}&=&2\ov{Q}^{mn}\dot{\ov{Q}}_{n(k}\frac{\partial}{\partial\tau}
-2\ov{D}_n\ov{D}_{(k}\ov{Q}^{mn}
=4\Big(\frac{\dot{A}}{A}\Big)\delta^m_{(k}\frac{\partial}{\partial\tau}-2\ov{D}
_n\ov{D}_{(k}\ov{Q}^{mn}\nonumber\\
{[C_Q]}^m_{jk}&=&\dot{\ov{Q}}_{jk}\frac{\partial}{\partial
x^n}\ov{Q}^{mn}+\frac{(\Qb)^{\bf\dot{}}}{\Qb}\Big(\ov{Q}_{nk}\frac{\partial}{
\partial x^j}\ov{Q}^{mn}+\ov{Q}_{mj}\frac{\partial}{\partial
x^k}\ov{Q}^{mn}\Big)
+\frac{1}{2}\ov{Q}_{jk}\dot{\ov{Q}}_{tu}[G^{-1}]^{mntu}\frac{\partial}{\partial
x^n}\nonumber\\
&&
+\dot{\ov{Q}}_{kn}\frac{\partial}{\partial
x^j}\ov{Q}^{mn}+\dot{\ov{Q}}_{jn}\frac{\partial}{\partial x^k}\ov{Q}^{mn}
+\frac{\partial}{\partial\tau}\Big(\ov{Q}_{kn}\frac{\partial}{\partial
x^j}\ov{Q}^{mn}+\ov{Q}_{jn}\frac{\partial}{\partial
x^k}\ov{Q}^{mn}\Big)\nonumber\\
&&
-\ov{Q}^{tu}\dot{\ov{Q}}_{tk}\Big(\dot{\ov{Q}}_{un}\frac{\partial}{\partial
x^j}\ov{Q}^{mn}
+\dot{\ov{Q}}_{jn}\frac{\partial}{\partial x^u}\ov{Q}^{mn}\Big)
-(\ov{Q}^{tu}\dot{\ov{Q}}_{tj}\Big(\dot{\ov{Q}}_{un}\frac{\partial}{\partial
x^k}\ov{Q}^{mn}
+\dot{\ov{Q}}_{kn}\frac{\partial}{\partial x^u}\ov{Q}^{mn}\Big)\nonumber\\
&=&\Big(\frac{\dot{A}}{A}\Big)\Big(\delta^m_j\frac{\partial}{\partial
x^k}+\delta^m_k\frac{\partial}{\partial x^j}\Big)
+\Big(\delta^m_j\frac{\partial}{\partial x^k}+\delta^m_k\frac{\partial}{\partial
x^j}\Big)\frac{\partial}{\partial\tau}\nonumber\\
{[C_Q]}^{mn}_{jk}&=&-\frac{1}{2}\ov{Q}_{jk}\frac{\partial}{\partial\tau}\ov{Q}^{
mn}-\ov{Q}^{mr}\ov{Q}^{ns}\dot{\ov{Q}}_{rj}\dot{\ov{Q}}_{sk}+\frac{1}{4}\ov{Q}_{
jk}\dot{\ov{Q}}_{rs}\dot{\ov{Q}}_{tu}\ov{Q}^{ns}[G^{-1}]^{turm}
-\frac{1}{4}\ov{Q}_{jk}\dot{\ov{Q}}_{rs}[G^{-1}]^{mnrs}\frac{\partial}{
\partial\tau}\nonumber\\
&&
+\ov{D}_j\ov{D}_k\ov{Q}^{mn}+\frac{1}{2}[G^{-1}]^{mnrs}\ov{D}_r\ov{D}
_s\nonumber\\
&=&-4\Big(\frac{\dot{A}}{A}\Big)\delta^m_j\delta^n_k+\ov{D}_j\ov{D}_k\ov{Q}^{mn}
+\frac{1}{2}[G^{-1}]^{mnrs}\ov{D}_r\ov{D}_s.
 \ea
For the derivation of ${[C_Q]}^{mn}_{jk}$ we used that $R^{mn}$
vanishes in the case of FRW and we also employed several times the
identity $Q_{mn}[G^{-1}]^{jkmn}=-2Q^{jk}$. Recall that the
coefficient $[B_Q]_{jk}$ is associated with the perturbation of the
last term on the right-hand side of the equation of motion for
$\delta Q_{jk}$ in equation (4.25) in \cite{1}. We mentioned already
that this term is the only deviation from the standard Einstein
equations that use the Hamiltonian constraint as a true Hamiltonian.
However, since this term is quadratic in the shift vector $N_j$, it
vanishes for both the background equations and the linear
perturbations, when specialised to FRW. Nevertheless, we will see
below that we still get small deviations from the standard treatment
due to the shift vector being dynamical. Note that since
$N=\sqrt{1+Q^{jk}N_jN_k}$ we get \be \delta
N=-\frac{\ov{N}}{2}\frac{\ov{N}^j\ov{N}^k}{\ov{N}^2}\delta
Q_{jk}+\ov{N}\frac{\ov{N}^j}{\ov{N}^2}\delta N_j=0. \ee Hence, the
perturbation of the lapse function also vanishes in the context of
an FRW background. Finally, we use the coefficients in equation
(\ref{QCoeffFRW}) and obtain the following equation: \ba
\label{DerQFRW}
\lefteqn{\Big[\frac{\partial^2}{\partial\tau^2}
-3\Big(\frac{\dot{A}}{A}\Big)^2-\Lambda+\frac{\kappa}{4\lambda}\big(\dot{\ov{\Xi}}^2-v(\ov{\Xi})\big)
-\ov{D}_m\ov{D}_n\ov{Q}^{mn}
+3\Big(\frac{\dot{A}}{A}
\Big)\frac{\partial}{\partial\tau}\Big]\delta
Q_{jk}}\\
&=&\Big[-\frac{1}{A^2}\frac{\kappa}{2\lambda}\dot{\ov{\Xi}}\frac{\partial}{
\partial\tau}+\frac{1}{A^2}\frac{\kappa}{4\lambda}v^{\prime}(\ov{\Xi})\Big]
\delta\Xi
+\Big[4\Big(\frac{\dot{A}}{A}\Big)\delta^m_{(k}\frac{\partial}{\partial\tau}
-2\ov{D}_n\ov{D}_{(k}\ov{Q}^{mn}\Big]\delta Q_{j)m}
\nonumber\\
&&
+\Big[\Big(\frac{\dot{A}}{A}\Big)\Big(\delta^m_j\frac{\partial}{\partial
x^k}+\delta^m_k\frac{\partial}{\partial x^j}\Big)
+\Big(\delta^m_j\frac{\partial}{\partial x^k}+\delta^m_k\frac{\partial}{\partial
x^j}\Big)\frac{\partial}{\partial\tau}\Big]\delta N_j\nonumber\\
&&
+\Big[-4\Big(\frac{\dot{A}}{A}\Big)\delta^m_j\delta^n_k+\ov{D}_j\ov{D}_k\ov{Q}^{
mn}+\frac{1}{2}[G^{-1}]^{mnrs}\ov{D}_r\ov{D}_s\Big]\delta
Q_{mn}.\nonumber
\ea
Applying the derivatives to the perturbation
terms, rearranging them and using equation (\ref{DerQFRW1}), we end up with \ba \label{PerQfinal}
\ddot{Q}_{jk} &=& 2\Big(\frac{\ddot{A}}{A}\Big)\delta
Q_{jk}+\Big(\frac{\dot{A}}{A}\Big)\delta\dot{Q}_{jk}
-\delta_{jk}A^2\frac{\kappa}{2\lambda}\Big(\dot{\ov{\Xi}}\delta\dot\Xi-\frac{1}{2}v^{\prime}
(\ov{\Xi})\delta\Xi\Big)
+2\Big(\frac{\dot{A}}{A}\Big)\delta N_{(j,k)}\\
&& +\ov{D}_m\ov{D}_n\ov{Q}^{mn}\delta
Q_{jk}+\Big[\ov{D}_j\ov{D}_k\ov{Q}^{mn}+\frac{1}{2}[G^{-1}]^{mnrs}\ov{D}_r\ov{D}_s\Big]\delta
Q_{mn}-2\ov{D}_n\ov{D}_{(k}\ov{Q}^{mn}\delta Q_{j)m}.\nonumber \ea
In the equation above we used that the perturbed shift vector
$\delta N_j$ is a constant of motion which was shown in our
companion paper \cite{1} in appendix E. Therefore we have
$\delta\dot{N}_j$=0 and thus can omit this term in equation
\ref{DerQFRW}. This equation agrees with the one for
$\delta\ddot{Q}_{jk}$ in equation (\ref{b.20}) if we consider that
the perturbed Riemann tensor can be expressed as \be \delta
R_{jk}=-\frac{1}{2}\ov{D}_m\ov{D}_n\ov{Q}^{mn}\delta
Q_{jk}-\frac{1}{2}\ov{D}_j\ov{D}_k\ov{Q}^{mn}\delta
Q_{mn}+\ov{D}_n\ov{D}_{(k}\ov{Q}^{mn}\delta Q_{j)m} \ee and the
perturbed Ricci scalar reduces for FRW to \be \delta
R=\ov{R}_{mn}\delta Q^{mn}+\ov{Q}^{mn}\delta
R_{mn}=\ov{Q}^{mn}\delta
R_{mn}=[G^{1-}]^{mnrs}\ov{D}_r\ov{D}_s\delta Q_{mn}. \ee Likewise,
in the case of the matter perturbation we obtain an additional term
containing the variation of the shift vector which does not occur in
the SCPT analysis. This finishes our discussion of the
specialisation of the perturbed equation of motion to an FRW
background. In the next section we will compare our results in
detail with the one obtained in SCPT.
\section{Comparison with SCPT}
\label{s3}
The aim of this section is to show that our invariant Hamiltonian
perturbation theory reproduces the SCPT results up to small
deviations. By this we mean the following:
\begin{itemize}
\item[1.] {\it SCPT Lagrangian Approach: Linearly gauge invariant,
linear perturbations}\\
SCPT uses the Lagrangian formulation. It takes as matter a scalar
field $\zeta$ coupled to the metric $g_{\mu\nu}$. These eleven
fields are of course not invariant under (infinitesimal) spacetime
diffeomorphisms of the manifold $M$ which are considered as gauge
transformations. However, one constructs seven functions which are
certain linear combinations built from the metric perturbations \be
\label{3.1} \delta g_{00}=2a^2 \phi,\; \delta g_{0a}=a^2
(S_a+B_{,a}),\; \delta g_{ab}=a^2 [2E_{,ab}+2\psi
\delta_{ab}+2F_{(a,b)}+h_{ab}] \ee and from the scalar field
perturbation $\delta\zeta$. Here $a$ is the scale factor of the
spatially flat FRW background, the vector fields $F_a,S_a$ are
transversal with respect to that flat Euclidean spatial metric and
the tensor $h_{ab}$ is transversal and tracefree. The seven
invariants are explicitly given by \be \label{3.2} \Phi:=\phi-{\cal
H}(B-E')-(B-E')',\; \Psi:=\psi+{\cal
H}(B-E'),\;V_a:=S_a-F_{,a},\;h_{ab},\;Z:=\delta\zeta+\overline{\zeta}'(B-E')
\ee where ${\cal H}=a'/a$, a prime denotes derivation with respect
to conformal time $d\eta=dt/a$ and $\overline{\zeta}$ is the
background scalar field.

One then expresses the ten perturbed Einstein equations directly in
terms of these seven invariants. Four of these equations, namely the
temporal -- temporal and the temporal -- spatial equations do not
contain second time derivatives of four of the seven fields, they are
constraints. They imply that four of the seven invariants can be
expressed in terms of the other three. The system has only three
independent, linearly gauge invariant degrees of freedom.
Specifically, the transversal part of the temporal -- spatial part
of those equations imposes $V_a=0$, while its longitudinal part
together with the temporal -- temporal equation allows us to
express, for instance, $\Phi,Z$ in terms of $\Psi$. Namely, since
the cosmological constant term and the energy momentum tensor have
no contribution of the form $f_{,ab}$ it follows
immediately\footnote{We assume here that the perturbations vanish
sufficiently fast at spatial infinity which is consistent with
making asymptotically FRW boundary conditions. Then the Laplacian
$\Delta$ has no zero modes and is therefore invertible. An equation
of the form $f_{,ab}+g\delta_{ab}=0$ then implies $\Delta f+3 g=0$
and $\Delta(\Delta f+g)=0$ that is $f=g=0$.} from the third equation
in  (\ref{a.37}) that $\Phi=\Psi$. From the longitudinal part of the
temporal -- spatial equations we then immediately find that
$Z+4\lambda(\Psi'+{\cal H}\Psi)/\kappa$ is a spatial constant which
then must vanish  due to the boundary conditions.

Thus the system of a scalar field coupled to the geometry has only
the three degrees of freedom $h_{ab},\;\Psi$ as independent,
linearly invariant, propagating degrees of freedom. Without $Z$ we
even have $\Phi=\Psi=0$.
\item[2.] {\it Hamiltonian Approach: Manifestly invariant,
linear perturbations}\\
Our approach is completely different: first of all we work in the
Hamiltonian framework. Secondly, we add additional four dust scalar
fields. These are altogether fifteen configuration degrees of
freedom to begin with and they are accompanied by the corresponding
canonical momenta. Thus, from the very outset the number of degrees
of freedom is very different compared to the SCPT framework. The
components $g_{00},\;g_{0a}$ of the metric have the same information
content as lapse and shift functions $n,n_a$ respectively. Their
conjugate momenta are constrained to vanish by the primary
constraints, hence $n,n_a$ are pure gauge and can be considered as
Lagrange multipliers in the Hamiltonian formalism. This is in
agreement with the Lagrangian formalism where $\delta
g_{00}=2a^2\phi,\;\delta g_{0a}=a^2(B_{,a}+S_a)$, which are also
pure gauge and are used to construct the gauge-invariant functions
$\Phi,\Psi,V_a,h_{ab}$. Thus, the number of physical configuration
degrees of freedom is reduced by four. In addition there are the
secondary Hamiltonian and spatial diffeomorphism constraints. These
reduce the number of physical configuration further by four. Thus,
we end up with $15-8=7$ physical degrees of freedom. Without dust we
also would have 3 physical degrees of freedom only, which shows that
with equal matter content the Hamiltonian and Lagrangian framework
are equivalent.

Indeed \cite{12} the temporal -- temporal and temporal -- spatial
components of the Einstein equations are proportional to the
Hamiltonian and spatial diffeomorphism constraints, respectively,
and the gauge transformation equations (in second order time
derivative form) generated by the constraints are equivalent to the
remaining spatial -- spatial Einstein equations. The gauge
transformations generated by the secondary constraints of the
variables different from lapse and shift and their conjugate momenta
(the secondary constraints, when written in terms of the momenta
conjugate to the spatial metric and the scalar field, do not depend
on them) are equivalent to spacetime diffeomorphisms, on shell
\cite{12}. In fact, this statement can be extended also to the lapse
and shift fields \cite{Pons} as we will recall in the next section.

Hence the difference between our framework and the usual one is not
in the usage of the Hamiltonian versus the Lagrangian formalism but
rather in the addition of the four dust fields. Now what one could
do is to consider the dust -- scalar -- geometry system, perform
linear perturbations in non gauge-invariant quantities and then
construct {\it linearly} gauge-invariant quantities from those. This
we will do in the next section. However, what we will do here is to
construct {\it quantities which are fully gauge-invariant to all
orders} and then look directly at perturbations of those manifest
invariants. Hence, gauge-invariance is treated here
non-perturbatively, and our perturbations themselves are manifestly
gauge-invariant. This is achieved by means of deparametrisation in
terms of the dust degrees of freedom, which in one stroke solves the
secondary constraints and spells out the {\it true} i.e.
gauge-invariant degrees of freedom. Namely, the dust momenta can be
expressed in terms of those invariants, and the dust fields
themselves together with the remaining fields form non linear
expressions which are fully gauge-invariant. This leaves us, before
performing perturbation theory, with seven degrees of freedom,
namely the invariants $\Xi,Q_{jk}$ corresponding to the scalar field
$\zeta$ (or rather its pull back $\xi$ to the spatial manifold $\cal
X$) and the spatial -- spatial components of the metric together
with the corresponding canonical momenta.

In terms of these seven degrees of freedom there are no longer any
secondary constraints, they have been reduced. However, our
equations of motion for the perturbations $\delta \Xi,\delta Q_{jk}$
of the seven perturbed manifest invariants take a form almost
completely identical to those equations that we obtain in the
Lagrangian formalism for the seven perturbed linear invariants
$\Phi,\Psi,V_a,h_{ab}$, when proper identifications are made. The
crucial difference arises from the fact that the latter quantities
are still subject to constraints while the former are unconstrained.
Yet, it turns out that while in our formalism there are no
constraints any more, the would -- be constraints are now constants
of the motion. These constants of the motion express the influence
of the dust on the system. In particular, in the limit of vanishing
influence (test dust) these conservation equations turn effectively
into constraint equations and then we get a precise match between
the two formalisms with proper identification of variables and
spatial manifolds ${\cal X},\;{\cal S}$, respectively. Thus, for
test dust our formalism just produces arbitrarily tiny
modifications, that is, source terms, of the usual formalism
although fundamentally the number of physical degrees of freedom was
changed. In particular, in the late universe these modifications
decay as compared to the usual terms.
\end{itemize}
It is quite remarkable that the two formalisms {\it with
fundamentally different numbers of degrees of freedom} can be
brought to such a close match. Looking at the details, it is hard to
imagine that any material reference system other than pressure free
dust can lead to such a modest modification of the usual formalism.
Indeed, in the appendix of \cite{1} we compute the modifications
that arise when using instead the phantom scalar field of \cite{5}
for purposes of deparametrisation. It is shown that the
modifications, in fact, grow in the late universe which is
qualitatively drastically different. However, the coupling constant
of the phantom and the constants of the motion can be tuned in such
a way that the moment of time when the modifications become
significant can be chosen to be arbitrarily late.

It is worth mentioning that since we can take the limit of 1. vanishing
dust
influence (i.e. both the background energy momentum
$(\overline{\epsilon},\overline{\epsilon}_j)$ and its perturbation
$(\delta \epsilon,\delta\epsilon_j)$ can be taken to zero), 2. vanishing
cosmological constant, 3. vanishing physical scalar field and finally 4.
vanishing time dependence of the background FRW scale factor $a\equiv 1$,
it follows
immediately from our formalism that in this limit we recover also the
two gravitational wave modes on Minkowski space, the other four modes
are frozen out by the equations of motion.\\
\\
The reader may now ask what the advantage of our more complicated
formalism is compared to the usual one. There are at least three
worth mentioning:
\begin{itemize}
\item[A.] {\it Higher order classical general relativistic perturbation
theory}\\
As far as standard cosmological perturbation theory is concerned,
the usual formalism cannot be easily extended in a gauge invariant
way to higher than linear order. The reason is that one has to
construct the relevant gauge-invariant quantities at each order from
scratch again. In our manifestly gauge-invariant formalism, by
contrast, those quantities are simply given by the relevant
higher-order perturbation of the metric and matter degrees of
freedom. This is a tremendous simplification. We expect that this
should also make it much easier to address general question of the
stability of cosmological perturbation theory \cite{stability}.
\item[B.] {\it Conceptual Improvement}\\
General relativity is a generally covariant field theory without true
Hamiltonian. This leads to many conceptual problems such as the problem
of time (gauge invariant functions do not evolve with respect to the
Hamiltonian constraint(s)). The reason that in perturbation theory one
still has non trivial evolution of the linear invariants is only due to
the fact that one has chosen a background spacetime which provides a
preferred notion of time and what one computes is evolution with respect
to that background. However, fundamentally general relativity is a
background independent field theory and since when going to higher order
both
classical perturbation theory (due to reasons of gauge invariance) and
quantum perturbation theory (due to reasons of non renormalisability)
fail, one is forced to adopt a non perturbative approach so that the
problem of time reappears. The relational formalism solves the problem
of time by determining a physical Hamiltonian (which is not constrained
to vanish) from a chosen material reference system which evolves the
manifest invariants. This brings the analysis back to the
conceptually safe realm of
a dynamical Hamiltonian system, albeit a technically complicated one.
\item[C.] {\it Quantum Theory}\\
At least for the pressure free dust chosen here, the Poisson algebra
of the linear invariants is as simple as for their non invariant
analogs. Therefore it is possible to adopt a reduced phase space
quantisation approach \cite{13} and to find Hilbert space
representations of that algebra. This has the advantage that, for
instance, in the framework of Loop Quantum Gravity the kinematical
Hilbert space used there becomes now a {\it physical} Hilbert space.
The constraints have completely disappeared from the screen. What
remains is to quantise the complicated physical Hamiltonian
\cite{14} and to analyse its spectrum. This may be technically
complicated but, conceptually, it is crystal clear. In particular,
the fact that one has a physical Hamiltonian at one's disposal which
by construction does not depend explicitly on some background time
may improve the vacuum problem that one encounters in QFT on curved
(time dependent) backgrounds. The problem with explicitly time
dependent Hamiltonians is that the notion of vacuum (ground state)
is time dependent which may lead to (infinite) particle production
and, in particular, means that there is no unique ground state. The
physical Hamiltonian, in principle, selects a preferred class of
states as ground states, namely its zero eigenvalue states. It may
be that zero or the minimum of the spectrum is not in the discrete
(more precisely, pure point) spectrum or that the zero eigenvalue is
vastly degenerate, however, at least conceptually, this appears to
be an improvement.
\end{itemize}
~\\
In what follows we now turn to the proof of the above claims.

\subsection{Conservation Equations}
\label{s3.1}

As shown in the appendix of \cite{1}, for any fully conserved
quantity $F$ of a Hamiltonian sytem with Hamiltonian $H$, when we
expand both the equations of motion and $F$ to order $n$ then $F$ is
still a constant of motion up to terms of order $n+1$. In
particular, for $n=1$ the coefficients of the correction vanish.
This means that we can derive conservation equations for manifestly
gauge-invariant perturbation theory by expanding the four times
infinitely many constants of motion $C_j(\sigma),\;H(\sigma)$ where
\ba \label{3.3} C_j(\sigma) &=&-\frac{2}{\kappa} D_k
P^k_j+\frac{1}{\lambda} \Pi D_j\Xi
\nonumber\\
H(\sigma)&=& \sqrt{C(\sigma,\tau)^2-Q^{jk}(\sigma,\tau) C_j(\sigma)
C_k(\sigma)}
\nonumber\\
C &=& C_{\rm geo}+C_{\rm matter}
\nonumber\\
C_{\rm geo} &=&
\frac{1}{\kappa}\Big[\frac{1}{\sqrt{\det(Q)}}G_{jkmn}P^{jk}P^{mn}-
\sqrt{\det(Q)} [R[Q]-2\Lambda]\Big]
\nonumber\\
C_{\rm matter} &=& \frac{1}{2\lambda}\Big[\frac{\Pi^2}{\sqrt{\det(Q)}}
+\sqrt{\det(Q)}(Q^{jk}\Xi_{,j}\Xi_{,k}+v(\Xi))\Big].
 \ea Let
$\overline{\epsilon}_j(\sigma)=-\overline{C}_j(\sigma),\;
\overline{\epsilon}(\sigma)=\overline{H}(\sigma)$ be the zeroth
order values of the constants of motion which are obtained
explicitly by inserting the FRW solution of appendix \ref{sb.1} into
the expressions (\ref{3.3}). We now expand $\delta
\epsilon_j:=\epsilon_j-\overline{\epsilon_j},\; \delta
\epsilon:=\epsilon-\overline{\epsilon}$ to first order, which then
are still constants of the linearised equation of motion. We find
after a short calculation (remember that $D_k P^k_j=\partial_k
P^k_j-\Gamma^l_{kj} P^k_l$, since $P^{jk}$ is a tensor density)  and
using the perturbed gauge-invariant variables defined in appendix
\ref{sb.2} \ba \label{3.4} -\delta \epsilon_j &=&
-\frac{2}{\kappa}\Big(A^2 \delta P^{jk}_{,k}-2\dot{A}\big(\delta
Q_{jk,k}-\frac{1}{2}\delta
Q_{kk,j}\big)\Big)+\frac{1}{\lambda}\overline{\Pi}\delta\Xi_{,j}
\nonumber\\
\delta \epsilon &=& \frac{1}{\overline{H}}\Big[\overline{C}\delta C
-\overline{Q}^{jk}
\overline{C}_k \delta C_j-
\overline{C}_k \overline{C}_j \delta Q^{jk}\Big]=\delta C
\nonumber\\
\kappa \delta C_{\rm geo} &=& -\Big[\frac{\dot{A}^2}{A} \delta^{jk}\delta
Q_{jk} -2\dot{A}A \delta_{jk}\delta P^{jk}+ A\delta^{jk} \delta
R_{jk}\Big]+\Lambda A \delta^{jk} \delta Q_{jk}
\nonumber\\
\delta R_{jk} &=& \frac{1}{2A^2}\Big[2\delta Q_{l(j,k)l}-\Delta \delta
Q_{jk}-\delta Q_{ll,jk}\Big]
\nonumber\\
\lambda \delta C_{\rm matter} &=&
\frac{1}{2}\Big[2\frac{\overline{\Pi}}{A^{3}}\delta \Pi-\frac{1}{2}
A \overline{p} \delta^{jk} \delta Q_{jk}+A^{3} v'(\overline{\Xi})
\delta\Xi\Big]. \ea Here we used that that the quantities $L,I$
introduced in appendix \ref{sb.1} are explicitly given by $L=A^2$
and $I=-2\dot{A}$. We will discuss the momentum and energy
conservation equations separately in what follows.

\subsubsection{Momentum conservation equation}
\label{s3.1.1}

We substitute the perturbed configuration variables for the
perturbed momenta by using the equations of motion, see equation
(\ref{b.19}). Notice that $N_j=-C_j/H$ is also a constant of motion
and that $\delta N_j=\delta \epsilon_j/\overline{\epsilon}$. Then we
find from the first relation in (\ref{3.4}) \ba \label{3.5} &&
\delta \epsilon_j+\frac{\kappa A}{\overline{\epsilon}}\big[\Delta
\delta\epsilon_j-\delta \epsilon_{k,kj}\big]
\nonumber\\
&=& \frac{1}{\kappa}\big[A(\delta\dot{Q}_{jk,k}-\delta Q_{kk,j})
-2\dot{A}(\delta Q_{jk,k}-\delta Q_{kk,j})\big] -\frac{1}{\lambda}
\overline{\Pi}\delta \Xi_{,j}. \ea In order to connect (\ref{3.5})
with the SCPT equations we parametrise our metric perturbations in
the following way: First of all, SCPT employs conformal time
$x^0=\eta$ for which $\bar{g}_{00}=-a^2$. On the other hand, we
always have $g_{\tau\tau}:=-N^2+Q^{jk} N_j N_k=-1$, which means that
we automatically work with cosmological time, that is, the eigentime
of the dust. We use the letter $A$ instead of $a$ for the scale
factor in what follows in order to emphasise that $A$ is an
observable, rather than a gauge-dependent function. From the
relation $dx^0=d\tau/A$ we find $g_{\tau\tau}=g_{00}/A^2,\;g_{\tau
j}=g_{0j}/A=N_j,\;g_{jk}=Q_{jk}$. Now in analogy to SCPT we
introduce the following ten functions \be \label{3.6}
g_{00}=(-1+2\phi)A^2,\;g_{0j}=(B_{,j}+S_j)A^2,\;g_{jk}=A^2(2\psi
\delta_{jk}+2 E_{,jk}+2F_{(j,k)}+h_{jk}), \ee where the tensors
$S_j,F_j,h_{jk}$ are transversal with respect to the Euclidean
metric $\delta_{jk}$ and $h_{jk}$ is trace free. We can also express
$\psi,E,F_j,h_{jk}$ as \ba \label{3.7} \psi &=&
\frac{1}{4A^2}\left(\delta Q_{kk}-\Delta^{-1} \delta
Q_{jk,jk}\right)
\nonumber\\
E &=& -\frac{1}{4A^2} \Delta^{-1}\left(\delta Q_{kk}-3\Delta^{-1} \delta
Q_{jk,jk}\right)
\nonumber\\
E &=& -\frac{1}{4A^2} \Delta^{-1}\left(\delta Q_{kk}-3\Delta^{-1} \delta
Q_{jk,jk}\right)
\nonumber\\
F_j &=& -\frac{1}{A^2} \Delta^{-1}\left(\delta Q_{jk,k}-\Delta^{-1} \delta
Q_{kl,jkl}\right)
\nonumber\\
h_{jk} &=& \delta Q_{jk}-2\left(\psi
\delta_{jk}+E_{,jk}+F_{(j,k)}\right). \ea From $g_{\tau\tau}=-1$ we
immediately see that automatically \be \label{3.8} \phi=0, \ee that
is, we are forced to work in partly synchronous\footnote{In the
usual not manifestly gauge invariant formalism, synchronous gauge
means $\phi=B=0$ while longitudinal gauge means $B=E=0$.} ``gauge''.
We write this expression in inverted commas because there is no
gauge involved here, all the quantities are manifestly gauge
invariant. We use this term only to make the analogy clear. In
particular, since $N_j=-C_j/H$ can be expressed in terms of
$Q_{jk},\;\Xi$ and (via the equations of motion) their velocities,
in our formalism the functions $B,S_j$ are not independent variables
from the outset.

Let us decompose (\ref{3.5}) into longitudinal $\delta
\epsilon^{||}_j=\Delta^{-1}\delta\epsilon_{k,kj}$ and transversal
$\delta \epsilon^\perp_j=\epsilon_j-\epsilon^{||}_j$ parts,
respectively \ba \label{3.9}
 \delta \epsilon^\perp_j+\frac{\kappa L^{1/2}}{\overline{\epsilon}}\Delta
\delta\epsilon^\perp_j &=&
\frac{1}{\kappa}\left[A\left(\delta\dot{Q}_{jk,k}-\Delta^{-1}\delta
Q_{kl,klj}\right) -2\dot{A}\left(\delta Q_{jk,k}-\Delta^{-1}\delta
Q_{kl,klj}\right)\right].
 \nonumber \ea Then we find \ba \delta
\epsilon_{j,j} &=&
\frac{1}{\kappa}\left[A\left(\delta\dot{Q}_{jk,jk}-\Delta \delta
Q_{kk}\right) -2\dot{A}\left(\delta Q_{jk,jk}-\Delta \delta
Q_{kk}\right)\right] -\frac{1}{\lambda} \overline{\Pi}\Delta \delta
\Xi. \ea Next we invoke (\ref{3.6}), (\ref{3.7}) remembering that
$\delta N_j=A(S_j+B_{,j})=-\delta\epsilon_j/\overline{\epsilon}$ and
that $\dot{(.)}=d(.)/d\tau=\frac{1}{A}(.)'$. Then the transversal
part of (\ref{3.9}) becomes simply \be \label{3.10} \Delta
V_j=-\kappa\frac{\delta \epsilon^\perp_j}{A^2} \ee where we have
introduced the variable $V_j=S_j-F'_j$ which is the analog to the
variable $V_a$ in the SCPT framework, see appendix \ref{sa}. We also
used $f'=a \dot{f}$ several times and, of course, $L=A^2$ and
$I=-2\dot{A}$. Equation (\ref{3.10}) should be compared\footnote{In
order to avoid confusion of the reader we point out that the
individual contributions of geometry and matter to the Hamiltonian
and spatial diffeomorphism constraints (without dust) $c$ and $c_a$
respectively are {\it not exactly} the temporal -- temporal and
temporal -- spatial components of the Einstein and energy momentum
tensor respectively, rather we have the identities \be \label{3.10a}
\kappa c^{grav}_a=-\sqrt{\det(q)} X^\mu_{,a} n^\nu G_{\mu\nu},\;\;
\kappa c^{grav}=-2\sqrt{\det(q)} n^\mu n^\nu G_{\mu\nu} \ee where
$n^\mu$ is the unit future normal to a foliation $X$.
Noticing that $n^\mu=(X^\mu_{,t}-n^a X^\mu_{,a})/n$
and that  $\delta n_a=\delta(q_{ab} n^b)=a(B_{,a}+S_a)\not=0$ one
finds \be \label{3.10b} \kappa \delta c_a=-a^2 [\delta
G_{0a}+(B_{,a}+S_a)(2{\cal H}'+{\cal H}^2),\;\; \kappa c=-2a [\delta
G_{00}+3{\cal H}^2(3\psi+\Delta E)] \ee The additional terms cancel
when we replace $G_{\mu\nu}$ by $E_{\mu\nu}=G_{\mu\nu}+\Lambda
g_{\mu\nu}-\frac{\kappa}{2}T_{\mu\nu}$ since
$\overline{E}_{\mu\nu}=0$. } with the vector contribution of the
perturbation of the temporal -- spatial part of the Einstein
equation, written in linearly gauge-invariant form, which we derived
in the 2nd equation of (\ref{a.28a}), the second relation of
(\ref{a.16b}) and the first relation of (\ref{a.33a}). That is \ba
\label{3.11} 0&=& \delta\tilde{G}_{0a}+\Lambda
\delta\tilde{g}_{0a}-\frac{\kappa}{2}\tilde\delta T_{0a}
\nonumber\\
&=& \left(-2(\Psi'+{\cal H}\Phi)_{,a}-\frac{1}{2}\Delta V_a-(2{\cal
H}'+{\cal H}^2) V_a\right)+\Lambda a^2
V_a-\left(\frac{\kappa}{2\lambda}(\overline{\zeta}'Z_{,a}
+a^2\lambda \overline{p} V_a)\right)
\nonumber\\
&=& \left(-2(\Psi'+{\cal H}\Phi)_{,a}-\frac{1}{2}\Delta V_a\right)
-\left(\frac{\kappa}{2\lambda}\overline{\zeta}'Z_{,a}\right) \ea
where we used the spatial -- spatial part of the background FRW
equations (\ref{a.7}), (\ref{a.7a}) in the second step, that is \be
\label{3.12} 2{\cal H}'+{\cal H}^2=a^2\left(\Lambda-\frac{\kappa}{2}
\overline{p}\right) \ee and where
$\overline{p}=\frac{1}{2\lambda}(\dot{\overline{\Xi}}^2-
v(\overline{\Xi}))$ (this holds with or without dust). The
transversal piece of (\ref{3.12}) is simply \be \label{3.13}
-\frac{1}{2}\Delta V_a=0.
 \ee If we compare this to (\ref{3.10}) and
if we identify the SCPT variable $V_a$ with $V_j$ then we see that,
instead of a Laplace equation without source, we have a
non-vanishing source. However, recall that $\delta\epsilon^\perp_j$
is a constant of motion. Hence, the source term decays as $1/A^2$
and thus plays no role anymore in the late universe. This happens
the sooner the smaller $\delta\epsilon^\perp_j$.

Next we turn to the longitudinal part of (\ref{3.9}) which can be
written as \be \label{3.14} \psi'_{,j}=-\frac{\kappa}{4}
\left(\frac{1}{A^2}\delta
\epsilon^{||}_j+\frac{1}{\lambda}\overline{\Xi}'\delta
\Xi_{,j}\right) \ee where we used $\overline{\Pi}=A^3
\dot{\overline{\Xi}}=A^2 \overline{\Xi}'$. A direct comparison
between (\ref{3.14}) and SCPT should be in terms of variables which
are direct analogs of the linearly invariant SCPT variables
$\Psi,\Phi,Z$. These are given by $\Psi=\psi+{\cal
H}f,\;\Phi=\phi-{\cal H}f-f',\;Z=\zeta+\overline{\zeta}'f$ where we
of course identify $\zeta$ with our $\Xi$ and where $f:=B-E'$. Hence
we should rewrite (\ref{3.14}) in terms of these variables,
recalling that in our formalism $\phi\equiv 0$ so that our $\Phi$
actually reads $\Phi=-{\cal H}f-f'$. This leads to \be \label{3.15}
\left[\Psi'+{\cal H} \Phi
+\frac{\kappa}{4\lambda}\overline{\Xi}'\delta Z\right]_{,j}
=\frac{\kappa}{4A}
\left(-\frac{1}{A}\delta\epsilon^{||}_j+\overline{\epsilon}
\left[B-E'\right]_{,j}\right). \ee In order to arrive at this form
of (\ref{3.14}), we simply have rewritten $\psi,\Xi$ in terms of
$\Psi,Z$ and made use of the gauge-invariant background FRW
equations with dust (\ref{b.11}), (\ref{b.12}). They are an exact
solution to our invariant Hamiltonian equations and allowed us to
write $\kappa(\overline{\Xi}')^2/(4\lambda)=-({\cal H}'-{\cal
H}^2-\overline{\epsilon}\kappa)/(4A)$ so that certain terms really
combine to $\Phi$. We now compare this to the longitudinal part of
the linearly invariant projection of the temporal -- spatial part of
the Einstein equations (\ref{3.11}) in the SCPT framework which can
be rewritten as \be \label{3.16} \left[\Psi'+{\cal
H}\Phi+\frac{\kappa}{4\lambda}\overline{\zeta}' Z\right]_{,a}=0. \ee
We see that (\ref{3.16}) and (\ref{3.15}) agree with each other
under the made identification $\zeta=\Xi$ up to the corrections on
the right hand side of (\ref{3.15}). The first of these terms decays
in time as $1/A^2$ since $\delta\epsilon^{||}_j$ is a constant of
motion. The second term decays only as $1/A$ under the assumption
that $B-E'$ remain small during the evolution, an assumption that
one always makes in perturbation theory. Notice that in the SCPT
framework the quantities $B,E$ and also the combination $B-E'$ is
not linearly gauge-invariant. However, in our framework, all those
quantities are gauge invariant to all orders from the outset so that
(\ref{3.15}) is a consistent relation among manifestly
gauge-invariant variables. We just have written it in a form as
close as possible to the equations in the usual formalism.

\subsubsection{Energy conservation equation}
\label{s3.1.2}

Finally, we consider the second conservation equation in (\ref{3.4})
which after some algebra can be brought into the form \ba
\label{3.17} \delta\epsilon &=& \delta C_{\rm geo}+\delta C_{\rm matter}
\nonumber\\
\delta C_{\rm geo} &=& -\frac{4A}{\kappa}\left(-\Delta\Psi+3{\cal
H}'\psi'\right)
+\frac{3\psi+\Delta E}{\kappa}\left(2\Lambda A^3-6{\cal H}^2A\right)
\nonumber\\
\delta C_{\rm matter} &=&
\frac{A^3}{2\lambda}\left(2\dot{\overline{\Xi}}\delta\dot{\Xi}+v'(\overline{\Xi}
)\delta\Xi\right) +A^3\left(3\psi+\Delta E\right)\overline{\rho} \ea
where
$\overline{\rho}=\frac{1}{2\lambda}(\dot{\overline{\Xi}}^2+v(\overline{\Xi}))$.
Using our FRW background equations (\ref{b.11}) and (\ref{b.12}), we
can simplify (\ref{3.17}) to \ba \label{3.18} \delta\epsilon
-\overline{\epsilon}(3\psi+\delta E) &=&
\frac{A^3}{2\lambda}\left(2\dot{\overline{\Xi}}\delta\dot{\Xi}+
v'(\overline{\Xi})\delta\Xi\right)
-\frac{4A}{\kappa}\left(-\Delta\Psi+3{\cal H}'\psi'\right) \ea which
now needs to be written in terms of the SCPT-like variables \be
\Psi:=\psi+{\cal H}f,\quad\Phi=-{\cal H}f-f',\quad
Z:=\Xi+\overline{\Xi}'f\quad\mathrm{with}\quad f=(B-E'). \ee We do
this by substituting $\psi,\Xi$ in (\ref{3.18}) by $\Psi,\;Z$,
respectively, thereby picking up the following correction term just
involving $f$ \be \label{3.19} A\left[\frac{12}{\kappa}{\cal
H}({\cal H}f)' -\frac{1}{2\lambda}(2\overline{\Xi}'(\overline{\Xi}'
f)'+A^2v'(\overline{\Xi})\overline{\Xi}' f)\right]. \ee We want to
massage this into a form that involves $\Phi$. We have, using
$A^2\overline{\rho}=\frac{1}{2\lambda}((\overline{\Xi}')^2
+v(\overline{\Xi}))$, \ba \label{3.20}
-\frac{1}{2\lambda}\left(2\overline{\Xi}'\left(\overline{\Xi}'
f\right)'+A^2v'\left(\overline{\Xi}\right)\overline{\Xi}' f\right)
&=& \left(A^2\overline{\rho}\right)' f+2\left(A^2
\overline{\rho}\right) f'-\frac{A}{\lambda}\left(Af\right)'
v(\overline{\Xi})
\\
&=& \left(A^2\overline{\rho}\right)' f+2\left(A^2 \overline{\rho}\right)
f'+\frac{A^2}{\lambda}\Phi
v(\overline{\Xi})\nonumber
\ea
where we used $\Phi=-(Af)'/A$. Now we use our background FRW equation
in the form
\be \label{3.21}
3{\cal
H}^2=A^2\left[\Lambda+\frac{\kappa}{2}\left(\frac{\overline{\rho}-\overline{
\epsilon}}{A^3}\right)\right]
\ee
from which we can compute $(A^2\overline{\rho})'$ by taking its derivative.
Combining then (\ref{3.20}) and (\ref{3.19}) we see that the
$\frac{12}{\kappa}{\cal H}({\cal H}f)'$ term in (\ref{3.19}) is
cancelled and that the remaining terms are either proportional to $\Phi$
or to $\overline{\epsilon}$. The end result is
\ba \label{3.22}
&& \frac{1}{A}\left[\delta\epsilon-\overline{\epsilon}\left(3\psi+\Delta
E-2(B-E')'+{\cal H}(B-E')\right)\right]
\nonumber\\
&=& \frac{1}{2\lambda}\left(2\overline{\Xi}' Z'+A^2
v'(\overline{\Xi})Z\right) -\frac{4}{\kappa}\left(-\Delta\Psi+3{\cal
H}'\Psi'\right)
-A^2\left(\frac{v(\overline{\Xi})}{\lambda}+\frac{4\Lambda}{\kappa}\right)\Phi.
\ea This should now be compared with the temporal -- temporal
Einstein equation, written in terms of the SCPT variables, derived
in the first equation of (\ref{a.28a}), the first equation of
(\ref{a.16ba}) and (\ref{a.37}), that is \ba \label{3.23} 0 &=&
\delta\tilde{G}_{00}+\Lambda\delta\tilde{g}_{00}-\frac{\kappa}{2}
\delta \tilde{T}_{00}
\nonumber\\
&=& 2(-\Delta\Psi+3{\cal
H}\Psi')+2a^2\Lambda\Phi-\frac{\kappa}{4\lambda}\left(2\overline{\zeta}'
Z'+a^2 v'(\overline{\zeta})Z-2a^2 v(\overline{\zeta})\Phi\right).
\ea Comparing (\ref{3.22}) and (\ref{3.23}), we see that the two
equations agree, with proper identification of the variables, up to
the correction term on the left hand side of (\ref{3.22}) which is
proportional to $\delta\epsilon,\;\overline{\epsilon}$ and the
perturbations $B,E,\psi$, which in our formalism are gauge
invariant. Assuming that these terms stay small during the cosmic
evolution as one always does, the correction term decays as $1/A$
and thus is negligible in the late universe.

\subsection{Evolution Equations}
\label{s3.2}

We derive the following perturbed evolution equation in
(\ref{b.24}):
\ba \label{3.24} \delta \ddot{Q}_{jk} &=&
\frac{\dot{A}}{A} \delta \dot{Q}_{jk}+2 \frac{\dot{A}}{A} (\delta
N_{(j})_{k)}+2 \frac{\ddot{A}}{A} \delta Q_{jk} -2\delta R_{jk}
\nonumber\\
&&+\delta_{jk}\left(\frac{1}{2} \delta R_{mm}
-\frac{\kappa}{2\lambda}
A^2(\dot{\overline{\Xi}}\delta\dot{\Xi}-\frac{1}{2}v'(\overline{\Xi})\delta
\Xi)\right).
 \ea
 First of all, we change the time variable again to
$dx^0=d\tau/A$ and write $\delta Q_{jk}=A^2 H_{jk}$, which gives \ba
\label{3.25} \delta \dot{Q}_{jk} &=& A\left(2{\cal H}
H_{jk}+H'_{jk}\right)
\nonumber\\
\delta \ddot{Q}_{jk} &=& 2\left({\cal H}'+{\cal H}^2\right) H_{jk}+3{\cal
H} H'_{jk}+H^{\prime\prime}_{jk}
\nonumber\\
\frac{\dot{A}}{A} \delta \dot{Q}_{jk} &=& {\cal H}\left(2{\cal H}
H_{jk}+H'_{jk}\right)
\nonumber\\
\frac{\ddot{A}}{A} \delta Q_{jk} &=& {\cal H}' H_{jk}
\nonumber\\
\delta R_{jk} &=& \frac{1}{2}\left(2H_{l(j,k)l}-\Delta
H_{jk}-H_{ll,jk}\right).
 \ea
 Inserting (\ref{3.25}) into
(\ref{3.24}) and using $\delta N_j=A[B_{,j}+S_j]$ yields
\ba
\label{3.26} 2{\cal H} H'_{jk}+H^{\prime\prime}_{jk} &=& 2 {\cal H}
\left[B_{,(j}+S_{,(j}\right]_{,k)}-2\delta R_{jk}
\nonumber\\
&&+\delta_{jk}\left(\frac{1}{2} \delta R_{mm}
-\frac{\kappa}{2\lambda}
A^2(\dot{\overline{\Xi}}\delta\dot{\Xi}-\frac{1}{2}v'(\overline{\Xi})\delta
\Xi)\right).
 \ea
 Equation (\ref{3.26}) should be compared to the
spatial -- spatial part of the Einstein equations written in
linearly gauge invariant form. Using the third relation in
(\ref{a.28a}), the third relation in (\ref{a.16ba}), as well as the
spatial -- spatial components of (\ref{a.30}), (\ref{a.33a}) and
(\ref{a.37}) gives \ba \label{3.27} \delta\tilde{G}_{ab}+\Lambda
\tilde{g}_{ab}-\frac{\kappa}{2}\delta\tilde{T}_{ab} &=&
\frac{1}{2}\left(h^{\prime\prime}_{ab}+2{\cal H} h_{ab}-\Delta
h_{ab}\right) -\left(2{\cal H}'+{\cal
H}^2\right)h_{ab}-\left(V'_{(a,b)}+2{\cal H}V_{(a,b)}\right)
\nonumber\\
&&
+\left[\Phi-\Psi\right]_{,ab}
-\left[\Delta(\Phi-\Psi)+2\left(\Psi^{\prime\prime}+(2{\cal H}'+{\cal
H}^2)(\Phi+\Psi)
+{\cal H}(2\Psi+\Phi)'\right)\right]\delta_{ab}
\nonumber\\
&& +\Lambda a^2\left(h_{ab}+2\Psi \delta_{ab}\right)
\nonumber\\
&& -\frac{\kappa}{2\lambda}\left(\lambda \overline{p} a^2 h_{ab}+2\lambda
\overline{p}
a^2
\Psi\delta_{ab}+\frac{1}{2}(2\overline{\zeta}'\left(\Phi+Z'\right)
-v'(\overline{\zeta}) a^2
Z)\delta_{ab}\right)
\nonumber\\
&=&
\frac{1}{2}\left(h^{\prime\prime}_{ab}+2{\cal H} h_{ab}-\Delta h_{ab}\right)
-\left(V'_{(a,b)}+2{\cal H}V_{(a,b)}\right)
\nonumber\\
&&
+\left[\Phi-\Psi\right]_{,ab}
-\left[\Delta(\Phi-\Psi)+2\left(\Psi^{\prime\prime}+(2{\cal H}'+{\cal
H}^2)\Phi
+{\cal H}(2\Psi+\Phi)'\right)\right]\delta_{ab}
\nonumber\\
&& -\frac{\kappa}{4\lambda}\left(
2\overline{\zeta}'\left(\Phi+Z'\right)-v'(\overline{\zeta}) a^2
Z\right)\delta_{ab}.
 \ea Here we used (\ref{3.12}) twice in the
second step.

In what follows we decompose (\ref{3.26}) into the various
irreducible pieces. We will see that, in contrast to the
conservation equations, maybe not very surprisingly, the evolution
equations do not adopt any corrections.

\subsubsection{Tensor contribution}
\label{s3.2.1}

In this case $H_{jk}=h_{jk},\;\delta N_j=0$ and thus $\delta
R_{jk}=-\Delta h_{jk}/2$, so that the tracefree, transversal
contribution to (\ref{3.26}) reduces to \be \label{3.28} -\Delta
h_{jk}+2{\cal H} h'_{jk}+h^{\prime\prime}_{jk}=0. \ee This should be
compared to the corresponding tensor contribution to (\ref{3.27})
which is given by \be \label{3.29}
\frac{1}{2}\left(h^{\prime\prime}_{ab}+2{\cal H} h_{ab}-\Delta
h_{ab}\right)=0. \ee Thus we obtain an exact match upon identifying
$h_{jk}$ on the dust space $\cal S$ with $h_{ab}$ on $\cal X$ which
is diffeomorphic to $\cal S$.

\subsubsection{Vector contribution}
\label{s3.2.2}

In this case $H_{jk}=2A^2 F_{(j,k)}$ and $B=0$. A short computation
reveals that $\delta R_{jk}=0$. Hence, dropping the term
proportional to $\delta_{jk}$ in (\ref{3.26}) which is a scalar
contribution, we find \be \label{3.30} 4{\cal H}
F'_{(j,k)}+2F^{\prime\prime}_{(j,k)} = 2 {\cal H} S_{(j,k)}. \ee
Using the conservation equation $(\delta N_j)'=(A S_j)'=0$, we find
the identity \be \label{3.31} {\cal H} S_j+S'_j=0 \;\;\Rightarrow
\;\;{\cal H} S_{(j,k)}+S'_{(j,k)}=0. \ee Adding twice the second
zero in (\ref{3.31}) to (\ref{3.30}) and dividing the resulting
equation by two, we find \be \label{3.32} 2{\cal H}
V_{(j,k)}+V'_{(j,k)}=0, \ee where we have again introduced
$V_j=S_j-F'_j$.

On the other hand the vector contribution to (\ref{3.27}) is
obviously \be \label{3.33} V'_{(a,b)}+2{\cal H}V_{(a,b)}=0, \ee
which is again an exact match when identifying $V_a$ and $V_j$. It
is instructive how the conservation law (\ref{3.31}) found its way
into the equations in order to establish this match.

\subsubsection{Scalar contribution}
\label{s3.2.3}

This time $H_{jk}=2(\psi\delta_{jk}+E_{,jk})$ and $S_j=0$. A short
calculation reveals that
$\delta R_{jk}=-(\psi_{,jk}+\Delta\psi \delta_{jk})$
and thus $\delta R_{kk}=-4\Delta \psi$. Thus (\ref{3.26})
becomes
\ba \label{3.34}
4{\cal H} (\psi' \delta_{jk}+E'_{,jk})+2
(\psi^{\prime\prime}\delta_{jk}+E^{\prime\prime}_{,jk})
&=& 2 {\cal H} B_{,jk}+2(\psi_{,jk}+\Delta\psi \delta_{jk})
+\delta_{jk}\left(-2\Delta \psi
-\frac{\kappa}{2\lambda}
A^2(\dot{\overline{\Xi}}\delta\dot{\Xi}-\frac{1}{2}v'(\overline{\Xi})\delta
\Xi)\right).\nonumber\\
\ea Equation (\ref{3.34}) is of the form $f_{,jk}+g \delta_{jk}=0$.
Taking the trace and operating with $\partial_j \partial_k$,
respectively, we learn that $3g+\Delta f=0$ and $\Delta(g+\Delta
f)=0$, respectively. Since, due to our boundary conditions, there
are no zero modes of the Laplacian\footnote{The proof is standard:
Assuming that $f$ is an at least twice differentiable function which
decays at infinity at least as $1/r$ where $r$ is an asymptotic
radial variable. It follows that $f_{,j}$ is square integrable with
respect to $L_2({\cal S},d^3\sigma)$. Now suppose that $\Delta f=0$.
Then $0=<f,\Delta f>=-\sum_j ||f_{,j}||^2$ where the boundary term
at the sphere at infinity drops out since $f f_{,j}$ decays as
$1/r^3$. It follows that $f_{,j}=0$ a.e. and since $f$ is in
particular continuous, it follows $f=$const. But that constant must
vanish since $f$ decays.}, we conclude $f=g=0$. Therefore we can
study the $(.)\delta_{jk}$ and $(.)_{,jk}$ pieces in (\ref{3.34})
separately, which we note as \ba \label{3.35} \left(4{\cal H}
\psi'+2\psi^{\prime\prime}- 2 \Delta\psi\right)\delta_{jk} &=&
\left(-2\Delta \psi -\frac{\kappa}{2\lambda}
A^2(\dot{\overline{\Xi}}\delta\dot{\Xi}-\frac{1}{2}v'(\overline{\Xi})\delta
\Xi)\right)\delta_{jk}
\nonumber\\
\left[4{\cal H} E'+2E^{\prime\prime}-2 {\cal H}
B-2\psi\right]_{,jk}&=&0. \ea Consider first the second equation. We
note the conservation equation $\delta N_j'=(A B_{,j})'=0$, which
can be written in the form \be \label{3.36} {\cal H}
B_{,j}+B'_{,j}=0\;\;\Rightarrow \;\; {\cal H} B+B'=0. \ee Here the
second relation follows again from the fact that the Laplacian has
no zero modes. It follows that \be \label{3.37} \Phi:=-\left[{\cal
H}(B-E')+(B-E')'\right]={\cal H} E'+E^{\prime\prime}. \ee Now also
using the definition $\Psi=\psi+{\cal H}(B-E')$ and (\ref{3.36}), a
short calculation reveals that the second equation in (\ref{3.35})
simply becomes \be \label{3.38} 2\left[-\Psi+\Phi\right]_{,jk}=0.
\ee

Next we consider the first equation in (\ref{3.35}), which can be
simplified to \be \label{3.39} \left(4{\cal H}
\psi'+2\psi^{\prime\prime}\right)\delta_{jk} =
-\frac{\kappa}{2\lambda}
\left(\overline{\Xi}'\delta\Xi'-\frac{A^2}{2}v'(\overline{\Xi})\delta
\Xi\right)\delta_{jk}. \ee We write $\delta\Xi=Z-\overline{\Xi}'f$,
where $f=B-E'$, so that on the right hand side of (\ref{3.39})
$\delta \Xi$ is replaced by $Z$ plus the correction term \ba
\label{3.40} \frac{\kappa}{4\lambda} \left(2\overline{\Xi}'
(\overline{\Xi}' f)'-A^2 v'(\overline{\Xi})\overline{X}' f\right)
&=&\frac{\kappa}{4\lambda} \left( ([\overline{\Xi}']^2)'
f+2(\overline{\Xi}')^2 f' -A^2 (v(\overline{\Xi}))' f\right)
\nonumber\\
&=&\frac{\kappa}{4\lambda}
\left[ \left((\overline{\Xi}')^2-A^2 v(\overline{\Xi})\right)'
f+2(\overline{\Xi}')^2 f'
+2 A^2 {\cal H} v(\overline{\Xi}) f\right]
\nonumber\\
&=&\frac{\kappa}{4\lambda}
\left[ 2\lambda \left(A^2 \overline{p}\right)' f+2(\overline{\Xi}')^2 (f'+{\cal
H} f)
-2 {\cal H} \left((\overline{\Xi}')^2-A^2 v(\overline{\Xi})\right) f\right]
\nonumber\\
&=&\frac{\kappa}{4\lambda}
\left[ 2\lambda \left(A^2 \overline{p}\right)' f-2(\overline{\Xi}')^2 \Phi
-4\lambda {\cal H} A^2 \overline{p} f\right]
\nonumber\\
&=&
-\frac{\kappa}{2\lambda} (\overline{\Xi}')^2 \Phi
+\frac{\kappa}{2}\left[\left(A^2 \overline{p}\right)' -2 {\cal H} A^2
\overline{p}\right]f
\nonumber\\
&=&
-\frac{\kappa}{2\lambda} (\overline{\Xi}')^2 \Phi
+\left(A^2 \Lambda -(2{\cal H}'+{\cal H}^2)\right)' f -2\frac{\kappa}{2} {\cal
H}
A^2 \overline{p} f
\nonumber\\
&=&
-\frac{\kappa}{2\lambda} (\overline{\Xi}')^2 \Phi
-2\left({\cal H}^{\prime\prime}+{\cal H} {\cal H}'\right) f
+2\left(\Lambda-\frac{\kappa}{2} \overline{p}\right){\cal H}
A^2 f
\nonumber\\
&=&
-\frac{\kappa}{2\lambda} (\overline{\Xi}')^2 \Phi
-2\left({\cal H}^{\prime\prime}+{\cal H} {\cal H}'\right) f
+2\left(2{\cal H}'+{\cal H}^2\right){\cal H}  f
\nonumber\\
&=& -\frac{\kappa}{2\lambda} (\overline{\Xi}')^2 \Phi -2\left({\cal
H}^{\prime\prime}-{\cal H} {\cal H}'-{\cal H}^3\right)f \ea times
$\delta_{jk}$. Here we have used (\ref{3.12}) twice and the
definition $\Phi:=-(f'+{\cal H}f)$.

On the other hand, if we write $\psi=\Psi-{\cal H}f$ on the left hand
side of (\ref{3.39}) then we can replace $\psi$ by $\Psi$ plus the
correction term
\ba \label{3.41}
 -2\left(2{\cal H} ({\cal H}f)'+({\cal H}f)^{\prime\prime}\right)
&=& -2\left(2{\cal H} ({\cal H}f)'+{\cal H}^{\prime\prime}f+2{\cal H}'
f'+{\cal H} f^{\prime\prime}\right)
\nonumber\\
&=& -2\left({\cal H} ({\cal H}f)'+{\cal H}^{\prime\prime}f+2{\cal H}'
f'+{\cal H} (f'+{\cal H}f)'\right)
\nonumber\\
&=&
2{\cal H}\Phi'
-2\left({\cal H} {\cal H}'f+{\cal H}^{\prime\prime}f+(2{\cal H}'+{\cal H}^2)
f'\right)
\nonumber\\
&=&
2\left({\cal H}\Phi'+(2{\cal H}'+{\cal H}^2)\Phi\right)
-2\left({\cal H} {\cal H}'+{\cal H}^{\prime\prime}f-(2{\cal H}'+{\cal H}^2)
{\cal H}\right)f
\nonumber\\
&=& 2\left({\cal H}\Phi'+(2{\cal H}'+{\cal H}^2)\Phi\right)
-2\left({\cal H}^{\prime\prime}f-{\cal H}' {\cal H}-{\cal H}^3)
{\cal H}\right)f \ea times $\delta_{jk}$. We have again used the
definition $\Phi:=-(f'+{\cal H}f)$ frequently. Combining
(\ref{3.40}) and (\ref{3.41}), we see that we can rewrite
(\ref{3.39}) in the form \ba \label{3.42} && \left[4{\cal H}
\Psi'+2\Psi^{\prime\prime} +2({\cal H}\Phi'+(2{\cal H}'+{\cal
H}^2)\Phi) -2({\cal H}^{\prime\prime}f-{\cal H}' {\cal H}-{\cal
H}^3) f\right]\delta_{jk}
\nonumber\\
&=& \left[-\frac{\kappa}{2\lambda}
(\overline{\Xi}'Z'-\frac{A^2}{2}v'(\overline{\Xi})Z+
(\overline{\Xi}')^2 \Phi) -2({\cal H}^{\prime\prime}-{\cal H} {\cal
H}'-{\cal H}^3)f\right]\delta_{jk} \ea or \ba \label{3.43}
\left[2{\cal H}(2\Psi+\Phi)'+2\Psi^{\prime\prime} +2(2{\cal
H}'+{\cal H}^2)\Phi\right]\delta_{jk} &=& -\frac{\kappa}{2\lambda}
\left[\overline{\Xi}'Z'-\frac{A^2}{2}v'(\overline{\Xi})Z+
(\overline{\Xi}')^2 \Phi\right] \delta_{jk}. \ea Now the scalar
contribution to (\ref{3.27}) is evidently given by \ba \label{3.44}
0&=&\left[\Phi-\Psi\right]_{,ab}
-\left[\Delta(\Phi-\Psi)+2(\Psi^{\prime\prime}+(2{\cal H}'+{\cal
H}^2)\Phi +{\cal H}(2\Psi+\Phi)'\right]\delta_{ab}
\nonumber\\
&& -\frac{\kappa}{4\lambda}\left[
2\overline{\zeta}'\left(\Phi+Z'\right)-v'(\overline{\zeta}) a^2
Z\right]\delta_{ab}. \ea This equation, of course, breaks into the
two independent relations \be \label{3.45} \Phi-\Psi=0 \ee and \be
\label{3.46} \left[2\Psi^{\prime\prime}+2(2{\cal H}'+{\cal H}^2)\Phi
+{\cal H}(2\Psi+\Phi)'\right]\delta_{ab} =
-\frac{\kappa}{4\lambda}\left[
2\overline{\zeta}'[\Phi+Z']-v'(\overline{\zeta}) a^2
Z\right]\delta_{ab}. \ee They are obviously equivalent to
(\ref{3.38}) and (\ref{3.43}) under the identification of the
obvious variables and $\overline{\zeta}=\overline{\Xi}$. Notice
again how the conservation equation ({\ref{3.35}) was used in order
to establish this result.

\section{Comparison with SCPT Coupled to Dust}
\label{s4}

In the previous section we showed that our manifestly
gauge-invariant formalism with dust reproduces the usual formalism
with linearly invariant
 quantities without dust, plus small corrections. In this section we
construct, on the one hand, linearly gauge-invariant quantities
 with dust a la SCPT which are certain linear combinations of the dust,
 scalar and geometry perturbations. On the other hand, we expand our
 manifestly gauge-invariant variables, which are certain non-linear
 aggregates made out of the non-invariant dust, scalar and geometry
 perturbations, to linear order. We will show that the sets of these two
quantities coincide precisely. \\
\\
 The original action that we started  with in the companion paper
 \cite{1} is the action of the physical system "gravity + scalar field +
 dust". Then we rewrote the Hamiltonian and diffeomorphism constraints
 in an equivalent form such that the Hamiltonian constraint could be
 deparametrised and constructed the manifestly gauge-invariant
 observables $Q_{ij},P^{ij}$ and $\Xi,\Pi$, corresponding to the
 gauge-variant three metric $q_{ab}$, the scalar field $\xi$ and the
 conjugated momenta thereof. As a second step, the equations of motion
 for $Q_{ij}$,$P^{ij}$ and $\Xi,\Pi$ were perturbed around a flat FRW
 background, resulting in second order time derivative equations of
 motion for the perturbations of the metric, denoted by $\delta
 Q_{ij}$,
and the scalar field $\delta\Xi$.
\\
 We discussed in the last section that, although starting with a system
 including the dust (and thus four more configuration degrees of
 freedom), the corresponding equation of motion for the physical degrees
 of freedom reproduce the results obtained by SCPT. As the latter does not employ
dust, it deals only with 3 physical degrees of freedom.
\\
\\
 In this section we want to consider the system "gravity + scalar field
 + dust" as our starting point again, but then follow the lines along the SCPT
 analysis and show that we can also use the perturbations of the dust
 fields, denoted by $\delta T, \delta S^k$, in order to construct an (up
 to linear order) gauge-invariant version of the perturbed four metric
 $\delta g_{\mu\nu}$. In this case, we have also 7 physical degrees of
 freedom in the system (6 gravitational and 1 matter).
 As a first step, we will use the dust fields $\delta T$ and $\delta S^k$
 in order to derive ten gauge-invariant components of $\delta
 g_{\mu\nu}$ and one for the scalar field $\delta\xi$. Throughout the
section we will denote an (up to linear order) gauge-invariant
 extension of a perturbed quantity $\delta f$ by $\delta\tilde{f}$.
 Secondly, we will demonstrate that 6 physical degrees of freedom
 contained in $\delta\widetilde{g}_{\mu\nu}$ and $\delta\widetilde{\xi}$
 agree with our  observables constructed in the previous sections, if
 these are expanded up to linear order in the non-invariant
perturbations of the configuration variables.
\\
\\
 As was done in the last section, we will denote an infinitesimal
 gauge transformation\footnote{Note, that in \cite{6} the letter $\xi$ is used for
 denoting their infinitesimal gauge transformation. Since $\xi$ is
 already taken by the gauge-variant version of our scalar field, we
 decided to use the letter $u^{\mu}$ here, which should not be confused
 with the dust four-velocity given by capital $U^{\mu}$.}
 by $x^{\mu}\mapsto x^{\mu}+u^{\mu}$  where $u=(u^0,u^a)$. Here
 $u^a=u^a_{\bot}+\delta^{ab}u_{,b}$ is a proper three vector and the
 scalar function $u$ is the solution of the equation
 $u_{,ab}\delta^{ba}=u^{a}_{,a}$. Let us recall the transformation
 behavior of the single components of the perturbed four metric $\delta
 g_{\mu\nu}$ under such an infinitesimal gauge transformation. Throughout
 the section we will indicate background quantities with a bar such that
 $\delta f:=f-\overline{f}$ for any quantity $f$. For the moment we will
 not decompose $\delta g_{\mu\nu}$ into scalar, vector and tensor modes,
 because the whole analysis done in this section can be applied to the
undecomposed tensor quantities.
 Denoting the gauge transformed quantities by $\delta g^{u}_{\mu\nu}$, we
have
\ba
\label{Trafodg}
 \delta g^{u}_{00}&=&\delta g_{00} + ({\cal
L}_{u}\overline{g})_{00}=\delta g_{00}-2a(au^0)^{\prime}\nonumber\\
 \delta g^{u}_{0a}&=&\delta g_{0a} + ({\cal
L}_{u}\overline{g})_{0a}=\delta g_{0a}
 +a^2\left(-u^0_{,a}+
\big(u^b_{\bot}+u^{,b}\big)^{\prime}\delta_{ab}\right)\nonumber\\
 \delta g^{u}_{ab}&=&\delta g_{ab} + ({\cal
 L}_{u}\overline{g})_{ab}=\delta
 g_{ab}+a^2\left(2\frac{a^{\prime}}{a}u^0\delta_{ab}
 +\big(u^c_{\bot}+u^{,c}\big)_{,a}\delta_{cb}
+\big(u^c_{\bot}+u^{,c}\big)_{,b}\delta_{ac}\right), \ea
 where the $f^{\prime}$ denotes a derivative with respect to conformal
time, denoted by $x_0$ and defined through $dx_0=a^{-1}dt$.

\subsection{Review of the SCPT Strategy for "Gravity + Scalar Field"}

 Before proceeding to the case of physical system plus dust, for the
 benefit of the reader let us briefly recall how the (up
 to linear order) gauge-invariant quantities are constructed in SCPT, in the case where no dust
 is present. A detailed revision of this construction can be found in
appendix A of this paper.
 As a first step, one decompose the metric into scalar components
 denoted by $\phi,\psi,B,E$, vector components $F_a,S_a$ and tensor
 components denoted by $h_{ab}$, where the latter is not of much
 concern,
 since it is already gauge-invariant. Their corresponding perturbed line
element is then given by
\ba
 \delta ds^2&=&\delta g_{\mu\nu}dx^{\mu}dx^{\nu}=a^2\Big[2\phi d\eta^2
 +2\big(B_{,a}+S_a\big) dx^ad\eta + \big(2\Psi\delta_{ab} + 2E_{,ab}) +
F_{a,b} + F_{b,a} + h_{ab}\big)dx^adx^b\Big],
\nonumber\\
&&
\ea
 from which we can read off the scalar, vector and tensor
components of $\delta g_{\mu\nu}$.
 Considering how $\delta g_{\mu\nu}$ transforms, we can derive the
 transformation behavior of $\phi,\psi,E,B,F_a,S_a$ and  (for
completeness) $h_{ab}$. It is given by \ba \phi^{u}&=&\phi -{\cal
H}u^0-(u^0)^{\prime},\quad
\psi^{u}=\psi + {\cal H}u^0\nonumber\\
E^{u}&=&E + u,\quad
B^{u}=B + u^{\prime} - u^0 \nonumber\\
S_a^{u}&=&S_a + (u^b_{\bot})^{\prime}\delta_{ab},\quad
F_a^{u}=F_a + u^b_{\bot}\delta_{ab}\nonumber\\
h^{u}_{ab}&=&h_{ab}
\ea
 where we introduced ${\cal H}:=a^{\prime}/a$. The strategy to construct
 gauge-invariant quantities in SCPT is (i) take the
 combination $B-E^{\prime}$, which undergoes a shift in $-u^0$ under a gauge
 transformation. This combination can then be used to
 construct gauge-invariant quantities for $\phi$ and $\psi$. Recalling that
 we denote the (up to linear order) gauge-invariant version of any
quantity $f$ by $\tilde f$, we obtain \be
 \widetilde{\phi}:=\Phi=\phi-{\cal
H}(B-E^{\prime})-(B-E^{\prime})^{\prime}\quad\mathrm{and}\quad
\widetilde{\psi}:=\Psi=\psi + {\cal H}(B-E^{\prime}). \ee (ii) One
constructs a gauge-invariant vector by using the combination \be
\widetilde{v}_a:=V_a=S_a - F_a^{\prime}. \ee
 Consequently, we see that for the construction of gauge-invariant
 quantities in SCPT it is convenient to decompose  $\delta g_{\mu\nu}$
 into scalar, vector and tensor parts, such that four unphysical
 components of $\delta g_{\mu\nu}$, namely $E,B,S_a$ can be used to
 make the remaining six gauge-invariant\footnote{Recall that the vectors
 $S_a,F_a$ are assumed to be divergenceless and therefore carry only
 two degrees of freedom.}. For the benefit of the reader we also mention
 the capital letters $\Phi,\Psi,V_a$, used in the last section and in \cite{6}.
  However, in this section we will use the $\tilde{}$  notation for
 (up to linear order) gauge-invariant quantities instead, because,
  when considering the gauge invariant quantity of
 $\delta g_{\mu\nu}$, we prefer not to use $\delta G_{\mu\nu}$ to avoid
possible confusion with the ordinary Einstein tensor.
\subsection{Analysis for "Gravity + Scalar field + Dust" in Analogy with SCPT}
\label{MFBDust}
 Let us now consider the system "gravity + scalar field + dust", for
 which we have 4 additional degrees of freedom, compared to the system
 "gravity + scalar field". In
 this case it is very natural to first keep all components of the
 perturbed metric $\delta g_{\mu\nu}$ and construct their gauge-invariant
 extension not by means of certain scalar/vector components
of the four metric, but taking the four dust fields $(\delta
T,\delta
 S^k)$ to perform this task. Since we want to derive
 $\delta\widetilde{g}_{\mu\nu}$, there is also no need to consider the
 decomposed expression of that tensor. Therefore, after having recalled
  SCPT for the case of no dust, we will go back to equation
 (\ref{Trafodg}), where the gauge-transformed components of $\delta
 g_{\mu\nu}$ are displayed. In the following, we will show that the
 behavior of $(\delta T,\delta S^j)$ under infinitesimal gauge
 transformations suggests in a natural way that these fields can be used
in order to make $\delta g_{\mu\nu}$ gauge-invariant.
 For this purpose, we first need an Ansatz for the background dust fields
 $(\overline{T},\overline{S}^k)$ compatible with the FRW symmetries.
 This can be derived from an Ansatz of the dust energy momentum tensor
 $\overline{T}^{\mathrm{D}}_{\mu\nu}$, specialised to the case of an FRW
background. Recall that the general energy-momentum tensor was given by
\be
 T^{\mathrm{D}}_{\mu\nu}\approx \rho
 U_{\mu}U_{\nu}\quad\mathrm{with}\quad U_{\mu}\approx -T_{,\mu}
+\Big(\frac{P_j}{P}\Big)S^j_{,\mu}, \ee
 where $\approx$ indicates weak equivalence. That means that this equations is only
 true up to second-class constraints. $U_{\mu}$ denotes the four-velocity
 of the dust, while $P_j,P$ are the  momenta conjugated to $T,S^j$,
 respectively. Note that here $x^{\mu}=(t,x^{a})$, since we have not yet
transformed to conformal time $x^{0}$.
\\
 By considering the most general Ansatz for
 $\overline{T}^{\mathrm{D}}_{\mu\nu}$ that respects the FRW symmetries, we
 get conditions on the dust four-velocity which carry over to the dust
fields $\overline{T},\overline{S}^k$. The most general Ansatz is
given by \be
\overline{T}^{\mathrm{D}}_{ta}=\rho^{\mathrm{D}}\overline{U}_0
\overline{U}_a\stackrel{!}{=}0\quad\mathrm{and}\quad
\overline{T}^{\mathrm{D}}_{ab}=\rho^{\mathrm{D}}\overline{U}_a
\overline{U}_b\propto \delta_{ab}. \ee
 In order to satisfy these requirements, we make an Ansatz for the dust
 fields $(\overline{T},\overline{S}^k)$. For
 $\overline{T}:=\overline{T}(t)$ we take a, so far, arbitrary function of
 $t$ so that $\overline{T}_{,a}=0$. For
 $\overline{S}^k=\overline{S}^k(x,y,z)$ we assume that it does not
 depend on time,
 so that $\overline{S}^k_{,t}=0$. Furthermore, we require that
 $\overline{S}^{k}_{,a}$ respects the homogeneity and isotropy
 requirements, that means $\overline{S}^{k}_{,ab}=0$. Therefore
 $\overline{S}^k=M^k_a x^a$ can only be linear in the spatial
 coordinates $x^a$, where $M^k_a$ is a matrix with constant entries.
 From the dust equation of motion shown in \cite{1} in section 2, we know
 that ${\cal L}_{\overline{U}}\overline{S}^k_{,\mu}=0$. Inserting the
 Ansatz for $\overline{T},\overline{S}^k$  yields
 $\overline{U}^{a}=1/a^2\overline{U}_a=0$. On the other hand, we have
 $\overline{U}_a:=-\overline{T}_{,a}-\frac{\overline{P}_k}{\overline{P}}
\overline{S}^{k}_{,a}=-\frac{\overline{P}_k}{\overline{P}}
\overline{S}^{k}_{,a}$.
 One of our basic assumptions is that the matrix $\overline{S}^k_{,a}$
 with $k,a=1,2,3$ is invertible and we denoted its inverse by
 $\overline{S}^a_{k}$. Requiring now that
 $\overline{U}_{a}=-\frac{\overline{P}_k}{\overline{P}}
 \overline{S}^{k}_{,a}\stackrel{!}{=}0$ and multiplying by the inverse
 $\overline{S}^a_{j}$ on both sides yields the condition
 $-\frac{\overline{P}_j}{\overline{P}}=0$ for all $j=1,2,3$. Hence, the
 conjugated momentum $\overline{P}_k$ of $\overline{S}^k$ vanishes in
 the case of FRW if $\overline{T}$ carries all the time dependence and
 $\overline{S}^k$ all the spatial dependence of $\overline{U}_{\mu}$.
 Furthermore, from the fact that $\overline{T}$ defines proper time along
 the dust lines, i.e. ${\cal L}_{\overline{U}}\overline{T}=1$
 as a consequence of the dust equation of motion, we get
 $(\overline{T}_{,t})^2=1$, that is $\overline{T}_{,t}=1$. Hence, we
obtain $\overline{T}=t$.
 Thus, the only non-vanishing component of $T^{\mathrm{D}}_{\mu\nu}$ is
the $tt$-component
\be
 \overline{T}^{\mathrm{D}}_{tt}=\rho^{\mathrm{D}},\quad
\overline{T}^{\mathrm{D}}_{ta}=\overline{T}^{\mathrm{D}}_{ab}=0. \ee
 Thus, the symmetry requirements for
 $\overline{T}^{\mathrm{D}}_{\mu\nu}$ lead to the usual four-velocity of
pressure-free dust given by $\overline{U}_{\mu}=(-1,0,0,0)$.
\\
\\
 Now, we want to analyse how the perturbed dust fields $(\delta T,
 \delta S^k)$, defined through $\delta T:=T-\overline{T}$ and $\delta
 S^k:=S^k-\overline{S}^k$, behave under a gauge transformation
 $x^{\mu}\to x^{\mu} + u^{\mu}$. The quantities  $(\delta T, \delta
 S^k)$ will, for instance, be included in the perturbed dust four-velocity
 $\delta U_{\mu}$ and thus in the perturbed energy momentum
 tensor\footnote{Note that, due to the vanishing of the background term
 $\overline{P}_k/\overline{P}$, the expression $\delta S^k_{,\mu}$ occurs
 in  $\delta U_{\mu}$ only formally, since it is multiplied by the
 vanishing background quantity $\overline{P}_k/\overline{P}$.}
 $T^{\mathrm{D}}_{\mu\nu}$. For this purpose we will work ,similarly to
 SCPT, in conformal time $dx^0=a^{-1}dt$. Then we have
 $\overline{T}_{,0}=a\overline{T}_{,t}=a$. Hence, the perturbations are
 given by  $\delta T=T(t(x^0))-\overline{T}(t(x^0))$ and $\delta
S^k=S^k-\overline{S}^k=S^k-M^k_ax^a$.
A gauge transformation $u$ acts on $\delta T$ as
\be
 \delta T^{u}=\delta T + {\cal L}_{u}\overline{T}=\delta T +u^0
\overline{T}_{,0}=\delta T + u^0\overline{T}^{\prime}=\delta T
+au^0, \ee
 where  $^\prime$ denotes derivative with respect to conformal time
 $x^0$ as introduced already above. We can see that $a^{-1}\delta T$ is
simply shifted by $u^0$.
 In the last step we used that
 $\overline{T}^{\prime}=\overline{T}_{,0}=a$. in \cite{6} the authors use the combination
 $B-E^{\prime}$ which undergoes a shift of $-u^0$ under gauge
 transformations. Hence, in order to use the perturbations of the dust
 field, we can simply take  $a^{-1}\delta T$  which is able to
 compensate any shift of $u^0$ in $\delta g^{u}_{\mu\nu}$ in equation
(\ref{Trafodg}).
 The other shift which occurs for the components $\delta g^{u}_{oa}$ and
 $\delta g^{u}_{ab}$ are certain derivatives of the three vector
 $u^{a}$. This shift can be compensated by the transformation behaviour
of $\delta S^k$ which is shown below
\be
 \delta (S^k_{,a})^u=\delta S^k_{,a}+({\cal
 L}_{u}\overline{S}^k)_{,a}=\delta
S^k_{,a}+u^b\overline{S}^k_{,ab}+u^b_{,a}\overline{S}^{k}_{,b}
=[\delta S^k + u^b\overline{S}^k_{,b}]_{,a}.
\ee
  Here we used that $\overline{S}^k$ has no conformal time dependence.
Consequently, the $\delta S^k$ themselves transform as
\be
 \delta (S^k)^u=\delta S^k + u^b\overline{S}^k_{,b}=\delta S^k +
(u^b_{\bot}+\delta^{bc}u_{,c})\overline{S}^k_{,b}, \ee
 which is the usual transformation rule for scalars under such a gauge
 transformation. Furthermore, we used the decomposition of $u^b$ into a
 longitudinal and transversal part mentioned before. From the equation
 above we can read off that the combination $\delta S^k
 \overline{S}^b_k\delta_{ab}$ will be shifted by the vector $u_a$ after
 having applied a gauge transformation. Therefore, this combination can
 be used to compensate for all shifts in $u_a$ and derivatives thereof which
 occur in the gauge-transformed components of $\delta g_{0a}$ and
 $\delta g_{ab}$.
 Hence, the perturbed dust fields $(\delta T, \delta S^k)$ suggest
 very naturally a way to construct the (up to linear order)  gauge-invariant metric
  perturbations $\delta\widetilde{g}_{\mu\nu}$, since
 $(a^{-1}\delta T, \overline{S}^b_k\delta S^k\delta_{ab})$ exactly
 provide the shifts in $u^0$ and $u^a$, respectively, which are needed. The
 components of $\delta\widetilde{g}_{\mu\nu}$ are, in particular, given by
\ba
\label{GIg}
 \delta\widetilde{g}_{00}&=&\delta g_{00} +2a(\delta T)_{,0}=\delta
g_{00} +2a(\delta T)^{\prime}\nonumber\\
 \delta\widetilde{g}_{oa}&=&\delta g_{0a} +a^2\big[a^{-1}(\delta T)_{,a}
- (\overline{S}^b_{k}\delta S^k)_{,0}\delta_{ab}\big]
=\delta g_{0a}
 +a^2\big[a^{-1}(\delta T)_{,a} - (\overline{S}^b_{k}\delta
S^k)^{\prime}\delta_{ab}\big]\nonumber\\
 \delta\widetilde{g}_{ab}&=&\delta g_{ab}
 -a^2\big[2\frac{a^{\prime}}{a^2}\delta T\delta_{ab} +
 (\overline{S}^c_k\delta S^k)_{,a}\delta_{bc} +
(\overline{S}^c_{k}\delta S^k)_{,b}\delta_{ac}\big]. \ea
 Note that all ten components of $\delta\widetilde{g}_{\mu\nu}$ are
 gauge-invariant now. These gauge-invariant combinations are different
 from the usual SCPT analysis where, as mentioned above, 4 out of the ten
 metric components are used to make the remaining six components gauge-invariant.
 Upon including the dust, the analysis above shows that
 the dust is a natural candidate for the unphysical degrees of freedom,
 because their changes under gauge transformation  can in a simple way
 compensate for the changes of the components of $\delta g_{\mu\nu}$. A
 consequence is that when decomposing $\delta\widetilde{g}_{\mu\nu}$
 into scalar, vector and tensor modes, we also obtain gauge-invariant
 extensions of the quantities $E,B,S_a,F_a$. For completeness and to facilitate
  comparison with the standard SCPT framework, we also list the gauge-invariant
   extensions of these decomposed
 quantities
 $\widetilde{\phi},\widetilde{\psi},\widetilde{E},\widetilde{B},
 \widetilde{S}_a,\widetilde{F}_a$ for the case where the dust fields are
 used to construct those gauge-invariant quantities. For this purpose, we
 need to decompose the vector $\delta w^c:=\overline{S}^c_k\delta S^k$
 into its longitudinal and transversal part, thus $\delta w^c=\delta
 w^c_{||}+\delta w^c_{\bot}=\delta^{cd} \delta w_{||,c}+\delta
 w^c_{\bot}$,
 where $\delta w_{||}$ is the solution of the equation $\Delta
 \delta w_{||} =\delta w^c_{,c}$. Explicitly, we have
\ba
 \widetilde{\phi}&=&\Phi=\phi +{\cal H}a^{-1}\delta T +(a^{-1}\delta
 T)^{\prime}\quad\quad
 \widetilde{\psi}=\Psi=\psi -{\cal H}a^{-1}\delta T\nonumber\\
 \widetilde{E}&=&E-\delta
w_{||} \quad\quad\quad\quad\quad\quad\quad\quad\quad\quad\quad
\widetilde{B}=B+a^{-1}\delta T-(\delta w_{||})^{\prime}\nonumber\\
 \widetilde{S}_a&=&S_a-(\delta
w^b_{\bot})^{\prime}\delta_{ab}\quad\quad\quad\quad\quad\quad\quad\,\,\,
\widetilde{F}_a=F_a-\delta w^b_{\bot}\delta_{ab}. \ea
 Finally, in order to derive the gauge-invariant extension of the scalar
 field perturbation $\delta\widetilde{\xi}$, we consider first the
 energy momentum tensor of the unperturbed scalar field $\xi$, given by \be
 \lambda
 T_{\mu\nu}^{\mathrm{matter}}=[\nabla_{\mu}\xi][\nabla_{\nu}\xi]
 -\frac{1}{2}g_{\mu\nu}\left[g^{\rho\sigma}[\nabla_{\rho}\xi]
[\nabla_{\sigma}\xi]+v(\xi)\right].
\ee
Here $v(\xi)$ denotes the potential of the scalar field.
 Specialising to an FRW background as discussed in appendix A, we obtain
for the components of $\overline{T}_{\mu\nu}$
\ba
 \overline{T}^{\mathrm{matter}}_{00}&=&\frac{1}{2}
[(\overline{\xi}^{\prime})^2+a^2v(\overline{\xi})]
=a^2\overline{\rho}^{\mathrm{matter}}\nonumber\\
\overline{T}^{\mathrm{matter}}_{0a}&=&0\nonumber\\
\overline{T}^{\mathrm{matter}}_{ab}&=&\frac{1}{2}\delta_{ab}
[(\overline{\xi}^{\prime})^2-a^2v(\overline{\xi})]=a^2
\overline{p}^{\mathrm{matter}}, \ea
 where we used
 $\overline{\rho}^{\mathrm{matter}}=\frac{1}{2}(\dot{\xi}^2+v(\xi))$ and
$\overline{p}^{\mathrm{matter}}=\frac{1}{2}(\dot{\xi}^2-v(\xi))$.
 The linear perturbation $\delta T_{\mu\nu}^{\mathrm{matter}}$ of the
$\overline{T}^{\mathrm{matter}}_{\mu\nu}$ components are
\ba
 \delta
 T^{\mathrm{matter}}_{00}
&=&\frac{1}{2\lambda}\Big[2\overline{\xi}^{\prime}\delta\xi^{\prime}
+a^2v'(\overline{\xi})\delta\xi-v(\overline{\xi})\delta g_{00}\Big]\nonumber\\
 \delta
T^{\mathrm{matter}}_{0a}
 &=&\frac{1}{\lambda}\Big[\overline{\xi}^{\prime}\delta\xi_{,a}
+\lambda\overline{p}\delta g_{0a}\Big]\nonumber\\
\delta T^{\mathrm{matter}}_{ab}
&=&\frac{1}{\lambda}\Big[\lambda\overline{p}\delta g_{ab}
+\frac{1}{2}\delta_{ab}\big[\frac{1}{a^2}(\overline{\xi}^{\prime})^2
\delta g_{00}  + 2\overline{\xi}^{\prime}\delta
\xi^{\prime}-a^2v_{,\overline{\xi}}(\overline{\xi})\delta\xi\big]\Big].
\ea
 Here we introduced $v_{,\overline{\xi}}:=dv/d\overline{\xi}$. The gauge-invariant
  extension of $\delta T^{\mathrm{matter}}_{\mu\nu}$, denoted by
 $\delta\widetilde{T}^{\mathrm{matter}}_{\mu\nu}$, can then be
 constructed by replacing the quantities $\delta g_{00},\delta g_{0a}$
 in equation (\ref{GIg}) and $\delta\xi$ by their gauge-invariant
 extension. We have not worked out the latter yet, so
  we need to consider how $\delta\xi$ transforms under a gauge
transformation. This is simply given by
\be
 \delta \xi^{u}=\delta\xi+({\cal
 L}_{u}\overline{\xi})=\delta\xi+u^0\overline{\xi}_{,0}=\delta\xi
+u^0\overline{\xi}^{\prime}. \ee
 Consequently, the (up to linear order) gauge-invariant extension of
$\delta\xi$ is simply \be \label{GIxi}
\delta\widetilde{\xi}=\delta\xi-a^{-1}\overline{\xi}^{\prime}\delta
T. \ee Thus, the components of
$\delta\widetilde{T}^{\mathrm{matter}}_{\mu\nu}$ are \ba
 \delta
 \widetilde{T}^{\mathrm{matter}}_{00}
&=&\frac{1}{2\lambda}\Big[2\overline{\xi}^{\prime}
\delta\widetilde{\xi}^{\prime}
+a^2v_{\overline{\xi}}(\overline{\xi})\delta\widetilde{\xi}
-v(\overline{\xi})\delta
\widetilde{g}_{00}\Big]\nonumber\\
\delta \widetilde{T}^{\mathrm{matter}}_{0a}
&=&\frac{1}{\lambda}\Big[\overline{\xi}^{\prime}\delta\widetilde{\xi}_{,a}
+\lambda\overline{p}\delta \widetilde{g}_{0a}\Big]\nonumber\\
\delta \widetilde{T}^{\mathrm{matter}}_{ab}
&=&\frac{1}{\lambda}\Big[\lambda\overline{p}\delta
\widetilde{g}_{ab}
+\frac{1}{2}\delta_{ab}\big[\frac{1}{a^2}(\overline{\xi}^{\prime})^2\delta
\widetilde{g}_{00} +
2\overline{\xi}^{\prime}\delta\widetilde{\xi}^{\prime}
-a^2v_{,\overline{\xi}}(\overline{\xi})\delta\widetilde{\xi}\big]\Big].
\ea This finishes our task to show that, by means of the four
(perturbed) dust fields $\delta T, \delta S^k$, (up to linear)
gauge-invariant extensions  of the four metric
$\delta\widetilde{g}_{\mu\nu}$, as well as of  the scalar field
energy momentum tensor
$\delta\widetilde{T}^{\mathrm{matter}}_{\mu\nu}$ can be constructed.

\subsection{Comparison between SCPT + Dust and our Relational Framework}
 As a last step, we will demonstrate that the (up to linear order) gauge-invariant
 quantities $\delta\widetilde{g}_{\mu\nu}$ and
 $\delta\widetilde{T}^{\mathrm{matter}}_{\mu\nu}$ derived in the last
 section agree with our (manifestly) gauge-invariant quantities $\delta
 Q_{ij}$ (+ corresponding lapse $\delta N$ and shift $\delta N_j$) and
 $T^{\mathrm{matter}}_{ij}(\delta Q,\delta \Xi)$, when these are expanded
 up to linear order. By this we mean that we will expand the formula
 in \cite{1} for the gauge-invariant quantities $F:=O_{f}$, associated with any quantity
 $f$, around a given phase point
 (background), but now in terms of the non-invariant variables. The
 perturbations are then defined through $\delta
 O_f:=O_f-O_{\overline{f}}=O_f-\overline{f}$, where we assume that the
 perturbation is around the configuration $T=\overline{T}$,
 $S^k=\overline{S}^k$. We consider only terms in which the
perturbations occur  at most linearly.
  Recall that in order to construct the (manifestly) gauge-invariant
 extensions of the perturbed three metric $\delta q_{ab}$, perturbed
 lapse function $\delta n$ and perturbed shift vector $\delta n_a$,
 we first rewrote the Hamiltonian and diffeomorphism constraint
 $c^{\mathrm{tot}}=c^{\mathrm{geo}}+c^{\mathrm{matter}}+c^{\mathrm{D}}=c
+c^{\mathrm{D}}$ and $c_a^{\mathrm{tot}}=c^{\mathrm{geo}}_a
+c^{\mathrm{matter}}_a+c^{\mathrm{D}}_a=c_a +c^{\mathrm{D}}$,
respectively, in the following equivalent form: \ba \label{Constr}
 \tilde{c}^{\mathrm{tot}}&=&P+h\quad\mathrm{with}\quad
h:=\sqrt{c^2-q^{ab}c_ac_b}\nonumber\\
 \tilde{c}^{\mathrm{tot}}_j&=&P_j+h_j\quad\mathrm{with}\quad
h_j:=S^a_j[-hT_{,a}+c_a].
\ea
 Note that the $\widetilde{}$ here has nothing to do with gauge
invariance. We just took the notation used before in \cite{1}.
 Since these constraints are mutually Poisson commuting, we could then
 perform our construction of the gauge-invariant extensions in two
 steps. First we construct diffeomorphism -invariant quantities
 $O^{(1)}_{f}$ and afterwards use those in order to construct
 quantities that are also gauge-invariant with respect to the
 Hamiltonian constraint denoted by $F:=O^{(2)}_{O^{(1)}_{f}}$. The
 corresponding perturbed quantity is then given by $\delta
 F:=\delta\big(O^{(2)}_{O^{(1)}_f}\big)$.
 Here, we are only interested in the terms of $\delta F$ which are
 linear in the perturbations, because we want to compare them with (up to
 linear order) gauge-invariant components of
 $\delta\widetilde{g}_{\mu\nu}$ in equation (\ref{GIg}). For this
 reason,
 the general expression for $\delta F$ simplifies a lot and can be
 easily calculated. For the perturbations of the diffeomorphism-invariant
 observables $\delta O^{(1)}_{\delta f}$ we find up to
linear order in the perturbations \ba
 \delta O^{(1)}_{f}&=&\delta f +\int
d^3y(\overline{S}^k-S^k)(y)\overline{\{\tilde{c}^{\mathrm{tot}}_k(y),f(x)\}}
\nonumber\\
&=&\delta f -\int d^3y\delta
S^k(y)\overline{\{\tilde{c}^{\mathrm{tot}}_k(y),f(x)\}}. \ea
 Here we used $\delta f=f-\overline{f}$ and, in notation introduced
 in \cite{1}, $\overline{S}^k=\sigma^k$, as well as the definition of $\delta
 S^k=S^k-\overline{S}^k$ in the last line. Note that up to linear order
only Poisson brackets
 evaluated on background quantities occur, which we indicated by the bar
 above the Poisson bracket. This is due to the reason that $\delta S^k$
is already linear in the perturbations.
 Using $\delta O^{(1)}_{f}$, we are now able to construct $\delta
 F$,
which up to linear terms in the perturbation given by \ba
\label{GObs1}
\Big[\delta F\Big]_{\mathrm{lin}}&=:&\delta\tilde{f}\nonumber\\
&=&
 \delta O^{(1)}_{f}+\int d^3y
 (\overline{T}-T)(y)\{\tilde{c}^{\mathrm{tot}}(y),\delta
O^{(1)}_{f}(x)\}\nonumber\\
 &=&\delta f-\int d^3y\delta
 S^k(y)\overline{\{\tilde{c}^{\mathrm{tot}}_k(y), f(x)\}}
-\int d^3y (T-\overline{T})(y)\overline{\{\tilde{c}^{\mathrm{tot}}(y),
f(x)\}}\nonumber\\
&=&\delta f-\int d^3y \delta
T(y)\overline{\{\tilde{c}^{\mathrm{tot}}(y),f(x)\}} -\int d^3y
\delta S^k(y)\overline{\{\widetilde{c}^{\mathrm{tot}}_k(y),f(x)\}}.
\ea Here we denoted $[\delta F]_{\mathrm{lin}}$, expanded up to
linear order,  by $\delta\tilde{f}$ as we did for $\delta
g_{\mu\nu}$, in order to emphasize that, in contrast to $\delta F$,
the former are not gauge-invariant up to all orders. Furthermore we
used that $\overline{T}=\tau$ in our case and $\delta
T=T-\overline{T}$. We want to apply the above formula to
$f=q_{ab},\xi$, respectively, both of which are Poisson commuting
with $P,P_k$, the momenta conjugate to $T$ and $S^k$, respectively.
Recalling the form of the constraints in equation (\ref{Constr}), we
see that equation (\ref{GObs1}) simplifies even more to \ba
\delta\tilde{f}&=&\delta f-\int d^3y \delta
T(y)\overline{\{h(y),f(x)\}} -\int d^3y \delta
S^k(y)\overline{S}^a_k\overline{\{c_a(y),f(x)\}} \ea for those $f$'s
which do not depend on the dust variables $(T,S^k)$. Furthermore, we
used that \be
\overline{\{(hT_{,a})(y),f(x)\}}=\overline{T_{,a}(y)\{h(y),f(x)\}}=\overline{T}_
{,a}(y)\overline{\{h(y),f(x)\}}=0, \ee
 because we have $\overline{T}_{,a}=0$. The computation of the remaining Poisson
bracket above yields
\ba
\int d^3y \delta T(y)\overline{\{h(y),f(x)\}}&=&
\int d^3y\delta T(y)\Big(\overline{n}(y)\overline{\{c(y),f(x)\}}
+\overline{n}_a(y)\overline{\{c_b(y),f(x)\}}\Big)\nonumber\\
&=&\int d^3y\delta T(y)\Big(\overline{n}(y)\overline{\{c(y),f(x)\}}\Big)
+\overline{\big({\cal L}_{\vec{n}} f\big)}(x)
\nonumber\\
\int d^3y \delta
S^k(y)\overline{S}^a_k\overline{\{c_a(y),f(x)\}}&=&\overline{({\cal
L}_{\overline{S}^a_k\delta S^k} f)}(x), \ea where we used the
definition of the lapse function $n:=c/h$ and the shift vector
$n_a:=-c_a/h$. Hence, the final version of the formula for
$\delta\tilde{f}$ has the form \be \label{finalEqn}
\delta\tilde{f}=\delta f -\int d^3y\delta
T(y)\Big(\overline{n}(y)\overline{\{c(y),f(x)\}}\Big)
-\overline{\big({\cal L}_{\vec{n}} f\big)}-\overline{({\cal
L}_{\overline{S}^a_k\delta S^k} f)}. \ee In order to compare the
results obtained from equation (\ref{finalEqn}) with the one for
$\delta\widetilde{g}_{\mu\nu}$ derived in the last section shown in
equation (\ref{GIg}), we need to express the perturbations of the
four metric in terms of the corresponding perturbations of the three
metric $q_{ab}$, lapse function $n$ and shift vector $n_a$. Recall
that $g_{00}=a^2(-n^2+q^{ab}n_an_b)$, $g_{0a}=an_a$ and
$g_{ab}=q_{ab}$. Consequently, the relation for the (up to linear
order) gauge-invariant perturbations is given by \ba
\label{Pergmunu}
\delta\widetilde{g}_{00}&=&a^2\big(-2\overline{n}\delta\widetilde n
-\delta q_{ab}\overline{n}^a\overline{n}^c
+2\overline{q}^{ab}\overline{n}_a\delta\widetilde{n}_b\big)=-2a^2\delta\widetilde{n}\nonumber\\
\delta\widetilde{g}_{0a}&=&a\delta\widetilde{n}_a\nonumber\\
\delta\widetilde{g}_{ab}&=&\delta\widetilde{ q}_{ab}, \ea where we
used that $\ov{n}=1$ and $\ov{n}^a=0$. Starting with the spatial -
spatial component of the metric $\delta q_{ab}$ and applying the
formula in equation (\ref{finalEqn}) to it, we obtain \ba
\label{Dqab1} \delta\widetilde{q}_{ab}&=&\delta q_{ab}-\int
d^3y\delta
T(y)\Big(\overline{n}(y)\overline{\{c(y),q_{ab}(x)\}}\Big)
-\overline{\big({\cal L}_{\vec{n}} q\big)}_{ab}-\overline{({\cal
L}_{\overline{S}^a_k\delta S^k}
q)}_{ab}\nonumber\\
&=& \delta q_{ab}-\int d^3y\delta
T(y)\Big(\overline{\{c(y),q_{ab}(x)\}}\Big) -\overline{({\cal
L}_{\overline{S}^a_k\delta S^k} q)}_{ab}. \ea Here we used in the
last line that for FRW the lapse function and the shift vector are
given by \be
\overline{n}=c/h=c/\sqrt{c^2-q^abc_ac_b}=c/c=1\quad\mathrm{and}\quad\overline{n}
_a=-c_a/h=-c_a/\sqrt{c^2-q^{ab}c_ac_b}=0.\ee The remaining integral
in equation (\ref{Dqab1}) yields \ba \int d^3y\delta
T(y)\Big(\overline{\{c(y),q_{ab}(x)\}}\Big) &=& \int d^3y\delta
T(y)\Big(\overline{\frac{2n}{\sqrt{\det{q}}}\big(q_{ac}q_{bd}-\frac{1}{2}q_{ab}
q_{cd}\big)p^{cd}}\Big)(x)\delta(x,y)\\
&=&\delta
T(x)\frac{2}{a^3}\big(-\frac{1}{2}a^4\big)\big(-2\dot{a}\Big)\delta_{ab}
\nonumber\\
&=&\delta T(x) 2a\dot{a}\nonumber\\
&=&2a^{\prime}\delta T(x).\nonumber \ea In the second line we
inserted the explicit form of $\overline{q}_{ab}$ for FRW and the
corresponding momentum $p^{ab}$ conjugate to it, given by
$\overline{q}_{ab}=a^2\delta_{ab}$ and
$p^{ab}=-2\dot{a}\delta^{ab}$, respectively. In the last step we
replaced the derivative with respect to $\overline{T}$ by one with
respect to conformal time. Considering the explicit form of
$\overline{q}_{ab}$ also for the Lie derivative occurring in
equation (\ref{Dqab1}), we end up with \ba \delta\widetilde{q}_{ab}
&=&\delta q_{ab} -2a^{\prime}\delta
T\delta_{ab}-a^2(\overline{S}^c_k\delta
S^k)_{,a}\delta_{cb}-a^2(\overline{S}^c_k\delta S^k)_{,b}\delta_{ac}\nonumber\\
&=&\delta q_{ab}-a^2\big[2\frac{a^{\prime}}{a^2}\delta T\delta_{ab}
+ (\overline{S}^c_k\delta
S^k)_{,a}\delta_{cb}+(\overline{S}^c_k\delta
S^k)_{,b}\delta_{ac}\big]. \ea The last line agrees exactly with the
expression in equation (\ref{GIg}) if we identify
$\delta\widetilde{g}_{ab}$ with $\delta\widetilde{q}_{ab}$. Next, we
want to discuss the case for the scalar field perturbation
$\delta\xi$. Since any spatial derivative of the background scalar
field $\overline{\xi}$ vanishes, the Lie derivative with respect to
$\overline{S}^c_k\delta S^k$ in equation (\ref{finalEqn}), evaluated
on the FRW background, vanishes as well. Therefore the (up to linear
order) gauge-invariant extension is simply \ba
\delta\widetilde{\xi}&=&
\delta\xi-\int d^3y\delta T(y)\Big(\overline{\{c(y),f(x)\}}\Big)\\
&=&\delta\xi - \int d^3y\delta
T(y)\Big(\overline{\frac{n}{\sqrt{\det{q}}}\pi}\Big)(x)\delta(x,y)\nonumber\\
&=&\delta\xi -\delta
T(x)\Big(\frac{1}{\sqrt{\det(\overline{q})}}\dot{\overline{\xi}}\sqrt{\det{
\overline{q}}}\Big)\nonumber\\
&=&\delta\xi-a^{-1}\overline{\xi}^{\prime}\delta T(x), \ea which
exactly coincides with the result in equation (\ref{GIxi}). Note
that we reexpressed the momentum $\overline{\pi}$ conjugated to
$\overline{\xi}$ in terms of $\dot{\overline{\xi}}$ in the third
line, which in the case of FRW has the simply form
$\overline{\pi}=\dot{\overline{\xi}}\sqrt{\det{\overline{q}}}$.\\
\\
\\
Finally we have to discuss the lapse function $\delta n$ and the
shift vector $\delta n^a$. We want to stress that in the rest of
these two papers we always have treated them as pure gauge with
respect to the primary constraints and have reduced the problem to
the remaining degrees of freedom. They never appeared anymore as
independent variables and in the effective, physical Hamiltonian
description lapse and shift were derived quantities as well, they
were not independent.


\subsection{Subtleties with Lapse and Shift Functions}

\label{s4.1}

In order to compare our framework with SCPT we must deal with the
following subtlety:\\
\\
As mentioned earlier, in the Hamiltonian framework we have ten
gravitational configuration and momentum degrees of freedom to begin
with. These are $n,\;n^a,\;q_{ab}$, denoting lapse function, shift
vector and three metric, respectively, as well as their conjugate
momenta $p,\;p_a,\;p^{ab}$, respectively. This phase space is
subject to four primary constraints $z=p=0,\;z_a=p_a=0$, as well as
four secondary constraints $c^{\rm tot}=c^{\rm dust}+c^{{\rm
geo}}+c^{{\rm matter}}$ and $c_a^{\rm tot}=c_a^{\rm dust}+c_a^{{\rm
geo}}+c_a^{{\rm matter}}$. These eight constraints are first-class
and play a dual role: on the one hand they constrain the phase space
to the constraint surface, on the other hand they generate gauge
transformations via canonical transformations. The constraint
equations can be used to eliminate eight of the momenta while the
gauge transformations eliminate eight of the configuration degrees
of freedom. Hence the physical or reduced phase space has eight
canonical pairs less than the unreduced one. The primary constraints
and the gauge transformations generated by them, which consist in
arbitrary changes of lapse and shift, completely decouple from the
rest of the equations, namely $c^{\rm tot},\;c_a^{\rm tot}$ do not
depend on $n,n^a,p,p_a$, in particular, the gauge transformations
generated by $c^{\rm tot},\;c_a^{\rm tot}$ do not affect lapse and
shift functions. It is therefore mathematically convenient to simply
forget about $p,p_a$ and to treat $n,n^a$ as Lagrange multipliers
which is customary.
\\
\\
Let us compare this with the situation in the Lagrangian framework.
Here we consider only the ten gravitational configuration
coordinates, momenta are never introduced. The system is subject to
Einstein's equations. Four of them, the temporal -- temporal and the
temporal -- spatial ones are equivalent to the secondary constraints
that we also find in the Hamiltonian formalism. Namely, in the
Hamiltonian formalism the equations of motion are generated by the
primary Hamiltonian

\be \label{4.1} H_{{\rm primary}}=\int_{{\cal X}}\; d^3x\;\left(\nu
\;z +\nu^a\; z_a+n\; c^{\rm tot}+n^a\; c_a^{\rm tot}\right), \ee
where $\nu,\;\nu^a$ are Lagrange multiplier fields. The equation of
motion for $q_{ab}$ following from (\ref{4.1}) enables us to express
$p^{ab}$ in terms of $q_{ab},\;\dot{q}_{ab},\;n,\;n^a$, specifically
\be \label{4.2} p^{ab}=\sqrt{\det(q)}\left(q^{ac} q^{bd}-q^{ab}
q^{cd}\right) K_{cd}\quad{\mathrm with}\quad
K_{ab}=\frac{1}{2n}\Big(\dot{q}_{ab}-2D_{(a}n_{b)}\Big). \ee Here we
introduced the extrinsic curvature $K_{ab}$, and $D$ is the torsion
free covariant differential compatible with $q_{ab}$. When inserting
(\ref{4.2}) into $c^{{\rm geo}},\,\;c_a^{{\rm geo}}$, one obtains
precisely the temporal -- temporal and temporal -- spatial
components of the Einstein tensor, respectively. Thus we can match
four of the equations of motion in the Hamiltonian and the
Lagrangian equations of motion. The remaining spatial -- spatial
components of the Einstein equations are obtained in the Hamiltonian
formalism by inserting (\ref{4.2}) into the equations of motion for
$\ddot{q}_{ab}$ which also uses the equation of motion for $p^{ab}$
as obtained from (\ref{4.1}) \cite{12}, so that we get an exact
match at the level of the equations of motion (one has to use that
the spacetime metric is expressed in terms of the configuration
coordinates as $g_{tt}=-n^2+q_{ab} n^a n^b,\;g_{ta}=q_{ab}
n^b,\;g_{ab}=q_{ab}$). Now in the Lagrangian formalism the secondary
constraints do not generate gauge transformations because one does
not have a phase space formulation. Rather one notices that the
Lagrangian and thus the Einstein equations are invariant under
spacetime diffeomorphisms. Thus one imposes spacetime diffeomorphism
invariance as a gauge symmetry. Now we have seen that the temporal
-- temporal and temporal -- spatial components of the Einstein
equations do not involve time derivatives of lapse and shift. Thus
it is natural to eliminate them via the secondary constraint in
terms of $q_{ab},\;\dot{q}_{ab}$ and to eliminate four components of
$q_{ab}$ by using spacetime diffeomorphism gauge transformations.

We see that in both formalisms we get the same number of physical
configuration degrees of freedom, however, the avenue to get there
is somewhat different. In particular, there is no counterpart of the
gauge symmetry associated with the primary constraints at the
Lagrangian level. Furthermore, while the secondary constraints
generate spacetime diffeomorphisms on $q_{ab}$ \cite{12}, they
preserve lapse and shift fields, whereas in the Lagrangian formalism
lapse and shift do change under spacetime diffeomorphisms. Thus,
there is a mismatch in the number and action of the gauge
transformations of the Lagrangian and the Hamiltonian formalism at
the level of the unreduced configuration space. It is clear that the
physical degrees of freedom that both formalisms produce are the
same, since both formalisms tell us that one should extract the
spacetime diffeomorphism invariant information contained in $q_{ab}$
(as far as the gravitational degrees of freedom are concerned).
However, before using the secondary constraints, the diffeomorphism
invariant quantities that one constructs in the Hamiltonian theory
do not involve lapse and shift because they are invariant under the
secondary constraints, while in the Lagrangian formalism one has to
use them. This has the consequence that in the Lagrangian formalism
one has in principle six rather than two physical configuration
degrees of freedom whereas in the Hamiltonian formalism one has only
two. In the Lagrangian formalism, four of them are eliminated via
the secondary constraints and thus, at the end of the day, one gets
an exact match at the level of the physical degrees of freedom. But
before doing that it is somewhat difficult to compare the gauge
invariant degrees of freedom in the two formalisms. The reason for
why that happens lies deeper and has to do with the question whether
Lagrangian (Noether) symmetries have canonical generators in the
Hamiltonian formalism. This is analysed in all detail in
\cite{Pons}.

Fortunately, there is an elegant solution. The idea \cite{Pons} is
to add, in the Hamiltonian formalism, to the secondary constraints a
linear combination of the primary constraints with carefully chosen
coefficients such that the new constraints generate spacetime
diffeomorphisms on the full phase space, including lapse and shift.
This leads to an equivalent set of constraints which leads to the
same final set of physical degrees of freedom but which has, after
reducing with respect to the new secondary constraints only, the
same gauge-invariant degrees of freedom as in the Lagrangian
formalism. Then afterwards one must impose the secondary constraints
in the Lagrangian formalism and reduce with respect to the primary
constraints in the Hamiltonian formalism in order to arrive at the
true degrees of freedom. We will now sketch how
this is done as we will need this in the subsequent subsections.\\
\\
Fortunately, we do not need all the machinery developed in
\cite{Pons}, but can rather develop the required formulae by basic
methods. The starting point is to recall \cite{12} that for
arbitrary test functions $b,\; b^a$ we have \be \label{4.4} \{c^{\rm
tot}(b)+\vec{c}^{\rm tot}(\vec{b}),q_{\mu\nu}(x)\}={\cal L}_u
q_{\mu\nu}(x)\quad\mathrm{where}\quad u^\mu=b n^\mu+X^\mu_{,a} b^a
\ee with $c^{\rm tot}(f)=\int_{{\cal X}}\; d^3x \; b(x) \; c^{\rm
tot}(x)$ and $\vec{c}^{\rm tot}(\vec{b})=\int_{{\cal X}}\; d^3x \;
b^a(x) \; c_a^{\rm tot}(x)$. Here ${\cal L}_u$ denotes the spacetime
Lie derivative with respect to the vector field $u$ and
$q_{\mu\nu}=g_{\mu\nu}+n_\mu n_\nu$ is the three metric on the
leaves ${\cal X}_t$ of the foliation, with unit normal $n^\mu$,
parametrised by the one parameter family of embeddings
$Y_t=Y(t,.):\;{\cal X}\to {\cal X}_t$. In deriving (\ref{4.4}) the
equations of motion have been used and thus we can say that the
secondary constraints generate infinitesimal spacetime
diffeomorphisms parametrised by $u$. Notice that any vector field
can be split into components tangential and orthogonal to the
foliation so that this is no loss of generality.

We now would like to add to $c^{\rm tot}(b)$ and $\vec{c}^{\rm
tot}(\vec{b})$, respectively, a term linear in $p$ and $p_a$,
respectively, with smearing functions as coefficients that are
linear in both $b,\; b^a$ such the extended constraints also
generate infinitesimal spacetime diffeomorphisms on lapse and shift
functions parametrised by the above vector field $u$. In order to
determine those coefficients, we first need to determine the
transformation behaviour of lapse and shift functions under
infinitesimal spacetime diffeomorphisms. Starting from the
identities $g_{tt}=-n^2+q_{ab} n^a n^b,\;g_{ta}=q_{ab}
n^b,\;g_{ab}=q_{ab}$, we can solve for lapse and shift in terms of
spacetime metric components \be \label{4.5} n^a=g^{ab} g_{tb},\;
n^2=-g_{tt}+g^{ab} g_{ta} g_{tb}, \ee where $g^{ab}$ is the inverse
of the spatial metric $g_{ab}$. Now using that the change of the
metric under infinitesimal diffeomorphisms is given by the Lie
derivative \be \label{4.6} \delta_u g_{\mu\nu}=({\cal L}_v
g)_{\mu\nu}=u^\rho g_{\mu\nu,\rho}+ 2 g_{\rho(\mu} u^\rho_{,\nu)},
\ee we can work out the infinitesimal transformation of lapse and
shift. The result is, in the frame where $X^0=t$ and $X^a=x^a$
(frame adapted to the embedding),

\ba \label{4.7}
\delta_u n^a &=& u^b n^a_{,b}-n^b u^a_{,b}+u^a_{,t}+(n^a u^t)_{,t}-
\left(n^2 q^{ab}+n^a n^b\right) u^t_{,b}
\nonumber\\
2 n \delta_u n &=&
-n^a n^b \left(u^c g_{ab,c}+2 u^c_{,(a} g_{b)c}+u^t g_{ab,t}+2 n_a
u^t_{,b}\right)
\nonumber\\
&& + 2 n^a\left(u^t g_{ab,t}+u^t_{,t} g_{at}+g_{tt} u^t_{,a}
+u^b g_{at,b}+g_{ab} u^b_{,t}+u^b_{,a} g_{bc} n^c\right)
\nonumber\\
&& -\left(g_{tt,t} u^t +2 g_{tt} u^t_{,t}+g_{tt,a} +2 n_a
u^a_{,t}\right). \ea Here we have used the explicit form of
(\ref{4.6}). Formula (\ref{4.7}) is rather complicated. Worse than
that, it contains time derivatives of $n,\; n^a$. We cannot use the
equations of motion in order to replace those time derivatives by
the momenta $p,\;p_a$ because the very reason for the appearance of
the primary constraints $p=p_a$ is that the Legendre transform is
singular, as the Einstein Hilbert action does not contain time
derivatives of lapse and shift. This is different for time
derivatives of $q_{ab}$. Hence, in order to write (\ref{4.7}) in
terms of canonical coordinates, we must get rid of the time
derivatives of lapse and shift functions. Miraculously, the
equations (\ref{4.7}) simplify drastically when we split $u$ into
normal and tangential pieces, $u^\mu=b n^\mu+b^a Y^\mu_{,a}$.
Noticing that in the chosen frame we have from $g_{\mu\nu} n^\mu
n^\nu=-1,\;Y^\mu_{,a} g_{\mu\nu} n^\nu=0$ that
$n^{\mu=t}=1/n,\;n^{\mu=a}=-n^a/n$, as well as $n_t=-n,\;n_a =0$, we
find \be \label{4.8} u^t=\frac{b}{n},\;u^a=b^a-b\frac{n^a}{n}. \ee
Substituting (\ref{4.8}) into (\ref{4.7}), almost everything cancels
due to the explicit lapse and shift dependence of the vector field
$u$ and yields after some calculus \ba \label{4.7a} \delta_u n^a &=&
({\cal L}_{\vec{b}} \vec{n})^a+q^{ab}\left(b n_{,b}-n b_{,b}\right)+
b^a_{,t}
\nonumber\\
2n \delta_u n &=& 2n\left(b_{,t}+b^a n_{,a}-n^a b_{,a}\right). \ea
The time derivatives of lapse and shift have disappeared as desired.
The required extension of $c^{\rm tot}(b)+\vec{c}^{\rm
tot}(\vec{b})$ that generates spacetime diffeomorphisms on all ten
components of the spacetime metric and not only its spatial --
spatial components is therefore given by \be \label{4.8a}
\int_{{\cal X}} \; d^3x \; \left(p_a\left[({\cal L}_{\vec{b}}
\vec{n})^a +q^{ab}(b n_{,b}-n b_{,b})+ b^a_{,t}\right]
+p\left[b_{,t}+b^a n_{,a}-n^a b_{,a}\right]\right). \ee The
coefficients appearing in (\ref{4.8a}) are no accident, but directly
derive from the Dirac algebra of secondary constraints as shown
explicitly in \cite{Pons}. It follows that the primary constraints
have to be there in order to make the formalism manifestly spacetime
covariant. We observe that when choosing $b=n,\;b^a=n^a$ the
modified expression $c^{\rm tot}(f)+\vec{c}^{\rm
tot}(\vec{f})+$(\ref{4.8a}) turns {\it precisely} into the primary
Hamiltonian (\ref{4.1}), because the equation of motion for lapse
and shift can be used to eliminate $\nu=b,\;\nu^a=b^a$. Thus we may
say that for those special smearing fields the generator of
spacetime diffeomorphisms is nothing but the primary Hamiltonian.
Thus, the extended generator of spacetime diffeomorphisms is nothing
but the natural extension of the primary Hamiltonian to arbitrary
test functions different from $b=n,\;b^a=n^a$.

Performing integrations by parts (using suitable fall-off properties
of $b, \;b^a$), we collect the coefficients of $b^a$ and $b$ in
(\ref{4.8a}) (but not of $\dot{b}^a,\;\dot{b}$), which give the
lapse and shift (LS) contributions to the spatial diffeomorphism and
Hamiltonian constrain,t respectively: \ba \label{4.9} c_a^{LS} &=&
p\; n_{,a}+p_b \; n^b_{,a}+\left[p_b\; n^b\right]_{,a}
\nonumber\\
c^{LS} &=& \left[q^{ab} p_b n+n^a p\right]_{,a} +q^{ab} p_b n_{,a}.
\ea The first equation tells us that $n,n^a$ transform as scalar and
vector, respectively, under spatial diffeomorphims, while $p,p_a$
are correspondingly scalar and covector densities, respectively. For
this reason all the derivatives in the second equation in
(\ref{4.9}) can be replaced by covariant ones.

Since we have extended the phase space by
lapse and shift, we now must include the contributions (\ref{4.9}) into
the secondary constraints which thus read
\ba \label{4.10}
c^{{\rm tot}\prime} &=& \left[c^{{\rm LS}}+c^{{\rm geo}}+c^{{\rm
matter}}\right]+c^{\rm dust}
=:c'+c^{\rm dust}
\nonumber\\
c_a^{{\rm tot}\prime} &=& \left[c_a^{{\rm LS}}+c_a^{{\rm
geo}}+c_a^{{\rm matter}}\right]+c_a^{\rm dust} =:c_a'+c_a^{\rm
dust}. \ea They have to be used in our formula for gauge-invariant
completions with respect to the secondary spatial diffeomorphism and
Hamiltonian constraints, respectively. Now, as before, we solve
(\ref{4.10}) for the dust momenta and obtain

\ba \label{4.11}
\tilde{c}^{{\rm tot}\prime}&=& P+\sqrt{\left[c'\right]^2+q^{ab} c'_a c'_b}=:P+h'
\nonumber\\
\tilde{c}_a^{{\rm tot}\prime}&=& P_j+S_j^a\left(-h' T_{,a}+c'_a\right)
\ea
which is completely analogous to the formalism in which the primary
constraints have been reduced already.

The $\dot{b}$ and $\dot{b}^a$ terms in (\ref{4.8a}) cannot be taken
care of in $c^{LS},c_a^{LS}$. In principle, we have to consider
$\nu^0:=\dot{b},\; \nu^a:=\dot{b}^a$ as independent test functions.
Therefore, the $\dot{b} p+\dot{b}^a p_a=\nu^\mu z_\mu$ contributions
in (\ref{4.8a}) have to be attributed to the primary constraints
$z_\mu=p_\mu$, that is, $z=p,\;z_a=p_a$. Since the primary
constraints have not yet been reduced, we must also perform the
gauge-invariant completion with respect to the primary constraints.
 Notice that since $z,\;z_a,\;c^{{\rm tot}},\;c^{{\rm
tot}}_a$ are weakly Poisson commuting, so are the $z,\;z_a,\;c^{{\rm
tot}\prime},\;c^{{\rm tot}\prime}_a$ because (\ref{4.9}) is a linear
combination of the $z,z_a$. Hence the system remains first class.
Since $z,z_a$ are presented in deparametrised form, we know that the
constraints $z,z_a,\tilde{c}^{{\rm tot}\prime},\tilde{c}_a^{{\rm
tot}\prime}$ are mutually Poisson commuting. Hence, in addition to
what we did in our companion paper, we must supplement the
gauge-invariant extension $F$ of some function $f$ with respect to
the secondary constraints by gauge-invariant extensions with respect
to the primary constraints $z,z_a$, for which of course we use as
clocks the lapse and shift functions $n,n^a$. Specifically, that
last extension is given by the formula

\be \label{4.12} O^{(3)}_F[\nu]=F+\sum_{n=1}^\infty\;\frac{1}{n!}\;
\int_{{\cal X}} \; d^3x_1 \; .. \int_{{\cal X}}\; d^3x_n\;
\left(\nu^{\mu_1}-n^{\mu_1}\right)(x_1)\;..
\left(\nu^{\mu_n}-n^{\mu_n}\right)(x_n)\;..
\{z_{\mu_1}(x_1),..,\{z_{\mu_n}(x_n),F\}..\} \ee with arbitrary
functions $\nu^\mu,\;\mu=0,1,2,3$ and we have set
$n^0:=n,\;n^{\mu=a}:=n^a,\;a=1,2,3$. This operation leaves functions
independent of $n,n^a$ unaffected, of course, and, more generally,
replaces any $n^\mu$ by $\nu^\mu$, in accordance with the fact that
$n,n^a$ are pure gauge with respect to the primary constraints.

Notice that the gauge-invariant projection $O^{(3)}$ has to be
performed {\it after} the operations $O^{(1)},\;O^{(2)}$ of our
companion paper, because $p,p_a$ Poisson commute with $T,S^j$, but
$n,n^a$ do not Poisson commute with $\tilde{c}^{{\rm
tot}\prime}_a,\;\tilde{c}^{{\rm tot}\prime}_a$. Since in the
complete projection the constraints never act on the
$\tau-T,\;\sigma^j-S^j,\;
\nu-n,\;\nu^a-n^a$ powers, (\ref{4.12}) has to applied in this order.\\
\\
With this machinery at our disposal we can now
continue our comparison with the SCPT framework.


\subsection{Comparison with the SCPT framework for Lapse and Shift}


As discussed in the last section, for the purpose of comparing with
the SCPT framework, we have to use the extended Hamiltonian
formalism. Thus we now must use the constraints shown in equation
(\ref{4.10}), as well as the additional projection in equation
(\ref{4.11}). All that we have said before remains valid because on
the constraint surface defined by the primary constraints the
constraints (\ref{4.10}) coincide with the old ones. Let us write
the perturbed expansion to linear order of a complete
gauge-invariant quantity in the extended phase space as \be
\label{4.12a} \delta O_f=\delta f-\int_{{\cal X}}\;d^3y\;
\left(\delta T(y)\;\overline{\{\tilde{c}^{{\rm tot}\prime}(y),f\}}
+\delta S^j(y)\;\overline{\{\tilde{c}_j^{{\rm tot}\prime}(y),f\}}
+\delta n(y)\;\overline{\{z(y),f\}} +\delta
n^a(y)\;\overline{\{z_a(y),f\}}\right) \ee where \be \delta
T=T-\tau,\quad \delta S^j=S^j-\sigma^j,\quad \delta
n=n-\nu,\quad\mathrm{and}\quad \delta n^a=n^a-\nu^a. \ee
 Evaluating (\ref{4.12a}) for the functions $f=n(x),\,n^a(x)$
yields \ba \label{4.13} \delta\tilde{n}(x):=\delta O_{n(x)} &=&
\left[\delta n - \ov{S^a_j}\delta S^j
\overline{n_{,a}}+\overline{n^a}
\left[\delta T\right]_{,a}-\delta n\right](x)\\
&=&0\nonumber\\
\nonumber\\
\delta \tilde{n}^a(x):=\delta O_{n^a(x)} &=& \left[\delta n^a
- \ov{S^b_j}\delta S^j\; \overline{n^a_{,b}}
+ \overline{n^b}\; \left[\ov{S^a_j}\delta S^j\right]_{,b}-\overline{q^{ab}}\left(\delta T
\overline{n}_{,b}-\overline{n} \left[\delta T\right]_{,b}\right)-\delta n^a\right](x)\nonumber\\
&=&\frac{\delta^{ab}}{a^2}\left[\delta T\right]_{,b}(x)\nonumber\\
\ea where we used $\overline{n}=1,\;\overline{n^a}=0$. We want to
compare $\tilde{n}$ and $\tilde{n}_a$ with $\delta\tilde{g}_{00}$
and $\delta\tilde{g}_{0a}$, respectively, in equation (\ref{GIg}).
From equation (\ref{Pergmunu}) we can read off \be \delta
\tilde{g}_{00}=-2 a^2 \delta \tilde{n}\quad\mathrm{and}\quad \delta
\tilde{g}_{0a}=a^3 \delta_{ab} \delta \tilde{n}^b. \ee
 Hence the expressions for $\delta\tilde{g}_{00}$ and $\delta\tilde{g}_{0a}$ in equation (\ref{GIg}) are equivalent to
\be \label{4.16} \delta\tilde{n}=\delta
n-\left[\delta\dot{T}\right]\quad\mathrm{and}\quad
\delta\tilde{n}^a=\delta
n^a-\left[\ov{S}^a_j\delta\dot{S}^j-\frac{\delta^{ab}}{a^2}
\left[\delta T\right]_{,b}\right], \ee where we  used that
$\dot{\ov{S}^a_j}=0$. Comparing the Lagrangian result (\ref{4.16})
with the Hamiltonian result (\ref{4.13}), we do not seem to obtain a
match. The reason is the appearance of the time derivatives of
$\delta T$ and $\delta S^j$. On the other hand, notice that with the
identifications $b=\delta T$ and $b^a=\ov{S}^a_j\delta S^j$ the
square bracket terms on the right hand side of  (\ref{4.16})
precisely cancel the spacetime diffeomorphism transformation derived
in equation (\ref{4.7a}) via Hamiltonian methods. The difference is
that the fields $b,b^a$ in equation (\ref{4.7a}) were smearing
fields, while here $\delta T$ and $\delta \ov{S}^a_j\delta S^j$ are
phase space dependent functions. The only way in which the
Hamiltonian formalism can produce time derivatives of the canonical
fields is via the equations of motion. Now the equations of motion
in the Hamiltonian formalism which lead to an exact match with the
Euler -- Lagrange equations are with respect to the primary
Hamiltonian (\ref{4.1}). We find \ba \label{4.17} \dot{T}(x) &=&
\{H_{{\rm primary}},T(x)\}=\left[n^a T_{,a}-n
\frac{P}{\sqrt{P^2+q^{ab} c'_a c'_b}}\right](x)
\nonumber\\
\dot{S}^j(x) &=& \{H_{{\rm primary}},T(x)\} = \left[n^a
S^j_{,a}\right](x). \ea Perturbing equation (\ref{4.17}) around the
FRW background and using $\overline{c'_a}=\ov{T}_{,a}=0$, as well as
$\dot{\overline{S}^j_a}=0$, we find \be \label{4.18} \delta
\dot{T}=\delta n\quad\mathrm{and}\quad\overline{S^a_j} \delta
\dot{S}^j=\delta n^a. \ee Equation (\ref{4.18}) is evidently the
missing link to bring (\ref{4.13}) and (\ref{4.16}) to an exact
match.

By construction, the full, non-perturbative expression for
$O_{n^a(x)}$ and $O_{n(x)}$ will be expressed in terms of the
fundamental seven physical degrees of freedom corresponding to
$q_{ab}, \xi$ and their conjugate momenta. Hence in the Hamiltonian
framework (\ref{4.16}) are of no further interest. In the Lagrangian
formalism, the secondary constraints fulfill the task to express
those equations in terms of the other degrees of freedom. Hence in
both frameworks we end up with seven gauge-invariant degrees of freedom. \\
\\
Summarising, we showed that, using the four dust fields $\delta
T,\delta S^k$, we are able to construct (up to linear order)
gauge-invariant quantities along the lines of SCPT. We obtain 10
gauge-invariant components $\delta\widetilde{g}_{\mu\nu}$ and one
for the matter scalar field $\delta\widetilde{\xi}$. Out of these 11
degrees of freedom only 7 are physically relevant, because apart
from the non-physical dust degrees of freedom the system is reduced
by another four degrees of freedom due to the four primary
constraints of General Relativity. This leads to a reduction from
the 15 configuration degrees of freedom of the system "gravity +
scalar field + dust" down to 7 true degrees of freedom. These seven
degrees of freedom reside in the perturbations of the three metric
$\delta\widetilde{g}_{ab}=\delta\widetilde{q}_{ab}$ and the scalar
field $\delta\widetilde{\xi}$. Finally, we expanded the (manifestly)
gauge-invariant observables constructed by the method introduced in
\cite{1} up to linear order in the perturbations and compared them
with $\delta\widetilde{g}_{\mu\nu}$ and $\delta\widetilde{\xi}$. It
was shown that these two ways of constructing (up to linear order)
gauge-invariant quantities lead to exactly the same results. Four of
those variables are redundant in both formalisms. Thus, for the
physical degrees of freedom we get an exact match. In contrast to
section \ref{s3} where SCPT without dust was considered, we also
have an agreement on the number of physical degrees of freedom here,
7 for SCPT and 7 in our framework. However, as was discussed in
section \ref{s3}, the equations of motion derived for the 7 degrees
of freedom show that 4 of these degrees of freedom freeze out in the
late universe, so that the physics is mainly governed by the 3
degrees of freedom used in SCPT.

\section{Summary and Conclusions}
\label{s5} In this paper we applied a new framework for general
relativistic perturbation theory, developed in our companion paper
\cite{1}, to the important case of cosmological perturbations. The
central feature of our new approach is the use of a dynamically
coupled observer medium, given by pressureless dust. This allows for
a complete deparametrisation of the physical system of interest,
that is gravity coupled to whatever matter one wishes to include. As
a result, the usual gauge freedom of general relativity is
eliminated and true observables are obtained, together with a true
time evolution generated by a physical Hamiltonian. By specializing
our general framework to the case of an FRW background spacetime, we
developed a manifestly gauge-invariant cosmological perturbation
theory. As the quantities that are perturbed, namely the
three-metric and all the (non-dust) matter fields, are already fully
gauge-invariant by construction, the familiar problems with gauge
freedom, that had been troublesome for standard cosmological
perturbation theory (SCPT) for a long time, never arise here. In
particular, it is straightforward, though no doubt involved, to
derive perturbed equations of motions to arbitrarily high order.

In this paper, we limited ourselves to investigating the linear
order. We found that up to a small correction term, our formalism
reproduces the known SCPT equations. The correction term is an
imprint of the dust system. On physical grounds, this could have
been anticipated right from the start, as the observer dust adds to
the overall energy momentum balance. Our framework thus illustrates
the fact that including a realistic, that is non-idealized, observer
in the description of a physical system will always disturb the
original system to some extent. As this correction turned out to be
inversely proportional to the scale factor, however, it decays away
very quickly in the early universe. Consequently, the equations of
motion coming out of our framework are well within an acceptable
range of the standard cosmological model results.

There are many routes for further investigations that one can take
from here. Since the issue of gauge invariance has been entirely
settled to all orders in our approach to perturbation theory, the
obvious next step is to calculate the equations of motion for higher
order perturbations, in particular for the second order \cite{7}. As
mentioned in the introduction, the latter is a topic that currently
attracts a considerably amount of attention, due to its connection
with the issue of non-Gaussianity of cosmological perturbations.
This should also help to settle the stability issue of linear
perturbation theory \cite{stability}. Another interesting project is
to investigate the spherically symmetric sector of the theory and to
determine whether there are modifications of Newton's law at large
distances. In fact, it might not be totally surprising if there were
modifications because the dust is a perfect candidate for a WIMP,
one of the candidates for dark matter, since it only interacts
gravitationally.
\\
\\
\\
\\
{\large\bf Acknowledgements}\\
\\
K.G. thanks the Perimeter Institute for Theoretical Physics for
hospitality and financial support. Research performed at Perimeter
Institute for Theoretical Physics is supported in part by the
Government of Canada through NSERC and by the Province of Ontario
through MRI. O.W. was partially supported by the Province of Ontario
through an ERA award, ER06-02-298.

\begin{appendix}

\section{Review of Standard Cosmological Perturbation Theory}
\label{sa}

For the benefit of the reader we collect here the most important
formulae of the SCPT formalism, adapted to our notation. In
particular, notice that we are using the relativist's signature
$(-,+,+,+)$, which is opposite to the one used by cosmologists.

\subsection{Curvature and Energy Momentum Tensor}
\label{sa.1}

Our convention for the curvature tensor is \be \label{a.2}
R_{\mu\nu\rho}\;^\sigma\;\omega_\sigma:=
[\nabla_\mu,\nabla_\nu]\omega_\rho. \ee Its explicit expression in
terms of the  Christoffel symbols (we use the notation
$(.)_{,\mu}:=\partial (.)/\partial x^\mu$) \be \label{a.3}
\Gamma^\sigma_{\mu\nu}=g^{\sigma\rho}\Gamma_{\rho\mu\nu},\;\;
\Gamma_{\rho\mu\nu}=\frac{1}{2}\left( g_{\rho\mu,\nu}
+g_{\rho\nu,\mu} -g_{\mu\nu,\rho}\right), \ee corresponding to the
torsion free covariant differential $\nabla$ compatible with a
general metric $g_{\mu\nu}$ with inverse $g^{\mu\nu}$, is \be
\label{a.3a}
R_{\mu\nu\rho}\;^\sigma=2\left(-\partial_{[\mu}\Gamma^\sigma_{\nu]\rho}
+\Gamma^\lambda_{\rho[\mu}\;\Gamma^\sigma_{\nu]\lambda}\right). \ee
From this we infer the Ricci tensor, Ricci scalar and Einstein
tensor, respectively: \be \label{a.5}
R_{\mu\nu}:=R_{\mu\rho\nu}\;^\rho,\;R:=g^{\mu\nu} R_{\mu\nu},\;
G_{\mu\nu}=R_{\mu\nu}-\frac{1}{2} R g_{\mu\nu}. \ee ~\ The energy
momentum tensor for bosonic matter is given by \be \label{a.5a}
T_{\mu\nu}:=-\frac{2}{\sqrt{|\det(g)|}}\;\frac{\delta
S_{\rm matter}}{\delta g^{\mu\nu}} \ee which, for a minimally coupled
scalar field $\zeta$ with action \be \label{a.5b}
S_{\rm matter}=-\frac{1}{2\lambda}\int_M\; d^4X\;\sqrt{|\det(g)|}\;
\left(g^{\mu\nu}(\nabla_\mu \zeta)(\nabla_\nu \zeta)+v(\zeta)\right)
\ee (where $v$ is its potential), becomes \be \label{a.5c} \lambda
T_{\mu\nu}=\left[\nabla_\mu \zeta\right]\left[\nabla_\nu
\zeta\right]-\frac{1}{2}g_{\mu\nu}\Big(g^{\rho\sigma}(\nabla_\rho
\zeta)(\nabla_\sigma \zeta)+v(\zeta)\Big). \ee The Euler-Lagrange
equations for the gravity and scalar matter system (including a
cosmological term) \be \label{a.5d}
S_{\rm geo}=\frac{1}{\kappa}\int_M\;d^4X\;\sqrt{|\det(g)|}\;\left(R-2\Lambda\right)
\ee are the Einstein equations \be \label{a.5e} G_{\mu\nu}+\Lambda
g_{\mu\nu}=\frac{\kappa}{2} T_{\mu\nu} \ee and the matter field
equation \be \label{a.5f} g^{\mu\nu} \nabla_\mu\nabla_\nu
\zeta=\frac{1}{2} v'(\zeta). \ee

\subsection{FRW Background}
\label{sa.2}

The spatial constant curvature $k=0$ FRW line element with scale
factor $a(t)$ reads \be \label{a.1} ds^2=-dt^2+a(t)^2 \delta_{ab}
dx^a dx^b= a(x^0)^2 \eta_{\mu\nu} dx^\mu dx^\nu \ee where we have
switched to conformal time $dx^0:=d\eta:=dt/a(t)$ and
$\eta_{\mu\nu}$ denotes the Minkowski metric. In the second step we
should actually use a different symbol $\tilde{a}(x^0):=a(t(x^0))$,
but we will slightly abuse the notation as is customary.

All quantities that refer to a background metric will carry a bar.
For the FRW background metric $\overline{g}_{\mu\nu}=a^2\eta_{\mu\nu}$ a short
computation reveals
\ba \label{a.6}
\overline{\Gamma}^0_{00} &=& \frac{a'}{a}=:{\cal H}
\nonumber\\
\overline{\Gamma}^0_{0a} &=& 0
\nonumber\\
\overline{\Gamma}^0_{ab} &=& {\cal H}\; \delta_{ab}
\nonumber\\
\overline{\Gamma}^a_{00} &=& 0
\nonumber\\
\overline{\Gamma}^a_{0b} &=& {\cal H}\; \delta^a_b
\nonumber\\
\overline{\Gamma}^a_{bc} &=& 0. \ea Here and in what follows a prime
denotes derivation with respect to conformal time $x^0=\eta$, rather
than cosmological time $t$, and we have introduced the Hubble
function ${\cal H}=a'/a$. It is useful to note that (\ref{a.6})
implies that \be \label{a.6a} \overline{\Gamma}^\nu_{\nu\mu}=4\;
{\cal H}\; \delta_{\mu 0}. \ee

Carefully using the definitions, we find again after a short
calculation \ba \label{a.7} \overline{R}_{00} &=& -3{\cal H}'
\nonumber\\
\overline{R}_{0a} &=& 0
\nonumber\\
\overline{R}_{ab} &=& \left({\cal H}'+2{\cal H}^2\right)\;\delta_{ab}
\nonumber\\
\overline{R} &=& \frac{6}{a^2}\left({\cal H}'+{\cal H}^2\right)
\nonumber\\
\overline{G}_{00} &=& 3 \; {\cal H}^2
\nonumber\\
\overline{G}_{0a} &=& 0
\nonumber\\
\overline{G}_{ab} &=& -\left(2 {\cal H}'+{\cal
H}^2\right)\;\delta_{ab}. \ea
~\\
The non-vanishing components of the energy momentum tensor are \ba
\label{a.7a} \lambda \overline{T}_{00} &=&(\overline{\zeta}')^2
+\frac{1}{2}a^2\left(-\frac{1}{a^2}(\overline{\zeta}')^2+v(\overline{\zeta}
)\right) =\frac{1}{2}\left((\overline{\zeta}')^2+a^2
v(\overline{\zeta})\right)=:a^2 \lambda \overline{\rho}
\nonumber\\
\lambda \overline{T}_{ab} &=&
\frac{1}{2}\left((\overline{\zeta}')^2-a^2
v(\overline{\zeta})\right)\delta_{ab}=:a^2\lambda \delta_{ab}
\overline{p}, \ea where we have introduced background energy density
$\overline{\rho}$ and background pressure $\overline{p}$,
respectively. The matter field equation becomes \be \label{a.7b}
\frac{1}{\sqrt{|\det(\overline{g})|}}\partial_\mu
\overline{g}^{\mu\nu} \sqrt{|\det(\overline{g})|}\partial_\nu
\overline{\zeta}=-\frac{1}{a^4}(a^2 \overline{\zeta}')'=
-\frac{1}{a^2}\left(2{\cal
H}\overline{\zeta}'+\overline{\zeta}^{\prime\prime}\right)=\frac{1}{2}
v'(\overline{\zeta}). \ee

\subsection{Linear Perturbations}
\label{sa.3}

We consider perturbations $\delta
g_{\mu\nu}:=g_{\mu\nu}-\overline{g}_{\mu\nu}$ and $\delta
\zeta=\zeta-\overline{\zeta}$. Any function $F=F(g,\zeta)$ of the
metric $g$ and the scalar field $\zeta$ is expanded to linear order
in $\delta g,\;\delta\zeta$, that is, $\delta F$ will denote the
linear order term in the Taylor expansion of
$F(g,\zeta)-F(\overline{g},\overline{\zeta})$. For instance, \ba
\label{a.8} \delta \Gamma^\sigma_{\mu\nu} &=& \delta
g^{\sigma\rho}\; \overline{\Gamma}_{\rho\mu\nu}+
\overline{g}^{\sigma\rho}\; \delta \Gamma_{\rho\mu\nu}
\nonumber\\
&=&
-\overline{g}^{\sigma\lambda}\;\delta g_{\lambda\tau} \;\overline{g}^{\tau\rho}
\overline{\Gamma}_{\rho\mu\nu}+
\overline{g}^{\sigma\rho}\; \delta \Gamma_{\rho\mu\nu}
\nonumber\\
&=& -\overline{g}^{\sigma\lambda}\;\delta g_{\lambda\rho}
\overline{\Gamma}^\rho_{\mu\nu}+ \overline{g}^{\sigma\rho}\;
\frac{1}{2}\left( \delta g_{\rho\mu,\nu} +\delta g_{\rho\nu,\mu}
-\delta g_{\mu\nu,\rho}\right). \ea For the FRW background this
yields \be \label{a.9} \delta \Gamma^\sigma_{\mu\nu}
=\frac{1}{a^2}\left[-\eta^{\sigma\lambda}\;\delta g_{\lambda\rho}
\overline{\Gamma}^\rho_{\mu\nu}+ \eta^{\sigma\rho}\; \frac{1}{2}(
\delta g_{\rho\mu,\nu} +\delta g_{\rho\nu,\mu} -\delta
g_{\mu\nu,\rho})\right]. \ee Inserting (\ref{a.9}), we find
explicitly \ba \label{a.10} \delta \Gamma^0_{00} &=&
-\left(\frac{1}{2a^2} \delta g_{00}\right)'
\nonumber\\
\delta \Gamma^0_{0a} &=& \frac{1}{2a^2}\left[2{\cal H} \delta g_{0a}-\delta
g_{00,a}\right]
\nonumber\\
\delta \Gamma^0_{ab} &=& \frac{1}{2a^2}\left[2{\cal H} \delta g_{00}
\delta_{ab}-(2\delta g_{0(a,b)}-\delta g_{ab}')\right]
\nonumber\\
\delta \Gamma^a_{00} &=& \frac{1}{2a^2}\left(-2{\cal H} \delta g_{0a}
+2\delta g_{0a}'-\delta g_{00,a}\right)
\nonumber\\
\delta \Gamma^a_{b0} &=& \frac{1}{2a^2}\left(-2{\cal H} \delta g_{ab}
+\delta g_{ab}'+2\delta g_{0[a,b]}\right)
\nonumber\\
\delta \Gamma^a_{bc} &=& \frac{1}{a^2}\left(-{\cal H} \delta g_{0a}
\delta_{bc} +2\delta g_{a(b,c)}'-\delta g_{bc,a}\right).
 \ea It is
useful to note that (\ref{a.10}) implies \be \label{a.11} \delta
\Gamma^\nu_{\nu\mu}=\left[\frac{1}{2a^2}\left(\delta^{bc}\delta
g_{bc}-\delta g_{00}\right)\right]_{,\mu}. \ee Now we use
(\ref{a.3}) and (\ref{a.3a}) to find \be \label{a.12} \delta
R_{\mu\nu}=-2\partial_{[\mu} \delta\Gamma^\rho_{\rho]\nu}
+\overline{\Gamma}^\rho_{\rho\sigma} \delta\Gamma^\sigma_{\mu\nu}
+\overline{\Gamma}^\sigma_{\mu\nu} \delta \Gamma^\rho_{\rho\sigma}
-2 \overline{\Gamma}^\rho_{\sigma (\mu}
 \delta{\Gamma}^\sigma_{\nu)\rho}.
\ee Specialising (\ref{a.12}) to the temporal -- temporal, temporal
-- spatial and spatial -- spatial components, respectively, yields
\ba \label{a.13} \delta R_{00} &=& \left(-\frac{1}{2a^2} \delta
g_{cd} \delta^{cd}\right)^{\prime\prime}
+\frac{1}{a}\left(\frac{\delta^{cd}\delta
g_{0c,d}}{a}\right)'-\frac{1}{2a^2}\Delta \delta g_{00}
 -{\cal H}\left(\frac{1}{2a^2} \delta g_{cd} \delta^{cd}\right)^{\prime}
-3{\cal H}\left(\frac{1}{2a^2} \delta g_{00}\right)'
\nonumber\\
\delta R_{0a} &=&
\left(-\frac{1}{2a^2} \delta g_{cd,a} \delta^{cd}\right)^{\prime}
+\left(\frac{{\cal H}}{2a^2} \delta g_{0a}\right)^{\prime}
 +\left(\frac{1}{2a^2} \delta^{cd} \delta g_{ac,d}\right)^{\prime}
+\frac{1}{2a^2}\left(\delta^{cd} \delta g_{0c,da}-\Delta \delta g_{0a}\right)
\nonumber\\
&&
+\frac{{\cal H}}{a^2}\left( 4 {\cal H} \delta
g_{0a}-\delta g_{00,a}-\delta g_{0a}'\right)
\nonumber\\
\delta R_{ab} &=&
-\frac{1}{2a^2}\left[\delta^{cd} \delta g_{cd}-\delta g_{00}\right]_{,ab}
+\left[\frac{{\cal H}}{a^2} \delta g_{00} \delta_{ab}
-\frac{1}{2a^2}\left(2\delta g_{0(a,b)}-\delta g_{ab}'\right)\right]'
\nonumber\\
&& +2{\cal H}\left[\frac{{\cal H}}{a^2} \delta g_{00} \delta_{ab}
-\frac{1}{2a^2}\left(2\delta g_{0(a,b)}-\delta g_{ab}'\right)\right] +
\frac{1}{a^2}\left(-{\cal H} \delta^{cd} \delta_{ab} \delta g_{0c,d}
+\delta^{cd} \delta \Gamma_{cab,d}\right)
\nonumber\\
&& -2 {\cal H}\left(\frac{1}{2a^2} \delta g_{ab}\right)' +{\cal H}
\delta_{ab} \left(\frac{1}{2a^2}(\delta^{cd} \delta g_{cd}-\delta
g_{00})\right)'.
 \ea Here $\Delta=\delta^{ab} \partial_a
\partial_b$ denotes the flat background Laplacian and $2\delta
\Gamma_{cab}=2\delta g_{c(a,b)}-\delta g_{ab,c}$. This implies \ba
\label{a.14} \delta R &=& -\ov{g}^{\nu\sigma}\overline{g}^{\mu\rho}\; \delta
g_{\rho\sigma}\; \overline{R}_{\mu\nu} +\overline{g}^{\mu\nu} \delta
R_{\mu\nu}\nonumber\\
&=&
 -\frac{1}{a^4}\left(-3{\cal H}'\delta g_{00}+({\cal
H}'+2{\cal H}^2)\delta^{cd} \delta g_{cd}\right)
+\frac{1}{a^2}\left(-\delta R_{00}+\delta^{cd} \delta R_{cd}\right)
\nonumber\\
\delta^{cd} \delta R_{cd} &=&
-\frac{1}{2a^2}\Delta\left(\delta^{cd} \delta g_{cd}-\delta g_{00}\right)
+\left(3\frac{{\cal H}}{a^2} \delta g_{00}
-\frac{\delta^{cd}}{2a^2}(2\delta g_{0(c,d)}-\delta g_{cd}')\right)'
-2 {\cal H}\left(\frac{\delta^{cd}}{2a^2} \delta g_{cd}\right)'
\nonumber\\
&&+ \frac{1}{a^2}\left(-3{\cal H} \delta^{cd} \delta g_{0c,d}
+\delta^{cd} \delta^{ab} \delta \Gamma_{cab,d}\right)
+2{\cal H}\left(3\frac{{\cal H}}{a^2} \delta g_{00}
-\frac{\delta^{cd}}{2a^2}(2\delta g_{0(c,d)}-\delta g_{cd}')\right)\nonumber\\
&& +3{\cal H} \left(\frac{1}{2a^2}(\delta^{cd} \delta g_{cd}-\delta
g_{00})\right)'.
 \ea Since \be \label{a.15} \delta G_{\mu\nu}=\delta
R_{\mu\nu}-\frac{\overline{R}}{2}\delta g_{\mu\nu}
-\frac{\overline{g}_{\mu\nu}}{2} \delta R, \ee we find \ba
\label{a.16} \delta G_{00} &=& \frac{1}{2}\left(\delta
R_{00}+\delta^{cd} \delta R_{cd}\right)
-\frac{1}{a^2}\left(\frac{1}{2}({\cal H}'+{\cal H}^2)\delta^{cd}
\delta g_{cd}+3{\cal H}^2 \delta g_{00}\right)
\nonumber\\
\delta G_{0a} &=& \delta R_{0a} -\frac{3}{a^2}\left({\cal H}'+{\cal
H}^2\right)\delta g_{0a}
\\
\delta G_{ab} &=& \delta R_{ab}-\frac{3}{a^2}({\cal H}'+{\cal
H}^2)\delta g_{ab}
 -\frac{a^2}{2}\delta_{ab}\left[-\frac{1}{a^4}\left(-3{\cal H}' \delta g_{00}+
\left({\cal H}'+2{\cal H}^2\right)\delta^{cd} \delta g_{cd}\right)
+\frac{1}{a^2}\left(-\delta R_{00}+\delta^{cd} \delta
R_{cd}\right)\right]. \nonumber \ea
Next the perturbation of (\ref{a.5c}) yields after a short
computation \be \label{a.16a} \lambda \delta T_{\mu\nu}=2 \zeta'
\delta^0_{(\mu} (\delta \zeta)_{,\nu)} -\frac{1}{2} \lambda
\overline{p} \delta g_{\mu\nu}
+\frac{1}{2}\eta_{\mu\nu}\left(\frac{1}{a^2}(\overline{\zeta}')^2
\delta g_{00}+2\overline{\zeta}' \delta\zeta'-a^2
v'(\overline{\zeta})\delta\zeta\right) \ee or, in components, \ba
\label{a.16b} \delta T_{00} &=& \frac{1}{2\lambda}\left(2
\overline{\zeta}'\delta\zeta'+a^2v'(\overline{\zeta})
\delta\zeta-v(\overline{\zeta}) \delta g_{00}\right)
\nonumber\\
\delta T_{0a} &=& \frac{1}{\lambda}\left(
\overline{\zeta}'\delta\zeta_{,a}+\lambda
\overline{p}\delta g_{0a}\right)
\nonumber\\
\delta T_{ab} &=& \frac{1}{\lambda}\left[\lambda \overline{p}\delta
g_{ab}
+\frac{1}{2}\delta_{ab}\left(\frac{1}{a^2}(\overline{\zeta}')^2
\delta g_{00}+2\overline{\zeta}' \delta\zeta'-a^2
v'(\overline{\zeta})\delta\zeta\right)\right]. \ea

\subsection{Parametrisation of perturbations and linear invariants}
\label{sa.4}

The complicated equations of the previous section can be decomposed
into scalar, vector and tensor contributions. As in \cite{6} we
introduce four
scalar fields $\phi,\psi,B,E$, two transversal (with respect to the
flat 3d Euclidean background metric)
covector fields $S_a,\; F_a$ and a symmetric tracefree, transversal
tensor
$h_{ab}$. Thus
$\delta^{ab} F_{a,b}=\delta^{ab} S_{a,b}=
\delta^{bc} h_{ab,c}=
\delta^{ab} h_{ab}=0$. These fields encode the ten independent
components of
$\delta g_{\mu\nu}$ as follows: We write the perturbed line element
\be \label{a.17}
ds^2=a^2 \eta_{\mu\nu} dx^\mu dx^\nu+\delta g_{00} (dx^0)^2
+2 \delta g_{0a} dx^0 dx^a+\delta g_{ab} dx^a dx^b
\ee
in the form
\be \label{a.18}
ds^2=a^2\left[\left(-1+2\phi\right) (dx^0)^2+2\left(S_a+B_{,a}\right) dx^0 dx^a
+\left((1+2\psi) \delta_{ab} +2 E_{,ab}+2F_{(a,b)}+h_{ab}\right)
dx^a dx^b\right]
\ee
from which one reads off
\ba \label{a.19}
\delta g_{00} &=& +2 a^2 \phi
\nonumber\\
\delta g_{0a} &=& + a^2\left(S_a+B_{,a}\right)
\nonumber\\
\delta g_{0a} &=& +a^2 \left[2(\psi \delta_{ab} +
E_{,ab}+F_{(a,b)})+h_{ab}\right].
 \ea Notice that we use different
signs from the literature, due to our signature conventions: All
perturbations enter with a positive coefficient proportional to
$a^2$. One speaks of scalar, vector and tensor perturbations,
respectively, when only the respective field perturbations are
non-vanishing.

The fields $\phi,\psi,E,B,S_a,F_a,h_{ab}$ are not invariant under
(infinitesimal) gauge transformations, that is, spacetime
diffeomorphisms. In general, a metric changes under an infinitesimal
diffeomorphism generated by a spacetime vector field $u^\mu$ by the
corresponding Lie derivative \be \label{a.20} \delta_u g_{\mu\nu}
=\left[\frac{d}{dt}\left[(\varphi^u_t)^\ast
g\right]_{\mu\nu}\right]_{t=0} =:\left[{\cal L}_u g\right]_{\mu\nu}
=u^\rho g_{\mu\nu,\rho}+2u^\rho_{,(\mu} g_{\nu)\rho},
 \ee where
$t\mapsto \varphi^u_t$ is the one parameter family of
diffeomorphisms generated by the integral curves of the vector field
$u$ \cite{12}. If we take the order of $u$ to be the same as the
order of the general perturbations, then we see that we can gauge
away four of the ten perturbation fields. We define
$u_\mu:=\eta_{\mu\nu} u^\nu$, that is $u_0=-u^0,\;u_a=u^a$ and
$u:=\Delta^{-1} u^a_{,a},\;u^\perp_a=u_a-\partial_a u$, where
$\Delta^{-1}$ is the Green function associated to $\Delta$. We find
explicitly \ba \label{a.21} \delta_u g_{\mu\nu} &=& a^2\left(2{\cal
H} u_0 \eta_{\mu\nu}+2u_{(\mu,\nu)}\right)
\nonumber\\
\delta_u g_{00} &=& 2a^2 \left(\frac{1}{a} \left(a u\right)'\right)
\nonumber\\
\delta_u g_{0a} &=& a^2 \left(\left[u_0+u'\right]_{,a}+u^{\perp\prime}_a\right)
\nonumber\\
\delta_u g_{ab} &=& 2a^2 \left(-{\cal H} u_0 \delta_{ab}
+u_{,ab}+u^\perp_{(a,b)}\right).
 \ea Comparing with (\ref{a.19}), we
read off \ba \label{a.22} \delta_u \phi &=& \frac{1}{a}\left(a
u_0\right)'
\nonumber\\
\delta_u B &=& u_0+u'
\nonumber\\
\delta_u \psi &=& -{\cal H} u_0
\nonumber\\
\delta_u E &=& u
\nonumber\\
\delta_u S_a &=& u^{\perp\prime}_a
\nonumber\\
\delta_u F_a &=& u^{\perp}_a
\nonumber\\
\delta h_{ab} &=& 0.
 \ea From (\ref{a.22}) we immediately see that a
complete and linearly independent set of linear invariants under
spacetime diffeomorphisms is given by \be \label{a.23}
\Phi=\phi-\frac{1}{a}\left[a\left(B-E'\right)\right]',\;\;\Psi=\psi+{\cal
H}(B-E'),\;\; V_a=S_a-F_a',\;\;h_{ab}, \ee which are six gauge
invariant degrees of freedom. This follows from the fact that \be
\label{a.24} \delta_u (B-E')=u_0,\;\;\delta_u E= u,\;\; \delta_u
F_a=u_a^\perp. \ee
~\\
A scalar field transforms under diffeomorphisms as \be \label{a.24a}
\delta_u \zeta=u^\mu \zeta_{,\mu}, \ee which to linear order equals
$u^0 \overline{\zeta}'=-\overline{\zeta}' u_0$. Thus the linearly
invariant scalar field perturbation is simply given by \be
\label{a.24b} Z:=\delta\zeta+\overline{\zeta}'(B-E'). \ee
~\\
Just like the metric $g_{\mu\nu}$, any symmetric tensor such as the
Einstein tensor $G_{\mu\nu}$ or the energy momentum tensor
$T_{\mu\nu}$ transforms as in (\ref{a.20}) under infinitesimal
spacetime diffeomorphisms. Therefore, the Einstein equations with
cosmological constant, $\Lambda$ \be \label{a.25}
E_{\mu\nu}:=G_{\mu\nu}+\Lambda g_{\mu\nu}-\frac{\kappa}{2}
T_{\mu\nu}=0, \ee are not written just in terms of the invariants
(\ref{a.23}). However, notice that for any symmetric tensor
$t_{\mu\nu}$ \ba \label{a.26} \delta_u t_{\mu\nu} &=& u^0
\overline{t}'_{\mu\nu}+2u^\rho_{,(\mu} \overline{t}_{\nu)\rho}
\nonumber\\
&=& u^0 \overline{t}'_{\mu\nu} +2u^0_{,(\mu} \overline{t}_{\nu)0}
+2\delta^{ab} u_{,a(\mu} \overline{t}_{\nu)b} +2\delta^{ab}
u^\perp_{a,(\mu} \overline{t}_{\nu)b}. \ea Recalling (\ref{a.24}) we
see that the tensor \be \label{a.27} \tilde{t}_{\mu\nu}:=t_{\mu\nu}
+(B-E') \overline{t}'_{\mu\nu} +2(B-E')_{,(\mu} \overline{t}_{\nu)0}
-2\delta^{ab} E_{,a(\mu} \overline{t}_{\nu)b} -2\delta^{ab}
F_{a,(\mu} \overline{t}_{\nu)b} \ee is gauge invariant. Now if we
choose $t_{\mu\nu}=E_{\mu\nu}$ then $\overline{t}_{\mu\nu}=0$, since
the FRW metric solves the Einstein equations. Therefore
$E_{\mu\nu}=\tilde{E}_{\mu\nu}$ is gauge invariant, and in the
second form it can be written just in terms of invariants when
decomposing $\overline{E}_{\mu\nu}=\overline{G}_{\mu\nu}+\Lambda
\overline{g}_{\mu\nu}-\kappa \overline{T}_{\mu\nu}/2=0$. The same
holds for the perturbations. Explicitly, for instance, \ba
\label{a.28} \delta \tilde{G}_{00} &=& \delta
G_{00}+(B-E')\overline{G}_{00}'+2(B-E')' \overline{G}_{00}
\nonumber\\
\delta \tilde{G}_{0a} &=& \delta G_{0a}+ \left[B-E'\right]_{,a}
\overline{G}_{00}
-\frac{1}{3}
\delta^{cd} \left(F_a+E_{,a}\right)' \overline{G}_{cd}
\nonumber\\
\delta \tilde{G}_{ab} &=& \delta G_{ab}+ (B-E') \overline{G}'_{ab}
-\frac{2}{3} \delta^{cd} \left[F_{(a}+E_{,(a}\right]_{,b)}
\overline{G}_{cd}, \ea where we used $\overline{G}_{00}=3{\cal
H}^2,\;\overline{G}_{0a}=0,\;\overline{G}_{ab}= -(2{\cal H}'+{\cal
H}^2)\delta_{ab}$. Equation (\ref{a.28}) equals (\ref{a.27}) with
$G_{\mu\nu}$ replaced by $t_{\mu\nu}$, whenever
$\overline{t}_{\mu\nu}$ is diagonal. Notice that the additional
terms in (\ref{a.28}) only contribute to scalar and vector
perturbation contributions. Hence $\delta \tilde{G}_{\mu\nu}=\delta
G_{\mu\nu}$ for tensor perturbations. We will see this explicitly.

Applying (\ref{a.27}) to $t_{\mu\nu}=g_{\mu\nu}$ we find \be
\label{a.28a} \delta\tilde{g}_{00}=2a^2 \Phi,\;\;
\delta\tilde{g}_{0a}=a^2 V_a,\;\; \delta\tilde{g}_{ab}=2a^2\Psi
\delta_{ab} \ee while, applying it to $t_{\mu\nu}=T_{\mu\nu}$, we
find after some algebra \ba \label{a.16ba} \delta \widetilde{T}_{00}
&=& \frac{1}{2\lambda}\left(2 \overline{\zeta}' Z'+a^2
v'(\overline{\zeta}) Z-2 v(\overline{\zeta}) a^2 \Phi\right)
\nonumber\\
\delta \widetilde{T}_{0a} &=& \frac{1}{\lambda}\left(
\overline{\zeta}' Z_{,a}+a^2 \lambda \overline{p} V_a\right)
\nonumber\\
\delta \widetilde{T}_{ab} &=& \frac{1}{\lambda}\left[\lambda
\overline{p} a^2h_{ab}+2\Psi\delta_{ab}
+\frac{1}{2}\delta_{ab}\left(2 (\overline{\zeta}')^2 \Phi
+2\overline{\zeta}' Z'-a^2 v'(\overline{\zeta})Z\right)\right]. \ea
The computation for the Einstein tensor itself is more complicated
and we divide it into modes.

\subsection{Tensor perturbations}
\label{sa.5}

For tensor perturbations we have $\delta g_{00}=\delta
g_{0a}=\delta^{cd}\delta g_{ac,d}=\delta^{cd}\delta g_{cd}=0$ and
$\delta g_{ab}=a^2 h_{ab}$. It immediately follows from (\ref{a.13})
that $\delta R_{00}=\delta R_{0a}=0$ and \be \label{a.29} \delta
R_{ab} = \left(\frac{1}{2a^2} \delta g_{ab}'\right)' -2{\cal H}
\left(\frac{1}{2a^2} \delta g_{ab}\right)' +\frac{{\cal H}}{a^2}
\delta g_{ab}'-\frac{1}{2a^2}\Delta \delta g_{ab}, \ee from which we
immediately infer $\delta R=0$, so that $\delta(g_{ab}
R)=\overline{R} \delta g_{ab}$. Using $\overline{R}= 6({\cal
H}'+{\cal H}^2)/a^2$, we find $\delta G_{00}=\delta G_{0a}=0$ and
for $\delta G_{ab}=\delta R_{ab}-\frac{1}{2}\overline{R}\delta
g_{ab}$ with $\delta g_{ab}=a^2 h_{ab}$ after some algebra \be
\label{a.30} \delta G_{ab}=\frac{1}{2}\left(h^{\prime\prime}_{ab}+2
{\cal H} h'_{ab}-\Delta h_{ab}\right)-\left(2{\cal H}'+{\cal
H}^2\right)h_{ab}. \ee As already mentioned, $\delta
G_{\mu\nu}=\delta \tilde{G}_{\mu\nu}$ for tensor perturbations.

The term in the second square bracket of (\ref{a.30}) is not
displayed in \cite{6}. However, notice that there one perturbs the
mixed components $G^\mu_\nu=g^{\mu\rho} G_{\rho\nu}$. Thus \be
\label{a.31} \delta G^\mu_\nu=-\overline{g}^{\mu\sigma}\delta
g_{\sigma\lambda} \overline{g}^{\lambda \rho}
\overline{G}_{\rho\nu}+\overline{g}^{\mu\rho} \delta G_{\rho\nu}.
\ee Hence $\delta G^0_0=\delta G^0_a=0$ and \be \label{a.31a} \delta
G^a_b= -\frac{1}{a^4} \delta g_{a c} \overline{G}_{c
b}+\frac{1}{a^2} \delta G_{a b}=\frac{1}{a^2}\left(-
h_{ab}\delta^{cd}\frac{\overline{G}_{c d}}{3}+ \delta G_{a
b}\right)=\frac{1}{2a^2}\left(h^{\prime\prime}_{ab}+2 {\cal H}
h'_{ab}-\Delta h_{ab}\right), \ee where we used that $\delta^{cd}
\overline{G}_{cd}/3=-(2{\cal H}'+{\cal H}^2)$. This is the equation
we find in \cite{6}. The reason why we display here the twice
covariant tensor components is that it is this form that one finds
more directly in the Hamiltonian formulation.

\subsection{Vector perturbations}
\label{sa.6}

In this case $\delta g_{00}=0,\;\delta g_{0a}=a^2 S_a,\;\delta
g_{ab}= 2 a^2 F_{(a,b)}$. Since both $S_a,\; F_a$ are transversal,
we have $\delta^{cd}\delta g_{0c,d}=\delta^{cd} \delta g_{cd}=0$, so
that we immediately find from (\ref{a.13}) that $\delta R_{00}=0$.
The remaining equations in (\ref{a.13}) simplify to \ba \label{a.32}
\delta R_{0a} &=& -\frac{1}{2} \Delta V_a+\left({\cal H}'+2{\cal
H}^2\right)S_a
\nonumber\\
\delta R_{ab} &=& -\left(V'_{(a,b)}+2{\cal H}V_{(a,b)}\right)+2
\left({\cal H}'+2{\cal H}^2\right)F_{(a,b)}, \ea where we have used
the linearly gauge invariant variable $V_a=S_a-F'_a$. Next from
(\ref{a.14}) we immediately see that $\delta R=0$ so that $\delta
(g_{\mu\nu} R)=\overline{R}\delta g_{\mu\nu}$. Hence $\delta
G_{00}=0$ and \ba \label{a.33} \delta G_{0a} &=& -\frac{1}{2} \Delta
V_a +\left(\delta^{cd}\frac{\overline{G}_{cd}}{3}\right) S_a
\nonumber\\
\delta G_{ab} &=& -\left(V'_{(a,b)}+2{\cal H}V_{(a,b)}\right)+2
\left(\delta^{cd}
\frac{\overline{G}_{cd}}{3}\right) F_{(a,b)}
\ea
where $\delta^{cd} \overline{G}_{cd}/3=-2{\cal H}'+{\cal H}^2$. Comparing
with (\ref{a.28}) we see that $\delta \tilde{G}_{00}=0$ and
\ba \label{a.33a}
\delta \tilde{G}_{0a} &=& -\frac{1}{2} \Delta V_a
+\delta^{cd}\frac{\overline{G}_{cd}}{3} V_a
\nonumber\\
\delta \tilde{G}_{ab} &=& -V'_{(a,b)}-2{\cal H}V_{(a,b)}.
 \ea These
are also the equations for the mixed components of the gauge
invariant Einstein tensor that we find in \cite{6}.

\subsection{Scalar perturbations}
\label{sa.7}

Now we have $\delta g_{00}=2\phi a^2,\;\delta g_{0a}=B_{,a}
a^2,\;\delta g_{ab}=2a^2 (\psi \delta_{ab}+E_{,ab})$. Hence
$\delta^{cd}\delta g_{0c,d}=a^2 \Delta B,\; \delta^{cd}\delta
g_{cd}=2a^2 (3\psi+\Delta E)$. Thus (\ref{a.13}) simplifies to \ba
\label{a.34} \delta R_{00} &=&
-\Delta\Phi-3\left(\psi^{\prime\prime}+{\cal
H}\left(\psi+\phi\right)'\right)
\\
\delta R_{0a} &=& -2\left[\Psi'+{\cal H}\Phi\right]_{,a}+3{\cal
H}'(B-E')_{,a}+({\cal H}'+2{\cal H}^2)E'_{,a}
\nonumber\\
\delta R_{ab} &=& \left[-\Psi+\Phi+({\cal H}'+2{\cal H}^2)
E\right]_{,ab} +\left(-\Delta \Psi+\psi^{\prime\prime}+2({\cal
H}'+2{\cal H}^2)(\psi+\phi) +5{\cal H}\psi'+{\cal H}\phi'\right)
\delta_{ab}\nonumber, \ea where we again used the gauge invariant
variables $\Psi=\psi+{\cal H}(B-E')$ and $\Phi=\phi-{\cal
H}(B-E')-(B-E')'$. As an intermediate result we have 
\ba
\label{a.35} \delta^{cd} \delta R_{cd} &=& \Delta
\left[-\Psi+\Phi+({\cal H}'+2{\cal H}^2) E\right] + 3\left(-\Delta
\Psi+\psi^{\prime\prime}+2({\cal H}'+2{\cal H}^2)(\psi+\phi) +5{\cal
H}\psi'+{\cal H}\phi'\right)\nonumber.\\ \ea Using (\ref{a.16}) we find after
some elaborate algebraic manipulations \ba \label{a.36} \delta
G_{00} &=& 2\left(-\Delta\Psi+3{\cal H} \psi'\right)
\\
\delta G_{0a} &=& -2\left[\Psi'+{\cal
H}\Phi\right]_{,a}-\overline{G}_{00}(B-E')_{,a}
+\frac{1}{3} \delta^{cd} \overline{G}_{cd} E'_{,a}
\nonumber\\
\delta G_{ab} &=& \left[-\Psi+\Phi-2(2{\cal H}'+{\cal H}^2)E\right]_{,ab}
\nonumber\\
&&+\left[\Delta(\Psi-\Phi)-2(\Psi^{\prime\prime}
+(2{\cal H}'+{\cal H}^2)(\Psi+\Phi)+{\cal H}(2\Psi+\Phi)')
+2({\cal H}^{\prime\prime}+{\cal H} {\cal H}')(B-E')\right]\delta_{ab}
\nonumber
\ea
where we used $\overline{G}_{00}=3{\cal
H}^2,\;\delta^{cd}\overline{G}_{cd}/3=-[2{\cal H}'+{\cal H}^2]$.

Hence, by (\ref{a.28}), we get \ba \label{a.37} \delta
\tilde{G}_{00} &=& \delta G_{00}+\overline{G}_{00}' (B-E')+
2\overline{G}_{00}
(B-E')'\nonumber\\
&=& \delta G_{00}+6{\cal H} \left[{\cal H}(B-E')\right]'
\nonumber\\
&=& 2\left(-\Delta\Psi+3{\cal H} \Psi'\right)
\nonumber\\
\delta \tilde{G}_{0a}
&=& \delta G_{0a}+\overline{G}_{00}'
\left[B-E'\right]_{,a}-\frac{1}{3}\delta^{cd} \overline{G}_{cd} E'_{,a}
\nonumber\\
&=& -2\left[\Psi'+{\cal H}\Phi\right]_{,a}
\nonumber\\
\delta \tilde{G}_{ab} &=& \delta G_{ab}+ (B-E') \overline{G}'_{ab}
-\frac{2}{3} \delta^{cd} E_{,ab} \overline{G}_{cd}
\nonumber\\
&=& \delta G_{ab}-2 \left(B-E'\right) \left({\cal H}^{\prime\prime}+{\cal
H}' {\cal H}\right) \delta_{ab}
+2\left(2{\cal H}'+2{\cal H}^2\right) E_{,ab}
\nonumber\\
&=& \left[-\Psi+\Phi\right]_{,ab}
-\left[\Delta(\Phi-\Psi)+2\left(\Psi^{\prime\prime} +\left(2{\cal
H}'+{\cal H}^2\right)(\Psi+\Phi)+{\cal
H}(2\Psi+\Phi)'\right)\right]\delta_{ab}. \ea The term $(2{\cal
H}'+{\cal H}^2)\Psi$ disappears in $\delta \tilde{G}^a_b$ of
\cite{6}, due to the variation of the additional metric contraction
involved, as one can explicitly check.

\section{Linear Perturbations following an Alternative Route}
\label{sb}

In the main text we computed the linear perturbations of the general
equations motion for our manifestly gauge-invariant configuration
observables, the metric $Q_{jk}$ and the scalar field $\Xi$. That
is, we used the second-order temporal derivative form of the
equations of motion, in which the canonical momenta were eliminated
by using the Hamiltonian equations of motion. This is rather
tedious, because one cannot use the special symmetries of the
background when eliminating the momenta, but has to assume a general
background. As shown in the appendix of our companion paper
\cite{1}, to linear order one can get those equations also by
perturbing the general Hamiltonian equations of motion and then
eliminating the perturbed momenta by using the perturbed equations
of motion. This is shorter because one can use the properties of the
background at an earlier stage in the computation. On the other
hand, the procedure followed in the main text quickly becomes more
economic for higher than linear order.

In this section we will carry out this
alternative derivation as a check of the result obtained in the main
text.

\subsection{Gauge invariant FRW equations}
\label{sb.1}

We will need the gauge invariant FRW equations in Hamiltonian form:\\
The metric takes the form $\overline{Q}_{jk}=A^2 \delta_{jk}$, where
$A$ is the {\it observable} scale factor. The corresponding
extrinsic curvature is given by \be \label{b.3}
\overline{K}_{jk}=\frac{1}{2\overline{N}}\left(\dot{\overline{Q}}_{jk}-{\cal
L}_{\vec{\overline{N}}} \overline{Q}_{jk}\right)=A\dot{A}
\delta_{jk}. \ee The momentum $\overline{P}^{jk}$ conjugate to
$\overline{Q}_{jk}$ is given by \be \label{b.4} \overline{P}^{jk}=
\sqrt{\det(\overline{Q})}\left(\overline{Q}^{jm}
\overline{Q}^{kn}-\overline{Q}^{jk}
\overline{Q}^{mn}\right)\overline{K}_{mn}=-2\dot{A}
\delta^{jk}=:I\delta^{jk}. \ee The background scalar fields
$\overline{\Xi}$, $\overline{\Pi}$ are simply spatially homogeneous
where $\overline{\Pi}$ is given by \be \label{b.4a}
\overline{\Pi}=\sqrt{\det(\overline{Q})}
\left(\dot{\overline{\Xi}}-\overline{N}^j \overline{\Xi}_{,j}\right)
=A^3\dot{\overline{\Xi}}. \ee
 The symplectic potential
is\footnote{This involves spatial averaging over the general
symplectic structure, or, alternatively, we only use the integrand.
The justification for this is that the corresponding equations of
motion derived via the Poisson brackets correctly reproduce the
connection between $\dot{A},\dot{\overline{\Xi}}$ and the momentum
conjugate to $A,\overline{\Xi}$ displayed in (\ref{b.4}) and
(\ref{b.4a}), respectively.} \be \label{b.5}
\Theta=\frac{1}{\kappa}\dot{\overline{Q}}_{jk} \overline{P}^{jk}
+\frac{1}{\lambda} \dot{\Xi}_0 \overline{\Pi}
=\frac{1}{\kappa}\dot{A}\; \left(6IA\right) +\frac{1}{\lambda}
\dot{\Xi}_0 \overline{\Pi}.
 \ee
Thus, with \be \label{b.6}
J:=6IA=-12A\dot{A},\;\;I=\frac{J}{6A}=-2\dot{A}, \ee we see that the
non-vanishing background Poisson brackets are \be \label{b.7}
\{J,A\}^-=\kappa,\;\;\{\overline{\Pi},\overline{\Xi}\}^-=\lambda.
\ee It follows that \be \label{b.8} \overline{C}_j=-2\overline{D}_k
\overline{P}^k_j +\overline{\Pi} \overline{D}_j\overline{\Xi}=0 \ee
and \ba \label{b.9} \kappa \overline{C}_{\rm geo} &=&
\frac{2}{\sqrt{\det(\overline{Q})}}\overline{G}_{jkmn}\overline{P}^{jk}
\overline{P}^{mn}-\sqrt{\det(Q)}R[\overline{Q}] +2\Lambda
\sqrt{\det(\overline{Q})}
\nonumber\\
&=&
-\frac{3}{2} A I^2 +2A^3 \Lambda\nonumber\\
&=&-\frac{J^2}{24A}+2A^3 \Lambda
\nonumber\\
\lambda C_{\rm matter} &=&
\frac{1}{2\lambda}\left[\frac{\overline{\Pi}^2}{\sqrt{\det(\overline{Q})}}+
\sqrt{\det(\overline{Q})}\left(\overline{Q}^{jk}\; \overline{\Xi}_{,j}\;
\overline{\Xi}_{,k}
+v(\overline{\Xi})\right)\right]
\nonumber\\
&=&\frac{1}{2\lambda}\left(\frac{\overline{\Pi}^2}{A^3}+
A^3v(\overline{\Xi})\right)\nonumber\\
&=:&\overline{\rho} A^3.
 \ea
Here we have introduced the background matter energy density
$\overline{\rho}$. Notice that
$\overline{H}=\overline{C}=\overline{C}_{\rm geo}
+\overline{C}_{\rm matter}$, so that $\overline{C}$ is also the
background Hamiltonian.

The Hamiltonian background equations of motion are
\ba \label{b.10}
\dot{\overline{\Xi}} &=& \lambda \frac{\partial \overline{C}}{\partial
\overline{\Pi}}
=\frac{\overline{\Pi}}{A^3}
\\
\dot{A} &=& \kappa \frac{\partial \overline{C}}{\partial J}
=-\frac{J}{12A}
\nonumber\\
\dot{\overline{\Pi}} &=& -\lambda
\frac{\partial \overline{C}}{\partial \overline{\Xi}}
=-\frac{1}{2}A^3 v'(\overline{\Xi})
\nonumber\\
\dot{J} &=& -\kappa \frac{\partial \overline{C}}{\partial A}=
-\Big[\Big(\frac{J^2}{24A^2}+6A^2\Lambda\Big)+\frac{3\kappa}{2\lambda}
(-\frac{\overline{\Pi}^2}{A^4}+A^2 v(\overline{\Xi}))\Big] =:
-\Big[\Big(\frac{J^2}{24A^2}+6A^2\Lambda\Big)-3A^2
\overline{p}\Big],\nonumber \ea where we have introduced the
background matter pressure $\overline{p}$.

The first two equations in (\ref{b.10}) correctly reproduce the
equations $\overline{\Pi}=A^3 \dot{\ov{\Xi}}$ and $J=-12 A\dot{A}$
or $I=-2\dot{A}$. Taking the second time derivatives of
$A,\overline{\Xi}$ and using the last two equations in (\ref{b.10}),
we find \ba \label{b.11} \ddot{\overline{\Xi}} &=& -\frac{1}{2}
v'(\overline{\Xi})-3\frac{\dot{A}}{A}\dot{\overline{\Xi}}
\nonumber\\
3 \frac{\ddot{A}}{A} &=&
\Lambda-\frac{\kappa}{4}\Big(\overline{\rho}+3\overline{p}
-\frac{\overline{\epsilon}}{A^3}\Big).
 \ea Here we have used the
conservation law that the total energy density
$\overline{H}=\overline{C}=:\overline{\epsilon}>0$ is a constant of
motion of the background Hamiltonian (it is {\it not} a constraint).
The conservation equation $C=\ov{\epsilon}$ can be solved for
$\dot{A}^2$ when using $J=-12 A\dot{A}$ and one finds \be
\label{b.12}
3\Big(\frac{\dot{A}}{A}\Big)^2=\Lambda+\frac{\kappa}{2}\Big(\overline{\rho}
-\frac{\ov{\epsilon}}{A^3}\Big), \ee which displays
$\overline{\rho}_{\rm dust}=-\overline{\epsilon}/A^3$ as background dust
energy density while $\overline{p}_{\rm dust}=0$, since the dust is
pressureless.

Equations (\ref{b.11}) and (\ref{b.12}) are the familiar FRW
equations. However, they now describe the physical evolution with
respect to the physical Hamiltonian $H$ of observable quantities,
rather than the gauge transformations of non-observables as is the
case in the usual formalism. This happens due to the
deparametrisation through the dust whose only fingerprint is in the
additional terms proportional to $\overline{\epsilon}$.

\subsection{Linear perturbations of the Hamiltonian equations of motion}
\label{sb.2}

We start from the general Hamiltonian equations of motion for our
physical degrees of freedom which we display once again below: \ba
\label{b.1} \dot{\Xi} &=&  \frac{N}{\sqrt{\det(Q)}}\Pi+{\cal
L}_{\vec{N}} \Xi
\nonumber\\
\dot{\Pi} &=&\left[N \sqrt{\det(Q)}\; Q^{jk} \Xi_{,k}\right]_{,j}
-\frac{N}{2}\sqrt{\det(Q)}\; v'(\Xi)
+ {\cal L}_{\vec{N}} \Pi
\nonumber\\
\dot{Q}_{jk} &=& \frac{2N}{\sqrt{\det(Q)}}\; G_{jkmn} \;
P^{mn}+ ({\cal L}_{\vec{N}} Q)_{jk}
\nonumber\\
\dot{P}^{jk} &=&
 N\Big[-\frac{Q_{mn}}{\sqrt{\det{Q}}}\Big(2 P^{jm} P^{kn}-P^{jk}
P^{mn}\Big)
+\frac{\kappa}{2}Q^{jk}\;C -
\Q\;Q^{jk}\big(2\Lambda +\frac{\kappa}{2\lambda}\big(\Xi^{,m}\Xi_{,m}
+v(\Xi)\big)\big)\Big]
\nonumber\\
&& +\Q [G^{-1}]^{jkmn}\Big((D_mD_n N)-NR_{mn}[Q]\Big)
+\frac{\kappa}{2\lambda}N\Q\;\Xi^{,j}\Xi^{,k}\nonumber\\
&&
-\frac{1}{2}H Q^{jm} Q^{kn} N_m N_n
 +({\cal L}_{\vec{N}} P)^{jk}
 \ea with $N=\sqrt{1+Q^{jk}N_j
N_k},\;N_j=-C_j/H,\;H=\sqrt{C^2-Q^{jk} C_j C_k}$. Notice that in our
convention $C=C_{\mathrm{geo}} +C_{\mathrm{matter}}$ and $C_{\rm geo}$ contains 
the cosmological constant term. The bimetric
$G_{jkmn}=Q_{j(m} Q_{n)k}-1/2Q_{jk} Q_{mn}$ has the inverse
$(G^{-1})^{jkmn}=[Q_{j(m} Q_{n)k}-Q_{jk} Q_{mn}]$, so that
$G_{jkpq}(G^{-1})^{pqmn}=\delta_{(j}^m \delta_{k)}^n$. In what
follows, we again adopt the notation of the main text and denote
background quantities by an overbar, while perturbations are denoted
by the symbol $\delta$. For instance, $\delta
Q_{jk}=Q_{jk}-\overline{Q}_{jk}$.

In perturbing equations (\ref{b.1}) around the FRW background we
make first the following observations which will drastically
simplify the subsequent analysis:
\begin{itemize}
\item[1.] In linear order the perturbation of the lapse vanishes
\be \label{b.2} \delta
N=\frac{1}{\overline{N}}\left(\overline{Q}^{jk} \overline{C}_j\delta
C_k+\frac{1}{2} \delta Q^{jk} \overline{C}_j
\overline{C}_k\right)=0, \ee since $C_j=-2D_k P^k_j+\Pi D_j\Xi$.
Hence $\overline{C}_j=0$ due to spatial homogeneity (in dust space
$\cal S$).
\item[2.] The same holds for the second covariant derivatives of the
lapse \be \label{b.2a} \delta D_j D_k N= \delta \left(\partial_j
\partial_k N-\Gamma_{jk}^m \partial_m N\right) =\partial_j
\partial_k \delta N-\overline{\Gamma}_{jk}^m \partial_m \delta
N+\delta \Gamma_{jk}^m \partial_m \overline{N}=0, \ee since both
$\overline{\Gamma}^m_{jk}=0,\;D_m\overline{N}=0$ due to spatial
homogeneity.
\item[3.] The terms quadratic in spatial derivatives of the scalar field
have vanishing variation because at least
one of the spatial derivatives is not varied and then vanishes.
\item[4.] In the variation of the Ricci curvature terms only the term
involving the variation of the Ricci curvature $R_{jk}$, which is
linear in the Christoffel symbol, survives.
\item[5.] As shown in the appendix of \cite{1}, the linear perturbations
of the conserved quantities $H(\sigma),\;C_j(\sigma)$ and thus
$N_j=-C_j/H$ are conserved with respect to the linear equations of
motion which are generated by the second order term of the perturbation
of the physical Hamiltonian $\HF$.
\end{itemize}
It follows that we can set everywhere $N=\overline{N}=1$. Notice
however, that, while $\delta N\equiv 0,\;\overline{N}_j\equiv 0$, we
have $\delta N_j\not=0$.

With these preparations out of the way, we can now perturb
(\ref{b.1}). In order to keep the formulae simple at intermediate
steps we define $\delta Q_{jk}=L\delta_{jk},\;\delta
P^{jk}=I\delta^{jk}$. Later we will substitute
$L=A^2,\;I=-2\dot{A}$. Remember that $P^{jk},\;\Pi$ are densities of
weight one, so that ($N^j=Q^{jk} N_k$) \be \label{b.13} {\cal
L}_{\vec{N}} P^{jk}=\partial_m (N^m P^{jk})-2 N^{(j}_{,m}
P^{k)m},\;\; {\cal L}_{\vec{N}} \Pi=\partial_m(N^m \Pi). \ee Since
$\overline{N}_j=0$, we have \ba \label{b.14} \delta {\cal
L}_{\vec{N}} Q_{jk} &=& {\cal L}_{\delta \vec{N}}
\overline{Q}_{jk}=2\delta N_{(j,k)}
\nonumber\\
\delta {\cal L}_{\vec{N}} \overline{\Xi} &=& 0
\nonumber\\
\delta {\cal L}_{\vec{N}} P^{jk} &=&
I \left(\left[\delta N^m\right]_{,m} \delta^{jk} -2\left[\delta
N^{(j}\right]_{,m}
\delta^{k)m}\right)
= \frac{I}{L} \left(\delta^{mn} \delta^{jk} -2(\delta^{m(j}
\delta^{k)n}\right) [\delta N_m]_{,n}
\nonumber\\
&=& -\frac{2I}{L} \overline{G}_{jkmn} \left[\delta N_m\right]_{,n}
\nonumber\\
\delta {\cal L}_{\vec{N}} \Pi &=& \partial_m\left(\delta N^m
\overline{\Pi}\right)=\frac{\overline{\Pi}}{L}\left[\delta
N_m\right]_{,m}. \ea Here we have introduced the flat bimetric
$\overline{G}_{jkmn}
=\delta_{j(m}\delta_{n)k}-1/2\delta_{jk}\delta_{mn}$ with inverse
$\overline{G}^{-1}_{jkmn}
=\delta_{j(m}\delta_{n)k}-\delta_{jk}\delta_{mn}$, so that
$\overline{G}_{jkpq}\overline{G}_{pqmn}=\delta_{j(m}\delta_{n)k}$.
Also we will use Einstein summation convention in what follows,
irrespective of index position on Kronecker $\delta$'s, that is, we
define $\overline{G}^{jkmn}:=\overline{G}_{jkmn}$ and
$[\overline{G}^{-1}]^{jkmn}:=[\overline{G}^{-1}]_{jkmn}$, as well as
$\delta^{jk}=\delta_{jk}$. However, notice that our fundamental
perturbation quantities are $Q_{jk},P^{jk},N_j$ with {\it that}
index position, and one has to take care of the additional metric
contractions involved when the index is a priori not in that
position. For instance, \be \label{b.15} \delta
Q^{jk}=-\left[\overline{Q}^{jm}\overline{Q}^{nk}\right]\delta
Q_{mn}= -\frac{1}{L^2} \delta Q_{jk}. \ee We also often need \ba
\label{b.16} \delta \sqrt{\det(Q)} &=&\left[ \frac{1}{2}
\sqrt{\det(\overline{Q})} \overline{Q}^{mn}\right]\delta
Q_{mn}=\frac{\sqrt{L}}{2} \delta Q_{mm}
\nonumber\\
\delta \frac{1}{\sqrt{\det(Q)}} &=& -\frac{1}{L^{5/2}} \delta
Q_{jj}. \ea Then we find after straightforward, but tedious
calculations (simply using the product rule for linear variations
all the time) \ba \label{b.17} \delta \dot{\Xi} &=& \frac{\delta
\Pi}{L^{3/2}}-\frac{\overline{\Pi}}{2L^{5/2}} \delta Q_{mm}
\nonumber\\
\delta \dot{\Pi} &=&
\frac{1}{2}\left(-\frac{L^{1/2}}{2}v'(\overline{\Xi})\delta
Q_{mm}-L^{3/2}v^{\prime\prime}(\overline{\Xi})\delta\Xi
+2L^{1/2}\Delta\delta\Xi\right)
+\frac{\overline{\Pi}}{L}[\delta N_m]_{,m}
\nonumber\\
\delta \dot{Q}_{jk} &=& \frac{2}{L^{1/2}} \overline{G}_{jkmn}\left(L \delta
P^{mn}
+\frac{I}{2} \delta Q_{mn}\right)+2\left[\delta N_{(j}\right]_{,k)}
\nonumber\\
\delta \dot{P}^{jk} &=& -\frac{2I}{L}\overline{G}_{jkmn} \left[\delta
N_{(m}\right]_{,n)}
-\frac{I}{L^{1/2}} \overline{G}_{jkmn} \delta P^{mn}
-\frac{1}{L^{1/2}} \overline{G}_{jkmn}\delta R_{mn}
\nonumber\\
&& +\frac{\delta_{jk}}{L^{1/2}}
\left[\frac{3}{8}\frac{I^2}{L}-\frac{\Lambda}{2}
-\frac{\kappa}{4}\overline{\rho}\right]
\delta Q_{mm}
+\frac{1}{L^{1/2}}
\left[-\frac{5}{4}\frac{I^2}{L}+\Lambda-\frac{\kappa}{2} \overline{p}\right]
\delta Q_{jk}
\nonumber\\
&& +\delta_{jk}
\frac{\kappa}{4\lambda}\left(2\frac{\overline{\Pi}}{L^{5/2}}
\delta\Pi- L^{1/2} v'(\overline{\Xi}) \delta\Xi\right).
 \ea Here
$\Delta=\delta^{mn}\partial_m\partial_n$ is the flat Laplacian and
all derivatives appearing are flat derivatives. As in the previous
subsection, we used
$\overline{\rho}=[\overline{\Pi}^2/L^3+v(\overline{\Xi})]/(2\lambda)$
and
$\overline{p}=[\overline{\Pi}^2/L^3-v(\overline{\Xi})]/(2\lambda)$.
The variation of the Ricci tensor is given explicitly by \be
\label{b.18} \delta R_{jk}=\frac{1}{2L}\left(2\delta
Q_{m(j,k)m}-\Delta \delta Q_{jk}-\delta Q_{mm,jk}\right). \ee

\subsection{Second time derivative form of the perturbed equations of
motion}
\label{sb.3}

The explicit inversion of the first and third relation in (\ref{b.17})
for the perturbed momenta in terms of the velocities
yields
\ba \label{b.19}
\delta \Pi &=& L^{3/2} \delta\dot{\Xi}+\frac{\overline{\Pi}}{2L}\delta
Q_{mm}
\nonumber\\
\delta P^{jk} &=& \frac{1}{2L^{1/2}}
[\overline{G}^{-1}]_{jkmn}\Big(\delta \dot{Q}_{mn}-2[\delta
N_{(m}]_{,n)}\Big)-\frac{I}{2L}\delta Q_{jk}.
 \ea Taking the second
derivative of the first and third relation in (\ref{b.17}), we
substitute for $\delta\dot{\Pi},\;\delta \dot{P}^{jk}$, using the
second and fourth relation in (\ref{b.17}), respectively. Afterwards
we substitute for $\delta \Pi,\;\delta P^{jk}$, using (\ref{b.19}).
In doing that, one has to remember that $d/d\tau \delta N_j=0$, i.e.
$\delta N_j$ is a constant of motion.

We find again after some tedious algebra\footnote{One should really
write $d^2/(d\tau)^2 \delta Q_{jk}$, rather than $\delta
\ddot{Q}_{jk}$ etc. However, these two quantities are numerically
identical.} \ba \label{b.20} \delta \ddot{\Xi} &=&
-\frac{3\dot{L}}{2L}\delta\dot{\Xi}-\frac{1}{2} v^{\prime\prime}
\delta\Xi+\frac{1}{L}\Delta \delta\Xi
+\left[\frac{\dot{L}\overline{\Pi}}{2L^{7/2}}-\frac{\dot{\overline{\Pi}}}{2
L^{5/2}}-\frac{v'(\overline{\Xi})}{4L}\right]\delta Q_{mm}
 -\frac{\overline{\Pi}}{2 L^{5/2}}\delta \dot{Q}_{mm}
+\frac{\overline{\Pi}}{L^{5/2}}
[\delta N_m]_{,m}
\nonumber\\
\delta \ddot{Q}_{jk} &=& \frac{\dot{L}}{2L}\Big(\delta \dot{Q}_{jk}-2[\delta
N_{(j}]_{,j)}\Big)
+\frac{1}{L^{1/2}} \overline{G}_{jkmn}\Big(\dot{I}\delta Q_{mn}+I\delta
\dot{Q}_{mn}-I\frac{\dot{L}}{L}\delta{Q}_{mn}\Big)
+ 2L^{1/2} \overline{G}_{jkmn}\delta \dot{P}^{mn}.\nonumber\\
\ea It is a good check to verify that the dimensionalities of the
various terms match: in our convention, the spatial coordinates
$\sigma^j$ are chosen to be dimensionfree, while $\tau$ has
dimension of length. Thus $L$ has dimension cm$^2$ and $A$ has
dimension cm$^1$. The scalar field $\Xi$ is dimensionfree, hence
$\Pi\propto A^3\dot{\Xi}$ has dimension cm$^2$. The potential term
$v(\overline{\Xi})$ has the same dimension as $\Pi^2/L^3$ which is
cm$^{-2}$. Likewise, $Q_{jk}$ has dimension cm$^2$ while
$K_{jk}\propto \dot{Q}_{jk}/N$ has dimension cm$^1$ so that
$P^{jk}\propto \sqrt{\det(Q)} Q^{jk} Q^{mn} K_{mn}$ is actually
dimensionless, just like $I=-2\dot{A}$.

In the last line of (\ref{b.20}) we still must insert the last
relation of (\ref{b.17}). Since there are various bimetric
contractions involved, we notice the identities \be \label{b.21}
\overline{G}_{jkpq}
\overline{G}_{pqmn}=\delta_{j(m}\delta_{n)k}-\frac{1}{4}\delta_{jk}\delta_{mn},
\; G_{jkmn} \delta_{mn}=-\frac{1}{2}\delta_{jk}. \ee The calculation
is very tedious but the result is rather simple. We directly
substitute $L=A^2$ and $I=-2\dot{A}$ and find \ba \label{b.22}
\delta \ddot{Q}_{jk} &=& \frac{\dot{A}}{A} \delta \dot{Q}_{jk}+2
\frac{\dot{A}}{A} [\delta
N_{(j}]_{k)}+2\left[-\Big(\frac{\dot{A}}{A}\Big)^2
-\frac{\ddot{A}}{A}+\big(\Lambda-\frac{\kappa}{2}\overline{p}\big)\right]
\delta Q_{jk}-2\delta R_{jk}
\nonumber\\
&&+\delta_{jk}\left(\Big[\frac{\ddot{A}}{A}
+\frac{1}{2}\Big(\frac{\dot{A}}{A}\Big)^2
-\frac{1}{2}\Lambda+\frac{\kappa}{4}
\overline{p}\Big]\delta Q_{mm} +\frac{1}{2} \delta R_{mm}
-\frac{\kappa}{2\lambda}
A^2\big(\dot{\overline{\Xi}}\delta\dot{\Xi}-\frac{1}{2}v'(\overline{\Xi})\delta
\Xi\big)\right).
 \ea We can simplify this expression further by
making use of the background equations (\ref{b.11}) and (\ref{b.12})
which imply \be \label{b.23}
2\frac{\ddot{A}}{A}+\Big(\frac{\dot{A}}{A}\Big)^2=\Lambda
-\frac{\kappa}{2} \overline{p}. \ee Therefore the $\delta Q_{mm}$
term in (\ref{b.22}) vanishes and the $\delta Q_{jk}$ term
simplifies, leading to \ba \label{b.24} \delta \ddot{Q}_{jk} &=&
\frac{\dot{A}}{A} \delta \dot{Q}_{jk} +2 \frac{\dot{A}}{A}[\delta
N_{(j}]_{k)}+2 \frac{\ddot{A}}{A} \delta Q_{jk} -2\delta R_{jk}
+\delta_{jk}\left(\frac{1}{2} \delta R_{mm} -\frac{\kappa}{2\lambda}
A^2\big(\dot{\overline{\Xi}}\delta\dot{\Xi}-\frac{1}{2}v'(\overline{\Xi}
\big)\delta \Xi)\right). \ea
~\\
Equations (\ref{b.20}) and (\ref{b.23}) are precisely equations
(\ref{PerXifinal}) and (\ref{PerQfinal}), derived by the more
general formalism of section \ref{s2}.

\end{appendix}

\end{document}